\documentclass[journal]{IEEEtran}

\usepackage{amsmath,amssymb} %for \therefore,\because,\lesssim,\gtrsim
\usepackage{graphicx}
\usepackage{bm}

\newtheorem{definition}{Definition}
\newtheorem{proposition}{Proposition} %\proof
\newtheorem{lemma}{Lemma}

\newtheorem{assumption}{Assumption}
\newtheorem{example}{Example}
\newcommand{\qed}{\hfill $\square$ \par}

\newcommand{\sgn}{\,\mbox{\rm sgn}\,}

\newcommand{\llangle}{\langle \kern -.23em \langle}
\newcommand{\rrangle}{\rangle \kern -.23em \rangle}

\newcommand{\rmi}{\mathrm{i}}
\newcommand{\rmd}{\mathrm{d}}
\newcommand{\rmD}{\mathrm{D}}
\newcommand{\rme}{\mathrm{e}}
\renewcommand{\vec}[1]{\boldsymbol{#1}}

\renewcommand{\baselinestretch}{1}
\usepackage{color}
\newcommand{\HLS}{\color{black}}
\newcommand{\HLE}{\color{black}}

\begin{document}

\title{Generating Functional Analysis for Iterative CDMA Multiuser Detectors}
\author{
  Kazushi~Mimura,~\IEEEmembership{Member,~IEEE,}
  and Masato~Okada 
  %Masato~Okada,~\IEEEmembership{Member,~IEEE}% <-this % stops a space
  \thanks{
    K. Mimura is with Faculty of Information Sciences, 
    Hiroshima City University, Hiroshima 731-3194, Japan. 
    e-mail: {\sf mimura@hiroshima-cu.ac.jp}.
  }
  \thanks{
    M. Okada is with 
    Graduate School of Frontier Sciences, University of Tokyo, Chiba 277-5861, Japan 
    and Brain Science Institute, RIKEN. 
    e-mail: {\sf okada@k.u-tokyo.ac.jp}. 
  }
  \thanks{
    This paper was presented in part at the IEEE International Symposium on Information Theory, Seattle, US and Nice, France." 
    Manuscript received Month Day, 2010; revised Month Day, 2010. 
  }
}
\markboth{Journal of IEEE xxx xxx,~Vol.~xx, No.~xx, August~2010}%
{Shell \MakeLowercase{\textit{et al.}}: Bare Demo of IEEEtran.cls for Journals}
%\IEEEpubid{0000--0000/00\$00.00~\copyright~2007 IEEE}
%\IEEEspecialpapernotice{(Invited Paper)}

%=====================================================================
\maketitle

%\tableofcontents \newpage

\begin{abstract}
  \par
  We investigate the detection dynamics of a soft parallel interference canceller (soft-PIC), 
  which includes a hard-PIC as a special case, for code-division multiple-access (CDMA) multiuser detection, 
  applied to a randomly spread, fully synchronous base-band uncoded CDMA channel model 
  with additive white Gaussian noise under perfect power control in the large-system limit. 
  We analyze the detection dynamics of some iterative detectors, namely soft-PIC, 
  the Onsager-reaction-cancelling parallel interference canceller (ORC-PIC) and the belief-propagation-based detector (BP-based detector), 
  by the generating functional analysis (GFA). 
  The GFA allows us to study the asymptotic behavior of the dynamics in the infinitely large system 
  without assuming the independence of messages. 
  We study the detection dynamics and the stationary estimates of an iterative algorithm. 
  \par
  We also show the decoupling principle in iterative multiuser detection algorithms 
  in the large-system limit. 
  For a generic iterative multiuser detection algorithm with binary input, 
  it is shown that the multiuser channel is equivalent to a bank of independent single-user additive non-Gaussian channels, 
  whose signal-to-noise ratio degrades due to both the multiple-access interference and the Onsager reaction, 
  at each stage of the algorithm. 
  If an algorithm cancels the Onsager reaction, 
  the equivalent single-user channels coincide with an additive white Gaussian noise channel. 
  We also discuss ORC-PIC and the BP-based detector. 
  %the equivalent single-user channels are given as an additive white Gaussian noise channel 
  %because these algorithms can cancel the Onsager reaction. 
\end{abstract}


\IEEEpeerreviewmaketitle

\section{Introduction}
%~~~~~~~~~~~~~~~~~~~~~~~~~~~~~~~~~~~~~~~~~~~~~~~~~~~~~~~~~~~~~~~~~~~~~
\par
\IEEEPARstart{D}{etection dynamics} of generic iterative 
%CDMA %DELETED.v3
\HLS code-division multiple-access (CDMA) \HLE %ADDED.v3
multiuser detectors, which utilize soft-decision in the large system limit, 
is presented in this paper. 
The CDMA is a digital modulation system that employs spreading codes to enable access 
to a mobile communication system by multiple users \cite{Varanashi1990, Verdu1998}. 
The statistical-mechanical approach has been applied 
to evaluate the performance of various wireless communication systems \cite{Tanaka2002}. 
This kind of systems is widely used in communications and signal processing, 
such as the code-division multiple-access, 
the multiple-input multiple-output channels \cite{Mueller2003, Guo2005}, 
and compressed sensing \cite{Claerbout1973, Santosa1986, Donoho1989, Donoho2006}. 
%In such a CDMA system, since many users share a common medium to communicate a single receiver, 
%the transmitted signal of each user is degraded by the other user's signal. 
\par
%In a multipoint-to-point communication framework, %DELETED
%CDMA allows several users to share a single communication channel to a base station. %DELETED
%Each user first modulates one's own information sequence using the spreading code assigned to the user, %DELETED
%and then the modulated sequence is transmitted to the base station. %DELETED
%The base station receives a mixture of the transmitted signals and additional channel noise. %DELETED
%A demodulator at the base station extracts the original information sequence %DELETED
%from the received noise-degraded mixture signal by using the users' spreading codes. %DELETED
%This process is called detection. %DELETED
\par
Various types of multiuser detectors utilizing soft-decision 
have been proposed so far \cite{Verdu1998, Varanashi1990, Divsalar1998, Kabashima2003}. 
Tanaka has first evaluated the properties of the maximum a posteriori detector 
and the marginal-posterior-mode detector 
% the minimum mean-squared error (MMSE) detector 
by the statistical-mechanical analysis, 
which is called the {\it replica analysis} (or the {\it replica method}) \cite{Tanaka2001, Tanaka2002, Nishimori2002}. 
The replica analysis has widely been applied 
to analyze communication systems \cite{Guo2005, Guo2007, Mueller2008} and other information theoretic problems. 
It can treat the properties of detection results, 
but cannot directly treat the detection dynamics of detectors. 
Since the optimal marginal-posterior-mode detector itself is known to be NP-hard \cite{Verdu1998}, 
it is important to construct suboptimal methods. 
Iterative algorithms are generally useful as such methods from the viewpoint of computational cost. 
Various kinds of iterative detection algorithms have been developed 
to date \cite{Nelson1996, Divsalar1998, Kabashima2003, Bickson2008, Rasmussen2008, Truhachev2009, Honig2009}. 
The analysis of iterative multiuser detection algorithms is therefore expected 
to play an important role in developments and improvements of detectors. 
\par
Recently, the state evolution to evaluate the dynamics of the approximate belief propagation has been proposed 
by Bayati and Montanari \cite{Bayati2010}. 
The detection dynamics of iterative algorithms which are characterized by a dense graph 
has attracted a great deal of attention from theoretical and practical viewpoints up to now \cite{Kabashima2003, Tanaka2005, Bayati2010}. 
Kabashima has proposed the belief-propagation-based detector and analyzed its performance \cite{Kabashima2003}. 
Tanaka and Okada have analyzed the detection dynamics \cite{Tanaka2005} 
of the {\it soft parallel interference canceller \HLS (soft-PIC) \HLE %ADDED.v3
} 
proposed by Divsalar et al \cite{Divsalar1998} by means of a dynamical theory for the Hopfield model \cite{Okada1995}. 
Bayati and Montanari have analyzed the reconstruction dynamics of approximate belief propagation algorithm for compressed sensing \cite{Bayati2010}. 
These existing studies \cite{Kabashima2003, Tanaka2005, Bayati2010} have succeeded 
in analyzing of various kinds of iterative algorithms including belief-propagation-based methods. 
\par
%However, these are justified only in the case where the correlation %DELETED
%between present estimates and past values can be neglected as well as density evolution \cite{Richardson2001, Tanaka2005}. %DELETED
However, these analyses, such as density evolution \cite{Richardson2001, Tanaka2005} and state evolution \cite{Bayati2010}, are justified %ADDED
only for the cases that the correlation between present estimates and their past values can be neglected. %ADDED
In other words, these can only be applied to the case where there is not a retarded self-interaction, 
which is caused by iterations and this is often called the Onsager reaction, 
and their predictions systematically deviate from computer simulation results in general. 
%When we apply these analysis, which cannot treat the retarded self-interaction, 
\par
We have already applied the {\it generating functional analysis}
\HLS {\it (GFA)} \HLE %ADDED.v3
\cite{DeDominicis1978} to the hard-PIC, i.e., 
%Varanasi's conventional PIC %DELETED.v3
\HLS Varanasi and Aazhang's conventional PIC \HLE %ADDED.v3
\cite{Mimura2005, Mimura2006} so far. 
This analysis can however treat the hard-decision only \cite{Mimura2005, Mimura2006}. 
The GFA, which uses the saddle-point method \cite{Copson1965, Merhav2009}, 
allows us to study the asymptotic behavior of the dynamics 
in the infinitely large system \cite{DeDominicis1978, Coolen2000, Coolen2005}. 
Since it is not based on the S/N analysis, 
% like density evolution based on statistical neurodynamics \cite{Tanaka2005}, %DELETED
it does not therefore need the Gaussian assumption of the noise part. 
In the S/N analysis, the signal part, that contains the user's information being estimated, is separated from the remaining noise part; 
besides the noise part is generally assumed to follow a given distribution such as the Gaussian distribution. %ADDED
%The saddle-point method is applied in GFA. 
%The GFA predictions are in good agreement with the computer simulation results %DELETED
%for any system loads and channel noise levels \cite{Mimura2005, Mimura2006, Mimura2007}. %DELETED
\par
In this paper, we investigate the detection dynamics of some iterative algorithms for CDMA multiuser detection, 
applied to a randomly spread, fully synchronous base-band uncoded CDMA channel model 
with additive white Gaussian noise under perfect power control. 
We here treat soft-PIC, 
the {\it Onsager-reaction-cancelling parallel interference canceller \HLS (ORC-PIC) \HLE %ADDED.v3
}, 
which is an analogue of soft-PIC, and the {\it belief-propagation-based detector \HLS (BP-based detector) \HLE %ADDED.v3
}. 
These models have the retarded self-interaction. 
To confirm the validity of our analysis, we have performed computer simulations 
under some typical system loads and channel noise conditions. 
%Especially, it is important to treat a BP-based algorithm for various applications, 
%e.g., low-density parity-check (LDPC) codes. 
\par
This paper is organized as follows. 
The next section explains the system model. 
Sections \ref{section:algorithm} and \ref{section:GFA} introduce some multiuser detection algorithms and the generating functional analysis, respectively. 
Sections \ref{section:GFA_softPIC} -- \ref{section:GFA_BP} present analyses of soft-PIC, ORC-PIC, and the BP-based detector. 
Section \ref{section:stationary_estimates} explains the stationary states of iterative algorithms and their stability. 
In section \ref{section:decouple}, we discuss the decoupling principle \cite{Guo2005, Guo2007} in iterative detectors. 
The final section is devoted to a summary.

\section{System Model and Notations \label{section:systemmodel}}
%~~~~~~~~~~~~~~~~~~~~~~~~~~~~~~~~~~~~~~~~~~~~~~~~~~~~~~~~~~~~~~~~~~~~~
\par
Let us focus on the basic fully synchronous $K$-user 
baseband direct-sequence / binary phase-shift-keying CDMA channel model 
with perfect power control as 
\begin{equation}
  y^\mu := \frac1{\sqrt{N}}\sum_{k=1}^K s_k^\mu b_k + \sigma_0 n^\mu,
  \label{eq:SystemModel}
\end{equation}
where $y^\mu$ is the received signal at chip interval $\mu\in\{ 1,\cdots,N\}$, 
and where $b_k\in\{-1,1\}$ and $s_k^\mu\in\{-1,1\}$ are the binary phase-shift-keying-modulated information bit 
and the spreading code of user $k\in\{1,\cdots,K\}$ at chip interval $\mu$, respectively. 
The Gaussian random variable $\sigma_0 n^\mu$, where $n^\mu \sim \mathcal{N}(0,1)$, 
represents channel noise whose variance is $\sigma_0^2$. 
The spreading codes are independently generated 
from the identical symmetric distribution $P(s_k^\mu=1)=P(s_k^\mu=-1)=1/2$. 
The factor $1/\sqrt{N}$ is introduced in order to normalize the power per symbol to 1. 
The signal-to-noise ratio is obtained as $E_b/N_0=1/(2\sigma_0^2)$ by using these normalizations. 
%the energy per bit per noise power spectral density $E_b/N_0$ 
The ratio $\beta := K/N$ is called system load. 
\par
In this paper, 
the letters $k, k'$ denote indices in $\{1,\cdots,K\}$ and 
the letters $\mu, \mu'$ denote indices in $\{1,\cdots,N\}$. 
The $(k,k')$ element of the matrix $\vec{W}$, whose index is a pair of the user numbers, is indicated as $W_{kk'}$. 
The elements of vectors $\vec{y}=(y_1,\cdots,y_N)$ and $\vec{b}=(b_1,\cdots,b_K)$, 
whose indices are the chip interval index or the user number, are indicated as $y^\mu$ and $b_k$, respectively. 
The $s$-th element of vector $\vec{x}=(x^{(-1)}, x^{(0)}, \cdots, x^{(t)})^\top$, whose index is the stage number, is indicated by $(\vec{x})^{(s)}$, 
e.g., $(\vec{x})^{(1)}=x^{(1)}$ (Note that $(\vec{x})^{(1)}$ does not mean $x^{(-1)}$). 
Here, $\vec{X}^\top$ denotes the transpose of $\vec{X}$. 
The $(s,s')$ element of the matrix 
\begin{align*}
  \vec{X}=\left(
  \begin{array}{cccc}
    X^{(-1,-1)} & X^{(-1,0)} & \cdots & X^{(-1,t)} \\
    X^{( 0,-1)} & X^{( 0,0)} & \cdots & X^{( 0,t)} \\
    \vdots      & \vdots     &        & \vdots     \\
    X^{( t,-1)} & X^{( t,0)} & \cdots & X^{( t,t)} \\
  \end{array}
  \right), 
\end{align*}
whose index $(s,s')$ is a pair of the stage numbers, is indicated as $(\vec{X})^{(s,s')}$, 
e.g., $(\vec{X})^{(1,1)}=X^{(1,1)}$ (Note that $(\vec{X})^{(1,1)}$ does not mean $X^{(-1,-1)}$). 
The notations are summarized in Appendix \ref{app:notations}.

\section{Detection Algorithms \label{section:algorithm}}
%~~~~~~~~~~~~~~~~~~~~~~~~~~~~~~~~~~~~~~~~~~~~~~~~~~~~~~~~~~~~~~~~~~~~~
\par
We discuss the detection dynamics of the following three kinds of iterative detection algorithms in this paper.

\subsection{Soft-PIC}
%~~~~~~~~~~~~~~~~~~~~~~~~~~~~~~~~~~~~~~~~~~~~~~~~~~~~~~~~~~~~~~~~~~~~~
\par
The soft parallel interference canceller has been proposed by Divsalar et al \cite{Divsalar1998}. 
Kaiser and Hagenauer have also proposed a similar algorithm \cite{Kaiser1997}. 
%-------------------------------------------------DEFINITION
\begin{definition}
  \label{def:soft-PIC}
  (Soft-PIC) The updating rule for tentative decision $\tilde{b}_k^{(t)}\in\mathbb{R}$ 
  of bit signal $b_k$ at stage $t$ is 
  \begin{equation}
    \tilde{b}_k^{(t)} = f \biggl( h_k - \sum_{k' =1, \ne k}^K W_{kk'}\tilde{b}_{k'}^{(t-1)} \biggr),
    \label{eq:deterministic_updating_rule}
  \end{equation}
  where $f:\mathbb{R}\to\mathbb{R}$, which is called a transfer function, that is arbitrary 
  and $h_k$ is the output of the matched filter for user $k$: 
  \begin{equation}
    h_k := \frac1{\sqrt{N}} \sum_{\mu=1}^N s_k^\mu y^\mu, 
    \label{eq:h_k}
  \end{equation}
  and $W_{kk'}$ is the $kk'$-element of sample correlation matrix $\vec{W}$ of the spreading code: 
  \begin{equation}
    W_{kk'} := \frac 1N \sum_{\mu=1}^N s_k^\mu s_{k'}^\mu. 
  \end{equation}
  The initial condition of iteration is $\tilde{b}_k^{(-1)}=0$. 
  When the transfer function chooses $f(x)=\tanh(x/\sigma^2)$, 
  this iterative detection algorithm is called soft-PIC. 
  Here, $\sigma^2$ is a control parameter representing the detector's estimate of channel noise variance. 
  \qed
\end{definition}
\par
M\"uller and Huber have improved Soft-PIC and have numerically evaluated its performance \cite{Mueller1998}.

\subsection{ORC-PIC}
%~~~~~~~~~~~~~~~~~~~~~~~~~~~~~~~~~~~~~~~~~~~~~~~~~~~~~~~~~~~~~~~~~~~~~
\par
In soft-PIC, matched filter output has a very complex correlation between all estimates. 
The correlation due to iterative calculation worsens performance of detection. 
The ORC-PIC is an analogue of soft-PIC, which has a term to cancel such correlation. 
The updating rule is modified to 
\begin{eqnarray}
  & & \tilde{b}_k^{(t)} = f \biggl( h_k - \sum_{k' \ne k} W_{kk'}\tilde{b}_{k'}^{(t-1)} \nonumber \\
  & & \qquad - \hat{\Gamma}_k^{(t,t-1)} \tilde{b}_k^{(t-1)} - \cdots - \hat{\Gamma}_k^{(t,-1)} \tilde{b}_k^{(-1)} \biggr), 
  \label{eq:def_ORC-PIC}
\end{eqnarray}
with $\tilde{b}_k^{(-1)}=0$. 
If we properly choose coefficients $\{ \hat{\Gamma}_k^{(s,s')} \}$ 
of the term $- \hat{\Gamma}_k^{(t,t-1)} \tilde{b}_k^{(t-1)} - \cdots - \hat{\Gamma}_k^{(t,-1)} \tilde{b}_k^{(-1)}$, 
the correlation can be cancelled. 
Tanaka and Okada derived the preceding parameter $\{ \hat{\Gamma}_k^{(s,s')} \}$ 
by applying density evolution \cite{Tanaka2005} based on the statistical neurodynamics \cite{Okada1995}. 
%-------------------------------------------------DEFINITION
\begin{definition}
  The ORC-PIC is defined by updating rule (\ref{eq:def_ORC-PIC}) with 
  \begin{align}
    (\hat{\vec{\Gamma}}_k \tilde{\vec{b}}_k)^{(t)} = \beta G^{(t,t-1)} [ \tilde{b}_k^{(t-1)} - (\hat{\vec{\Gamma}}_k \tilde{\vec{b}}_k)^{(t-1)} ],
    \label{eq:GammaHatRecursive}
  \end{align}
  and $(\hat{\vec{\Gamma}}_k \tilde{\vec{b}}_k)^{(-1)} = (\hat{\vec{\Gamma}}_k \tilde{\vec{b}}_k)^{(0)} = 0$. 
  Here, 
  $\tilde{\vec{b}}_k^{(t)}$ $=$ $(\tilde{b}_k^{(-1)},$ $\cdots,$ $\tilde{b}_k^{(t)})^\top$, 
  $\hat{\vec{\Gamma}}_k$ is a $(t+1) \times (t+1)$ matrix whose $(s,s')$ element is $\hat{\Gamma}_k^{(s,s')}$, 
  and $G^{(t,t-1)}$ denotes the average single-user response function. 
  \qed
\end{definition}
Detail of the parameter $G^{(t,t-1)}$ is introduced later (in Section \ref{sec:SPEandMOP}). 
Note that $G^{(t,t-1)}$ does not depend on the user index $k$.

\subsection{BP-based Detector}
%~~~~~~~~~~~~~~~~~~~~~~~~~~~~~~~~~~~~~~~~~~~~~~~~~~~~~~~~~~~~~~~~~~~~~
\par
Assuming that information bits $\vec{b}$ are independently generated from the symmetric distribution, 
the posterior distribution from received signals $\vec{y}$ is given as 
$p(\vec{b}|\vec{y})=p(\vec{y}|\vec{b})/\sum_{\vec{b}\in\{\pm 1\}^K} p(\vec{y}|\vec{b})$, where %MODIFIED
\begin{align}
  p(\vec{y}|\vec{b}) = \prod_{\mu=1}^N \frac 1{\sqrt{2\pi}\sigma} %MODIFIED
  \exp \biggl[ -\frac 1{2\sigma^2} \biggl( y^\mu \!\! - \frac1{\sqrt{N}} \sum_{k=1}^K s_k^\mu b_k \biggr)^2 \biggr], 
\end{align}
and $\sigma^2$ is a control parameter when true noise level parameter $\sigma_0^2$ is not known. 
The marginal-posterior-mode detector \cite{Tanaka2002} is represented by 
\begin{align}
  \tilde{b}_k = \mathop{\rm argmax}_{b_k \in \{-1,1\}} 
  \left( \sum_{\vec{b} \backslash b_k \in \{-1,1\}^{K-1}} p(\vec{b}|\vec{y}) \right). 
  \label{eq:MPM}
\end{align}
The BP-based detector is an iterative algorithm that employs the belief propagation 
to approximately calculate the posterior marginal included in (\ref{eq:MPM}). 
%-------------------------------------------------DEFINITION
\begin{definition} \label{def:BP-based_detector}
  (BP-based detector) 
  The BP-based detector \cite{Kabashima2003} is given by the following iterative equations: 
  \begin{align}
    \tilde{\vec{b}}^{(t+1)} &= \tanh ( R^{(t)} \vec{h} - \vec{U}^{(t)} + A^{(t)} \tilde{\vec{b}}^{(t)} ), \\
    \vec{U}^{(t)} &= A^{(t)} \vec{W} \tilde{\vec{b}}^{(t)} + A^{(t)} \beta ( 1-Q^{(t)} ) \vec{U}^{(t-1)},
  \end{align}
  where 
  \begin{align}
    & R^{(t)} = A^{(t)} + A^{(t)} \beta (1-Q^{(t)}) R^{(t-1)}, \\
    & A^{(t)} = \frac 1{\sigma^2+\beta (1-Q^{(t)})}, \\
    & Q^{(t)} = \frac 1K \sum_{k=1}^K (\tilde{b}_k^{(t)})^2, 
  \end{align}
  with initinal conditions: $R^{(-1)}=A^{(-1)}$ and $\tilde{b}_k^{(-1)}=0$. 
  Function $\tanh$ is applied componentwise. 
  From posterior average $\tilde{b}_k^{(t)}$, 
  the tentative decision at the $t^{\rm th}$ update is evaluated as $\hat{b}_k^{(t)} = \sgn ( \tilde{b}_k^{(t)} )$, 
  where function $\sgn(x)$ denotes the sign function taking $1$ for $x \ge 0$ and $-1$ for $x<0$. 
  \qed
\end{definition}
\par
This BP-based detector can be rewritten as 
\begin{align}
  \tilde{b}_k^{(t+1)} = \tanh \biggl( & R^{(t)} h_k  + A^{(t)} \tilde{b}_k^{(t)} \notag \\
  & - \sum_{s=-1}^t J^{(t,s)} \sum_{k'=1}^K W_{kk'} \tilde{b}_k^{(s)} \biggr), 
  \label{eq:BP-basedDetector}
\end{align}
where 
\begin{align}
  & J^{(t,s)} \! = \!\!
  \left\{ \!\!\!
  \begin{array}{ll}
    0, & s=-1 \\
    \displaystyle A^{(s)} \prod_{s'=s+1}^t A^{(s')} \beta [1-q^{(s')}], & 0 \le s \le t-1\\
    A^{(t)}, & s=t. \\
  \end{array}
  \right. 
\end{align}
%Since $Q^{(t)}$ can be regarded as a constant with respect to $\tilde{\vec{b}}$ 
%by utilizing self-averaging property, 
%the parameters $R^{(s)}$, $A^{(s)}$, $J^{(s,s')}$ are also regarded as a constant. 
%Therefore we can derive the following proposition. 
%The following proposition can be proved in a similar way of 
%Lemma \ref{lemma:DisorderAveragedGeneratingFunctional} and Proposition \ref{proposition:soft-PIC}. 

\section{Generating Functional Analysis \label{section:GFA}}
%~~~~~~~~~~~~~~~~~~~~~~~~~~~~~~~~~~~~~~~~~~~~~~~~~~~~~~~~~~~~~~~~~~~~~
\par
In this section, we briefly summarize methods on GFA \cite{DeDominicis1978}. 
\HLS %ADDED.v3 (FROM HERE)
Some books that introduces GFA are available, 
e.g., the analyses of minority games \cite{Coolen2005} and spin glasses \cite{Fischer1991, DeDominicis2006}. 
\par
First of all, let us compare GFA with the replica analysis \cite{Tanaka2001, Tanaka2002, Nishimori2002, Mezard2009}. 
Both have been developed in the literature of statistical mechanics \cite{Edwards1975, Martin1973} 
and have been applied to the analysis of problems in the field of information theory so far.

\subsection{Replica Analysis Versus Generating Functional Analysis}
%~~~~~~~~~~~~~~~~~~~~~~~~~~~~~~~~~~~~~~~~~~~~~~~~~~~~~~~~~~~~~~~~~~~~~
\par
In the replica analysis, the goal is to understand macroscopic {\it static} properties in the large-system limit, 
i.e., the number of interacting bodies becomes large. 
One evaluate the {\it free energy} that is proportinal to a logarithm of the partition function. 
The partition function is a logarithm of a Gibbs distribution. 
In the static computation based on the replica analysis, 
the average over disorder generates a coupling between distinct {\it replicas}. 
\par
%GFA has been introduced into the systems with quenched randomness \cite{DeDominicis1978}, 
%that is fixed random parameters such as spreading codes in communication systems. 
In GFA, the goal is, on the other hand, to understand macroscopic {\it dynamic} properties in the large-system limit. 
One evaluates not the free energy but the {\it generating functional} that is a kind of the characteristic function of the path probability. 
The generating functional is not a static object but contains dynamical information. 
In the dynamical computation, we do not have to introduce replicas. 
The effect of disorder is to generate a coupling between distinct {\it times}. 
Under some assumptions concerning stationary states, one can obtain static properties from GFA. 
We can, therefore, avoid the replica trick and obtain macroscopic static properties by GFA if the dynamics reaches a stationary state. 
From this point of view, GFA can be regarded as one of alternative methods for the replica analysis. 
\par
In the next two subsections, we briefly introduce both the replica analysis and GFA, and discuss the difference between them in more detail.

\subsection{Outline of Replica Analysis}
%~~~~~~~~~~~~~~~~~~~~~~~~~~~~~~~~~~~~~~~~~~~~~~~~~~~~~~~~~~~~~~~~~~~~~
\par
Suppose that we want to infer the {\it Marginal-Posterior-Mode (MPM)} performance of a system. 
Let vectors $\vec{x}=(x_k)$ $\in \mathcal{X}^K$ be the states of the model, 
where $\mathcal{X}$ denotes a set of values of each element. 
Let $p(\vec{x};\vec{w}) = [Z(\beta;\vec{w})]^{-1} \rme^{-\beta H(\vec{x};\vec{w})}$ 
be a posterior distribution in the Bayesian framework, 
where $\vec{w}$ denote parameters in it and $\beta$ denotes a non-negative parameter 
that is called the {\it inverse temperature} and commonly corresponds to a noise amplitude, 
i.e., a larger $\beta$ gives a smaller noise. 
(Note that $\beta$ denotes the system load except in Section \ref{section:GFA}.) 
A distribution of this form is called the {\it Gibbs distribution}. 
The function is referred to as a cost function or the {\it Hamiltonian}. 
The normalization constant of the posterior distribution 
\begin{align}
  Z(\beta;\vec{w}) = \sum_{\vec{x} \in \mathcal{X}^K} \rme^{-\beta H(\vec{x};\vec{w})}
\end{align}
is called the {\it partition function}. 
\par
The Gibbs distribution is derived by extremization (maximization) of the entropy 
$S(\beta;\vec{w})$ $:=$ $-$ $\sum_{\vec{x} \in \mathcal{X}^K}$ $p(\vec{x};\vec{w}) \ln p(\vec{x};\vec{w})$ 
under an average cost $U(\beta;\vec{w}) := \sum_{\vec{x} \in \mathcal{X}^K} H(\vec{x}) p(\vec{x};\vec{w})$ and a parameter $\vec{w}$ are fixed. 
When we fix the parameter $\beta$ instead of the average cost, this relationship can be rewritten as follows. 
Namely, for a fixed $\beta$ and a fixed $\vec{w}$, a functional 
$\tilde{F}[q(\vec{x};\vec{w})]$ $= \sum_{\vec{x}\in\mathcal{X}^k} H(\vec{x};\vec{w}) q(\vec{x};\vec{w})$ 
$+ \beta^{-1} \sum_{\vec{x}\in\mathcal{X}^k} q(\vec{x};\vec{w}) \ln q(\vec{x};\vec{w})$ 
with respect to $q(\vec{x};\vec{w})$ 
takes an extremum (minimum) value at $q(\vec{x};\vec{w})=p(\vec{x};\vec{w})$. 
The extremum (minimum) value of the functional is $\tilde{F}[p(\vec{x};\vec{w})]=-\beta^{-1} Z(\beta;\vec{w})=:F(\beta;\vec{w})$. 
This value is an important potential, called the {\it (Helmholtz) free energy}. 
From the free energy one can evaluate various quantities such as an average cost and the entropy for a fixed $\beta$, e.g., 
\begin{align*}
  & U(\beta;\vec{w}) = \frac {\partial [\beta F(\beta;\vec{w})]}{\partial \beta}, \\
  & S(\beta;\vec{w}) = \beta^2 \frac {\partial F(\beta;\vec{w})}{\partial \beta}. 
\end{align*}
by taking derivatives with respect to $\beta$. 
Since the free energy $F(\beta;\vec{w})$ is often proportional to $K$, 
the free energy density $f_K(\beta;\vec{w}):=F(\beta;\vec{w})/K$, that is the free energy per an interacting body, is defined. 
\par
For a given realization $\vec{w}$ it is, however, hard to calculate the free energy density, 
since the result must depend on detail of the realization. 
We therefore assume that the limit $\lim_{K\to\infty} f_K(\beta;\vec{w})=:\tilde{\mathcal{F}}$ exists 
and it is equal to its average $\lim_{K\to\infty} K^{-1} \mathbb{E}[F(\beta;\vec{w})]$ for almost all realizations, 
where the $\mathbb{E}_X$ denotes the expectation with respect to random variables $X$. 
Using the identity $\ln x=\lim_{n \to 0}\frac{\partial x^n}{\partial n}$, we have 
\begin{align}
  \tilde{\mathcal{F}} = \lim_{K\to\infty} 
  \biggl( 
    \lim_{n \to 0} \frac{\partial \mathbb{E}_{\vec{w}}[Z(\beta;\vec{w})^n]}{\partial n} 
  \biggr). 
\end{align}
The idea of the replica analysis is to calculate $\mathbb{E}_{\vec{w}}[Z(\beta;\vec{w})^n]$ as if $n$ were in integer. 
The $n$-th moment is 
\begin{align}
  Z(\beta;\vec{w})^n = \sum_{\vec{x}^1,\cdots,\vec{x}^n \in \mathcal{X}^K} \exp\biggl[ -\beta \sum_{a=1}^n H(\vec{x}^a;\vec{w}) \biggr], 
\end{align}
where $a$ is a replica index. 
We shall refer to copies $\vec{x}^1, \cdots, \vec{x}^n$ as {\it replicas}. 
It should be noted that the replicas are no longer statistically independent since the replicas have a common parameter $\vec{w}$. 
The average over the parameter $\vec{w}$ generates a coupling between distinct replicas. 
\par
In the replica analysis, one therefore has to assume 
(i) the self-averaging property applies, 
(ii) the ``replica trick'' is valid, and 
(iii) replica symmetry (or same kind of symmetries) holds, 
to keep tractability. 
The replica analysis, however, still does not have a rigorous justification.

\subsection{Outline of Generating Functional Analysis}
%~~~~~~~~~~~~~~~~~~~~~~~~~~~~~~~~~~~~~~~~~~~~~~~~~~~~~~~~~~~~~~~~~~~~~
\par
\HLE
%ADDED.v3 (TO HERE)
\par
The generating functional analysis, or the {\it path integral methods}, have been applied to the model 
which is described using realizations of random variables by de Dominicis \cite{DeDominicis1978}. 
In GFA, one can analyze the asymptotic behavior of the dynamics 
in the infinitely large system using the generating functional 
that can be regarded as a kind of the characteristic function in statistics. 
\par
We here consider the following model. 
Let vectors $\vec{x}^{(s)}=(x_k^{(s)})$ $\in \mathbb{R}^K$ be the states of the model at stage $s$ and 
let the initial state be $\vec{x}^{(-1)}$. 
Let the updating rule be 
\begin{align}
  \vec{x}^{(s+1)}=\mathcal{F}(\vec{x}^{(s)};\vec{w}) 
\end{align}
for $s \in \{-1,0,\cdots,t-1\}$, 
where $\mathcal{F}:\mathbb{R}\to\mathbb{R}$ denotes a function 
and $\vec{w}$ again denotes parameters in it. 
\par
If we know the probability of seeing a particular state at a given stage $p_s(\vec{x}^{(s)})$, 
we can evaluate the property of the system by using it. 
In GFA we consider the probability of observing a particular sequence or path of states, 
i.e., $p(\vec{x}^{(-1)}, \vec{x}^{(0)},\cdots,\vec{x}^{(t)})$ up to some finite time $t$ instead of $p_s(\vec{x}^{(s)})$. 
\HLS The probability $p(\vec{x}^{(-1)}, \vec{x}^{(0)},\cdots,\vec{x}^{(t)})$ is referred to as the {\it path probability}. \HLE %ADDED.v3
The way to do this is to introduce a generating functional which is defined as 
\begin{align}
  Z[\vec{\psi}]=\biggl\langle \exp \biggl[ - \rmi \sum_{s=-1}^t \sum_{k=1}^K \psi_k^{(s)} x_k^{(s)} \biggr] \biggr\rangle, 
\end{align}
where the bracket $\langle \cdot \rangle$ denotes the average over the path probability $p(\vec{x}^{(-1)}, \vec{x}^{(0)},\cdots,\vec{x}^{(t)})$ 
and we have introduced the dummy variables $\vec{\psi}^{(s)}=(\psi_k^{(s)})$ $\in \mathbb{R}^K$. 
\HLS Note that the generating functional differs from the partition function in the previous subsection. \HLE %ADDED.v3
Taking derivatives with respect to the dummy variables allow us to examine some averages, e.g., 
\begin{align}
  & \langle x_k^{(s)} \rangle = \rmi \lim_{\vec{\psi}\to\vec{0}} \frac{\partial Z[\vec{\psi}]}{\partial \psi_k^{(s)}}, \\
  & \langle x_k^{(s)} x_{k'}^{(s')} \rangle = - \lim_{\vec{\psi}\to\vec{0}} \frac{\partial^2 Z[\vec{\psi}]}{\partial \psi_k^{(s)} \partial \psi_k^{(s')}}. 
\end{align}
\HLS %ADDED.v3 (FROM HERE)
Since one does not have to average a ratio or a logarithm in this context,
one can compute correlations by entirely avoiding the replica trick. 
In the replica analysis, macroscopic quantities are obtained from derivatives with respect to the ``scalar'' parameter $\beta$. 
In GFA, averages are given from derivatives with respect to not a scalar parameter but some elements of the dummy variables (vectors) with same dimension as the original vectors $\vec{x}^{(-1)}, \vec{x}^{(0)},\cdots,\vec{x}^{(t)}$. 
\par
For a given realization $\vec{w}$ it is, again, hard to calculate the generating functional 
since the result must depend on detail of the realization. 
%It would not be possible to calculate the generating functional for a particular realization. %DELETED.v3
We, therefore, assume that the generating functional is concentrated to its average over the parameter $\mathbb{E}_{\vec{w}}(Z[\vec{\psi}])$ in the large system limit. 
The effect of the parameter $\vec{w}$ is to generate a coupling between distinct times. 
%, where the $\mathbb{E}_X$ denotes the expectation with respect to random variables $X$. %DELETED.v3
Averaging over the random variables, we will move to a saddle-point problem in the limit $K \to \infty$. 
It should be noted that the normalization relation $Z[\vec{0}]=1$ plays an important role in the elimination of spurious solutions to the saddle-point equations. 
(We see detail of this point in Section \ref{section:GFA_softPIC}-B.) 
\HLE %ADDED.v3 (TO HERE)
\par
The terms in the averaged generating functional can be split into three related parts. 
The first one is a signal part. 
The second one is a static noise part due to the random variables within the model. 
The last one is retarded self-interaction due to the influence of the state at the previous stage, which may be able to affect the present state. 
The GFA allows us to treat the last part.

\section{Generating Functional Analysis for Soft-PIC \label{section:GFA_softPIC}}
%~~~~~~~~~~~~~~~~~~~~~~~~~~~~~~~~~~~~~~~~~~~~~~~~~~~~~~~~~~~~~~~~~~~~~

\subsection{Averaged Generating Functional}
%~~~~~~~~~~~~~~~~~~~~~~~~~~~~~~~~~~~~~~~~~~~~~~~~~~~~~~~~~~~~~~~~~~~~~

The goal of multiuser detection is to simultaneously infer information bits $b_1,\cdots,b_K$ 
after base-band signals $y_1,\cdots,y_N$ are received. 
\par
If the transfer function takes $f(x)=\sgn(x)$, this updating rule coincides with the {\it hard-PIC} \cite{Varanashi1990}. 
%where function $\sgn(x)$ denotes the sign function taking 1 for $x \ge 0$ and -1 for $x<0$. %DELETED
Note that the hard-PIC, which means the 
%Varanasi's conventional PIC, %DELETED.v3
\HLS Varanasi and Aazhang's conventional PIC, \HLE %ADDED.v3
is also obtained by taking the limit $\sigma \to 0$ in soft-PIC. 
\par
We assume the matched filter stage, i.e., $\tilde{b}_k(0) = f(h_k)$, for initialization. 
This initialization is easily treated by formally assuming $p(\tilde{b}_k^{(-1)})=$ $\delta(\tilde{b}_k^{(-1)})$ for all $k$, 
where $\delta$ denotes the Dirac delta function. 
The widely used measure to determine the performance of a demodulator is the {\it bit error rate (BER)}. 
The BER $P_b^{(t)}$ of hard decisions $\hat{b}_k^{(t)}=\sgn(\tilde{b}_k^{(t)})$ 
at the $t^{\rm th}$ stage of soft PIC is given by $P_b^{(t)}=(1-m_h^{(t)})/2$, 
where $m_h^{(t)}=\frac1K\sum_{k=1}^K b_k \tilde{b}_k^{(t)}$ is 
the overlap between information vector $\vec{b}^{(t)}=(b_1,\cdots,b_K)^\top$ 
and tentative hard-decision vector $(\tilde{b}_1^{(t)},\cdots,\tilde{b}_K^{(t)})^\top$. 
%Operator $A^\top$ denotes the transpose of $A$. %DELETED
Without loss of generality, we can assume that the true information bits are all 1, 
i.e., $b_k=1$ for all $k$, because the spreading codes are symmetric. 
\par
Let us analyze the detection dynamics in the large system limit where $K,N\to\infty$, while the system load $\beta$ is kept finite. 
We introduce inverse temperature $\gamma$ for generating functional analysis. 
The stochastic updating rule for tentative decision $\tilde{b}_k^{(t)}\in\mathbb{R}$ 
of bit signal $b_k$ at stage $t$ is given by 
\begin{align}
  p(\tilde{b}_k^{(s+1)}|\tilde{\vec{b}}^{(s)}) 
  = \frac{\gamma}{\sqrt{2\pi}} e^{-\frac{\gamma^2}2 \{ \tilde{b}_k^{(s+1)}-f(u_k^{(s)}) \}^2}, 
  \label{eq:updating_rule}
\end{align}
with 
\begin{equation}
  u_k^{(t)} := h_k - \sum_{k'=1, \ne k}^K W_{kk'} \tilde{b}_{k'}^{(t)} + \theta_k^{(t)}, 
  \label{eq:def_of_SumOverAllMessage}
\end{equation}
which is a summation over all messages from other tentative decisions. 
Note that this updating rule coincides with $\tilde{b}_k^{(s+1)}=f(u_k^{(s)})+{\cal N}(0,\gamma^{-2})$. 
In the limit where $\gamma\to\infty$, this stochastic updating rule 
is equivalent to the deterministic rule (\ref{eq:deterministic_updating_rule}). 
Term $\theta_k^{(t)}$ is a stage-dependent external message that is introduced to define a response function. 
The inverse temperature and the external message 
are respectively set to $\gamma\to\infty$ and $\theta_k^{(t)}=0$ at the end of analysis. 
\par
The stochastic updating rule for tentative decision vector $\tilde{\vec{b}}^{(t)}=(\tilde{b}_k^{(t)})\in\mathbb{R}^K$ at stage $t$ is given 
by using (\ref{eq:updating_rule}), 
i.e., $p(\tilde{\vec{b}}^{(s+1)}|\tilde{\vec{b}}^{(s)}) =$ $\prod_{k=1}^K p(\tilde{b}_k^{(s+1)}|\tilde{\vec{b}}^{(s)})$. 
The dynamics is a Markov chain, since the present tentative decision depends only on the past decisions. 
A path probability (density) $p(\tilde{\vec{b}}^{(-1)},\cdots,\tilde{\vec{b}}^{(t)})$ is therefore simply given 
by the individual transition probability $p(\tilde{\vec{b}}^{(s+1)}|\tilde{\vec{b}}^{(s)})$ of the chain: 
\begin{equation}
  p(\tilde{\vec{b}}^{(-1)},\cdots,\tilde{\vec{b}}^{(t)})
  = p(\tilde{\vec{b}}^{(-1)}) \prod_{s=-1}^{t-1} p(\tilde{\vec{b}}^{(s+1)}|\tilde{\vec{b}}^{(s)}). 
  \label{eq:def_path_probability}
\end{equation}
The initial state probability becomes $p(\tilde{\vec{b}}^{(-1)})=\prod_{k=1}^K \delta(\tilde{b}_k^{(-1)})$. 
Therefore, we can calculate an expectation with respect to arbitrary function 
${\cal G}={\cal G}(\tilde{\vec{b}}^{(-1)},\cdots,\tilde{\vec{b}}^{(t)})$ of tentative decisions as 
\begin{align}
  \langle {\cal G} \rangle := \int_{\mathbb{R}^{(t+2)K}} 
  \biggl( \prod_{s=-1}^t d\tilde{\vec{b}}^{(s)} \biggr) 
  p(\tilde{\vec{b}}^{(-1)},\cdots,\tilde{\vec{b}}^{(t)}) {\cal G}. 
\end{align}
\par
To analyze the detection dynamics of the system, 
we define a generating functional as 
$
Z[\vec{\psi}] := 
    \langle \exp [ 
      -\rmi\sum_{s=-1}^t\tilde{\vec{b}}^{(s)} \cdot \vec{\psi}^{(s)} 
    ] \rangle
$,
where 
$\tilde{\vec{b}}^{(s)}=$ $(\tilde{b}_1^{(s)},$ $\cdots,$ $\tilde{b}_K^{(s)})^\top$ and 
$\vec{\psi}^{(s)}=$ $(\psi_1^{(s)},$ $\cdots,$ $\psi_K^{(s)})^\top$. 
The basic idea underlying generating functional formalism is very simple \cite{DeDominicis1978, Coolen2000, Coolen2005}. 
If the generating functional $Z[\vec{\psi}]$ can be evaluated as a functional with respect to dummy functions $\vec{\psi}^{(s)}$, 
one can obtain all averages of interest by differentiation 
%\begin{eqnarray} %DELETED
%  && \rmi \lim_{\vec{\psi}\to\vec{0}} %DELETED
%     \frac{\partial Z[\vec{\psi}]}{\partial \psi_k^{(s)}} %DELETED
%     = \langle \tilde{b}_k^{(s)}\rangle, \\ %DELETED
%  && -\lim_{\vec{\psi}\to\vec{0}} %DELETED
%     \frac{\partial^2 Z[\vec{\psi}]}{\partial \psi_k^{(s)} \partial \psi_{k'}^{(s')}} %DELETED
%     = \langle\tilde{b}_k^{(s)}\tilde{b}_{k'}^{(s')}\rangle , \\ %DELETED
%  && \rmi\lim_{\vec{\psi}\to\vec{0}} %DELETED
%     \frac{\partial^2 Z[\vec{\psi}]}{\partial \psi_k^{(s)} \partial \theta_{k'}^{(s')}} %DELETED
%     = \frac{\partial \langle\tilde{b}_k^{(s)}\rangle}{\partial \theta_{k'}^{(s')}}, %DELETED
%\end{eqnarray} %DELETED
from $Z[\vec{\psi}]$. 
The generating functional includes the random variables $\{s_k^\mu\}$ and $\{n^\mu\}$. 
%\begin{assumption}
%  \label{assumption:self-averaging}
%  The generating functional is self-averaging. 
%  Namely, in the large system limit, the generating functional is concentrated 
%  to its average over random variables $\{s_k^\mu\}$ and $\{n^\mu\}$. 
%  \qed
%\end{assumption}
%\par
%Under Assumption \ref{assumption:self-averaging}, 
%the typical behavior of the system only depends on the statistical properties of the random variables. 
%Since we assume Assumption \ref{assumption:self-averaging}, 
%we therefore evaluate the averaged generating functional: 
\par
We here assume that the generating functional is self-averaging, 
namely, in the large system limit, the generating functional is concentrated 
to its average over random variables $\{s_k^\mu\}$ and $\{n^\mu\}$ and 
the typical behavior of the system only depends on the statistical properties of the random variables. 
We therefore evaluate the averaged generating functional defined as follows. 
%-------------------------------------------------DEFINITION
\begin{definition}
  (Average Generating functional) The average generating functional $\bar{Z}[\vec{\psi}]$ is defined by 
  \begin{align}
    \bar{Z}[\vec{\psi}] := 
    \overline{ 
      \biggl\langle 
        \exp \biggl[ -\rmi\sum_{s=-1}^t\tilde{\vec{b}}^{(s)} \cdot \vec{\psi}^{(s)} \biggr] 
      \biggr\rangle 
    },
    \label{eq:def_barZ}
  \end{align}
  where $\overline{[\cdots]} := \mathbb{E}_{\vec{s}_1,\cdots,\vec{s}_K,\vec{n}}[\cdots]$ denotes 
  the average over spreading codes $\vec{s}_1,\cdots,\vec{s}_K$ and the noise $\vec{n}$. 
  \qed
\end{definition}
\par
From the averaged generating functional, we can obtain 
\begin{eqnarray}
  && \lim_{\vec{\psi}\to\vec{0}}\frac{\partial \bar{Z}[\vec{\psi}]}{\partial \psi_k^{(s)}} 
     = - \rmi \overline{\langle \tilde{b}_k^{(s)}\rangle}, 
     \label{eq:interest1} \\
  && \lim_{\vec{\psi}\to\vec{0}}\frac{\partial^2 \bar{Z}[\vec{\psi}]}{\partial \psi_k^{(s)} \partial \psi_{k'}^{(s')}} 
     = - \overline{\langle\tilde{b}_k^{(s)}\tilde{b}_{k'}^{(s')}\rangle}, 
     \label{eq:interest2} \\
  && \lim_{\vec{\psi}\to\vec{0}}\frac{\partial^2 \bar{Z}[\vec{\psi}]}{\partial \psi_k^{(s)} \partial \theta_{k'}^{(s')}} 
     = - \rmi \frac{\partial \overline{\langle\tilde{b}_k^{(s)}\rangle}}{\partial \theta_{k'}^{(s')}}. 
     \label{eq:interest3}
\end{eqnarray}
Calculating the average over spreading codes $\vec{s}_1,\cdots,\vec{s}_K$ 
and the noise $\vec{n}$, we have the following result. 
\\
%-------------------------------------------------LEMMA
\begin{lemma}
  \label{lemma:DisorderAveragedGeneratingFunctional}
  The averaged generating functional is simplified to % the saddle-point problem as 
  \begin{eqnarray}
    \bar{Z}[\vec{\psi}] 
    &=& \int 
    \rmd\vec{\eta} \rmd\hat{\vec{\eta}}
    \rmd\vec{k} \rmd\hat{\vec{k}}
    \rmd\vec{q} \rmd\hat{\vec{q}}
    \rmd\vec{Q} \rmd\hat{\vec{Q}}
    \rmd\vec{L} \rmd\hat{\vec{L}} \nonumber \\
    & & \quad \times \exp \biggl[ K(\Phi+\Psi+\Omega)+O(\ln K) \biggr], 
    \label{eq:barZ}
  \end{eqnarray}
  in which functions $\Phi$, $\Psi$, and $\Omega$ are given by 
  \begin{eqnarray}
    && \Phi :=
       \rmi \sum_{s=-1}^{t-1} \{ \hat{\eta}^{(s)} \eta^{(s)} + \hat{k}^{(s)}k^{(s)} \} \nonumber \\
    && \qquad + \rmi \sum_{s=-1}^{t-1} \sum_{s'=-1}^{t-1} \{ \hat{q}^{(s,s')} q^{(s,s')} \nonumber \\
    && \qquad \quad + \hat{Q}^{(s,s')} Q^{(s,s')} 
       + \hat{L}^{(s,s')} L^{(s,s')} \}, 
  \end{eqnarray}
  \begin{eqnarray}
    && \Psi :=
       \frac 1K \sum_{k=1}^K \ln \biggl\{ \int_{\mathbb{R}^{t+2}} 
       \biggl( \prod_{s=-1}^{t} \rmd \tilde{b}^{(s)} \biggr) p[\tilde{b}^{(-1)}] \int \underline{\rmd}u \underline{\rmd}\hat{u} \nonumber \\
    && \qquad \times \exp \biggl[ \sum_{s=-1}^{t-1} 
       \{ \ln \frac{\gamma}{\sqrt{2\pi}} - \frac{\gamma^2}2 [\tilde{b}^{(s+1)}-f(u^{(s)})]^2 \} \nonumber \\
    && \qquad - \rmi \sum_{s=-1}^{t-1} \sum_{s'=-1}^{t-1} \{ \hat{q}^{(s,s')} \tilde{b}^{(s)} \tilde{b}^{(s')} \nonumber \\
    && \qquad \quad + \hat{Q}^{(s,s')} \hat{u}^{(s)} \hat{u}^{(s')} + \hat{L}^{(s,s')} \tilde{b}^{(s)} \hat{u}^{(s')} \} \nonumber \\
    && \qquad + \rmi \sum_{s=-1}^{t-1} \hat{u}^{(s)} \{ u^{(s)} - \tilde{b}^{(s)} - \theta_k^{(s)} - \hat{k}^{(s)} \} \nonumber \\
    && \qquad - \rmi \sum_{s=-1}^{t-1} \tilde{b}^{(s)} \hat{\eta}^{(s)} - \rmi \sum_{s=-1}^t \tilde{b}^{(s)} \psi_k^{(s)} \biggr] \biggr\}, 
  \end{eqnarray}
  \begin{eqnarray}
    && \Omega := 
       \frac 1K \ln \int \underline{\rmd}\vec{v}\underline{\rmd}\hat{\vec{v}}\underline{\rmd}\vec{w}\underline{\rmd}\hat{\vec{w}} \nonumber \\
    && \qquad \times \exp \biggl[ \rmi \sum_{\mu=1}^N \sum_{s=-1}^{t-1} \{ \hat{v}_\mu^{(s)} v_\mu^{(s)} + \hat{w}_\mu^{(s)} w_\mu^{(s)} 
       - \beta v_\mu^{(s)} w_\mu^{(s)} \} \nonumber \\
    && \qquad - \frac 12 \sum_{\mu=1}^N \sum_{s=-1}^{t-1} \sum_{s'=-1}^{t-1} 
       \{ \beta \sigma_0^2 v_\mu^{(s)} v_\mu^{(s')} 
       + \hat{v}_\mu^{(s)} Q^{(s,s')} \hat{v}_\mu^{(s')} \} \nonumber \\
    && \qquad - \frac 12 \sum_{\mu=1}^N \sum_{s=-1}^{t-1} \sum_{s'=-1}^{t-1} 
       \{ \hat{v}_\mu^{(s)} [k^{(s)} - L^{(s',s)}] \hat{w}_\mu^{(s')} \nonumber \\
    && \qquad \quad + \hat{w}_\mu^{(s)} [k^{(s')} - L^{(s,s')}] \hat{v}_\mu^{(s')} \} \nonumber \\
    && \qquad - \frac 12 \sum_{\mu=1}^N \sum_{s=-1}^{t-1} \sum_{s'=-1}^{t-1} \{ \hat{w}_\mu^{(s)} \nonumber \\
    && \qquad \quad \times [1 - \eta^{(s)} - \eta^{(s')} + q^{(s,s')}] \hat{w}_\mu^{(s')} \} \biggr]. 
  \end{eqnarray}
  \qed
\end{lemma}
Details on the derivation and definitions of the notations are given in Appendix \ref{app:Z_soft-PIC}.

\subsection{Saddle-Point Equations and Meaning of Macroscopic Parameters} \label{sec:SPEandMOP}
%~~~~~~~~~~~~~~~~~~~~~~~~~~~~~~~~~~~~~~~~~~~~~~~~~~~~~~~~~~~~~~~~~~~~~
\par
One can deduce the meaning of macroscopic parameters 
by differentiating the averaged generating functional $\bar{Z}[\vec{\psi}]$ of (\ref{eq:barZ}) 
with respect to the external messages $\theta_k^{(s)}$ and dummy functions $\psi_k^{(s)}$. 
The averaged generating functional $\bar{Z}[\vec{\psi}]$ is dominated by a saddle-point for $K\to\infty$. 
We can thus simplify the saddle-point problem in Lemma \ref{lemma:DisorderAveragedGeneratingFunctional} in the large system limit. 
\HLS %MODIFIED.v3
Using the shorthand 
$\{\rmd\} := 
\rmd\vec{\eta} \rmd\hat{\vec{\eta}}
\rmd\vec{k} \rmd\hat{\vec{k}}
\rmd\vec{q} \rmd\hat{\vec{q}}
\rmd\vec{Q} \rmd\hat{\vec{Q}}
\rmd\vec{L} \rmd\hat{\vec{L}}$
and the normalization identity $\bar{Z}[\vec{0}]=\langle 1 \rangle=1$, 
\HLE %MODIFIED.v3
we now find derivatives of the averaged generating functional: 
\begin{eqnarray}
  & & \lim_{\vec{\psi}\to\vec{0}} \frac{\partial \bar{Z}[\vec{\psi}]}{\partial \psi_k^{(s)}} \nonumber \\
  & & \HLS \qquad = \lim_{\vec{\psi}\to\vec{0}} \int \{\rmd\} \rme^{K(\Phi+\Psi+\Omega)} 
      \frac{\partial (K \Psi)}{\partial \psi_k^{(s)}} \nonumber \\ \HLE %ADDED.v3
  & & \qquad = -\rmi \langle \tilde{b}^{(s)} \rangle_k, 
      \label{eq:field_dervatives1} \\
  & & \lim_{\vec{\psi}\to\vec{0}} \frac{\partial^2 \bar{Z}[\vec{\psi}]}{\partial \psi_k^{(s)} \partial \psi_{k'}^{(s')}} \nonumber \\
  & & \HLS \qquad = \lim_{\vec{\psi}\to\vec{0}} \int \{\rmd\} \rme^{K(\Phi+\Psi+\Omega)} \nonumber \\ \HLE %ADDED.v3
  & & \HLS \qquad \qquad \times 
      \biggl( 
        \frac{\partial^2 (K \Psi)}{\partial \psi_k^{(s)} \partial \psi_{k'}^{(s')}} + 
        \frac{\partial (K \Psi)}{\partial \psi_k^{(s)}} \cdot 
        \frac{\partial (K \Psi)}{\partial \psi_{k'}^{(s')}} 
      \biggr)
      \nonumber \\ \HLE %ADDED.v3
  & & \HLS \qquad = \delta_{k,k'} [ - \langle \tilde{b}^{(s)}\tilde{b}^{(s')}\rangle_k 
      + \langle \tilde{b}^{(s)}\rangle_k \langle \tilde{b}^{(s')}\rangle_{k'}] \nonumber \\ \HLE %ADDED.v3
  & & \HLS \qquad \qquad - \langle \tilde{b}^{(s)}\rangle_k \langle \tilde{b}^{(s')}\rangle_{k'} \nonumber \\ \HLE %ADDED.v3
  & & \qquad = -\delta_{k,k'} \langle \tilde{b}^{(s)}\tilde{b}^{(s')}\rangle_k 
      - (1-\delta_{k,k'}) \langle \tilde{b}^{(s)}\rangle_k \langle \tilde{b}^{(s')}\rangle_{k'} \nonumber \\
      \label{eq:field_dervatives2} \\
  & & \lim_{\vec{\psi}\to\vec{0}} \frac{\partial^2 \bar{Z}[\vec{\psi}]}{\partial \psi_k^{(s)} \partial \theta_{k'}^{(s')}} \nonumber \\
  & & \HLS \qquad = \lim_{\vec{\psi}\to\vec{0}} \int \{\rmd\} \rme^{K(\Phi+\Psi+\Omega)} \nonumber \\ \HLE %ADDED.v3
  & & \HLS \qquad \qquad \times 
      \biggl( 
        \frac{\partial^2 (K \Psi)}{\partial \psi_k^{(s)} \partial \theta_{k'}^{(s')}} + 
        \frac{\partial (K \Psi)}{\partial \psi_k^{(s)}} \cdot 
        \frac{\partial (K \Psi)}{\partial \theta_{k'}^{(s')}} 
      \biggr)
      \nonumber \\ \HLE %ADDED.v3
  & & \qquad = -\delta_{k,k'} \langle \tilde{b}^{(s)}\hat{u}^{(s')}\rangle_k 
      - (1-\delta_{k,k'}) \langle \tilde{b}^{(s)}\rangle_k \langle \hat{u}^{(s')}\rangle_{k'} \nonumber \\
      \label{eq:field_dervatives3} \\
  & & \lim_{\vec{\psi}\to\vec{0}} \frac{\partial \bar{Z}[\vec{\psi}]}{\partial \theta_k^{(s)}} 
      = -\rmi\langle \hat{u}^{(s)}\rangle_k \mathop{=}^{(a)} 0, 
      \label{eq:field_dervatives4} \\
  & & \lim_{\vec{\psi}\to\vec{0}} \frac{\partial^2 \bar{Z}[\vec{\psi}]}{\partial \theta_k^{(s)} \partial \theta_{k'}^{(s')}} 
      = -\delta_{k,k'}\langle \hat{u}^{(s)}\hat{u}^{(s')}\rangle_k \mathop{=}^{(b)} 0, 
      \label{eq:field_dervatives5}
\end{eqnarray}
where $\langle \; \rangle_k$ denotes the average as 
\begin{align}
  & \langle f(\{\tilde{b},u,\hat{u}\}) \rangle_k := \nonumber \\
  & \frac {\displaystyle \int_{\mathbb{R}^{t+2}} \biggl( \prod_{s=-1}^{t} \rmd \tilde{b}^{(s)} \biggr) \int \underline{\rmd}u \underline{\rmd} \hat{u} w_k(\{\tilde{b},u,\hat{u}\}) f(\{\tilde{b},u,\hat{u}\})}
          {\displaystyle \int_{\mathbb{R}^{t+2}} \biggl( \prod_{s=-1}^{t} \rmd \tilde{b}^{(s)} \biggr) \int \underline{\rmd}u \underline{\rmd} \hat{u} w_k(\{\tilde{b},u,\hat{u}\})}
  \label{eq:average_over_singleUserMeasure}
\end{align}
%\begin{equation}
%  \langle f(\{\tilde{b},u,\hat{u}\}) \rangle_k := 
%  \frac
%  {\displaystyle \int_{\mathbb{R}^{t+2}} \biggl( \prod_{s=-1}^{t} \rmd \tilde{b}^{(s)} \biggr) \int \underline{\rmd}u \underline{\rmd} \hat{u} w_k(\{\tilde{b},u,\hat{u}\}) f(\{\tilde{b},u,\hat{u}\})}
%  {\displaystyle \int_{\mathbb{R}^{t+2}} \biggl( \prod_{s=-1}^{t} \rmd \tilde{b}^{(s)} \biggr) \int \underline{\rmd}u \underline{\rmd} \hat{u} w_k(\{\tilde{b},u,\hat{u}\})}
%\end{equation}
with 
\begin{eqnarray}
  & & w_k(\{\tilde{b},u,\hat{u}\}) \nonumber \\
  & & := \delta [\tilde{b}^{(-1)}] \exp \biggl[ 
        \sum_{s=-1}^{t-1} \{ \ln \frac{\gamma}{\sqrt{2\pi}} - \frac{\gamma^2}2 [\tilde{b}^{(s+1)}-f(u^{(s)})]^2 \} \nonumber \\
  & & - \rmi \sum_{s=0}^{t-1} \sum_{s'=0}^{t-1} \{ \hat{q}^{(s,s')} \tilde{b}^{(s)} \tilde{b}^{(s')} 
      + \hat{Q}^{(s,s')} \hat{u}^{(s)} \hat{u}^{(s')} \nonumber \\
  & & + \hat{L}^{(s,s')} \tilde{b}^{(s)} \hat{u}^{(s')} \} \nonumber \\
  & & + \rmi \sum_{s=0}^{t-1} \hat{u}^{(s)} \{ u^{(s)} - \tilde{b}^{(s)} - \theta_k^{(s)} - \hat{k}^{(s)} \} \nonumber \\
  & & - \rmi \sum_{s=0}^{t-1} \tilde{b}^{(s)} \hat{\eta}^{(s)} \biggr] \biggl. \biggr|_{\mathrm{saddle}}, 
  \label{eq:def_Single-UserMeasure}
\end{eqnarray}
which is referred to as a {\it single-user measure}. 
Here, evaluation $f|_{\mathrm{saddle}}$ denotes an evaluation of function $f$ at the dominating saddle-point. 
To derive the $\mathop{=}^{(a)}$ in (\ref{eq:field_dervatives4}) and $\mathop{=}^{(b)}$ in (\ref{eq:field_dervatives5}), 
we use identities 
\begin{eqnarray}
  & & \lim_{\vec{\psi}\to\vec{0}} \frac{\partial \bar{Z}[\vec{\psi}]}{\partial \theta_k^{(s)}}
      = \frac{\partial \bar{Z}[\vec{0}]}{\partial \theta_k^{(s)}}
      = 0, \\
  & & \lim_{\vec{\psi}\to\vec{0}} \frac{\partial^2 \bar{Z}[\vec{\psi}]}{\partial \theta_k^{(s)} \partial \theta_{k'}^{(s')}}
      = \frac{\partial^2 \bar{Z}[\vec{0}]}{\partial \theta_k^{(s)} \partial \theta_{k'}^{(s')}}
      = 0, 
\end{eqnarray}
respectively. 
\HLS These are obtained from $\bar{Z}[\vec{0}]=1$. \HLE %ADDED.v3
Substituting (\ref{eq:interest1}) -- (\ref{eq:interest3}) 
and (\ref{eq:field_dervatives4}) -- (\ref{eq:field_dervatives5}) 
into (\ref{eq:field_dervatives1}) -- (\ref{eq:field_dervatives3}), 
\HLS the spurious solutions, that depend on $\langle \hat{u}^{(s)}\rangle_k$ or $\langle \hat{u}^{(s)}\hat{u}^{(s')}\rangle_k$, are eliminated and \HLE %ADDED.v3
we then have 
\begin{eqnarray}
  & & \overline{\langle \tilde{b}_k^{(s)}\rangle}
      = \langle \tilde{b}^{(s)} \rangle_k, 
      \label{eq:other_field_dervatives1} \\
  & & \overline{\langle\tilde{b}_k^{(s)}\tilde{b}_{k'}^{(s')}\rangle} 
      = \delta_{k,k'} \langle \tilde{b}^{(s)}\tilde{b}^{(s')}\rangle_k \nonumber \\
  & & \qquad \qquad \qquad     + (1-\delta_{k,k'}) \langle \tilde{b}^{(s)}\rangle_k \langle \tilde{b}^{(s')}\rangle_{k'}, 
      \label{eq:other_field_dervatives2} \\
  & & \frac{\partial \overline{\langle\tilde{b}_k^{(s)}\rangle}}{\partial \theta_{k'}^{(s')}} 
      = - \rmi \delta_{k,k'} \langle \tilde{b}^{(s)}\hat{u}^{(s')}\rangle_k. 
      \label{eq:other_field_dervatives3}
\end{eqnarray}
\par
In the large system limit $K\to\infty$, integral (\ref{eq:barZ}) will be evaluated 
by the dominating saddle-point of exponent $\Phi+\Psi+\Omega$. 
We can now derive the saddle-point equations by differentiation with respect to integral variables 
$\eta^{(s)}$, $\hat{\eta}^{(s)}$, 
$k^{(s)}$, $\hat{k}^{(s)}$, 
$q^{(s,s')}$, $\hat{q}^{(s,s')}$, 
$Q^{(s,s')}$, $\hat{Q}^{(s,s')}$, 
$L^{(s,s')}$, and $\hat{L}^{(s,s')}$. 
These equations will involve the average overlap $m^{(s)}$ (which measures the bit error rate), 
the average single-user correlation $C^{(s,s')}$ and the average single-user response function $G^{(s,s')}$: 
\begin{eqnarray}
  && m^{(s)} 
     := \lim_{K \to \infty} \frac 1K \sum_{k=1}^K     \overline{\langle \tilde{b}_k^{(s)} \rangle}, 
     % := \lim_{K \to \infty} \frac 1K \sum_{k=1}^K b_k \overline{\langle \tilde{b}_k^{(s)} \rangle}, 
     \label{eq:def_average_overlap} \\
  && C^{(s,s')} 
     := \lim_{K \to \infty} \frac 1K \sum_{k=1}^K \overline{\langle \tilde{b}_k^{(s)} \tilde{b}_k^{(s')} \rangle}, 
     \label{eq:def_average_SU_correlation_function} \\
  && G^{(s,s')} 
     := \lim_{K \to \infty} \frac 1K \sum_{k=1}^K \frac{\partial \overline{\langle \tilde{b}_k^{(s)} \rangle}}{\partial \theta_k^{(s')}}. 
     \label{eq:def_average_SU_response_function}
\end{eqnarray}
Using the identities (\ref{eq:field_dervatives1}) -- (\ref{eq:field_dervatives5}) 
and (\ref{eq:other_field_dervatives1}) -- (\ref{eq:other_field_dervatives3}), 
the straightforward differentiation of $\Phi+\Psi+\Omega$ with respect to 
$\eta^{(s)}$, $\hat{\eta}^{(s)}$, 
$k^{(s)}$, $\hat{k}^{(s)}$, 
$q^{(s,s')}$, $\hat{q}^{(s,s')}$, 
$Q^{(s,s')}$, $\hat{Q}^{(s,s')}$, 
$L^{(s,s')}$, and $\hat{L}^{(s,s')}$ 
leads us to the following saddle-point equations: 
\begin{eqnarray}
  & & \hat{\eta}^{(s)} = \rmi \left. \frac{\partial \Omega}{\partial \eta^{(s)}} \right|_{\mathrm{saddle}}, 
      \label{eq:spe_mhat} \\
  & & \eta^{(s)} = \lim_{K \to \infty} \frac 1K \sum_{k=1}^K \langle \tilde{b}^{(s)} \rangle_k, 
      \label{eq:spe_m}\\
  & & \hat{k}^{(s)} = \rmi \left. \frac{\partial \Omega}{\partial k^{(s)}} \right|_{\mathrm{saddle}}, 
      \label{eq:spe_khat} \\
  & & k^{(s)} 
      \HLS = \frac 1K \sum_{k=1}^K \langle \hat{u}^{(s)} \rangle_k \HLE %ADDED.v3
      = 0, 
      \label{eq:spe_k} \\
  & & \hat{q}^{(s,s')} = \rmi \left. \frac{\partial \Omega}{\partial q^{(s,s')}} \right|_{\mathrm{saddle}}, 
      \label{eq:spe_qhat} \\
  & & q^{(s,s')} = \lim_{K \to \infty} \frac 1K \sum_{k=1}^K \langle \tilde{b}^{(s)}\tilde{b}^{(s')} \rangle_k, 
      \label{eq:spe_q} \\
  & & \hat{Q}^{(s,s')} = \rmi \left. \frac{\partial \Omega}{\partial Q^{(s,s')}} \right|_{\mathrm{saddle}}, 
      \label{eq:spe_Qhat} \\
  & & Q^{(s,s')} 
      \HLS = \frac 1K \sum_{k=1}^K \langle \hat{u}^{(s)} \hat{u}^{(s')} \rangle_k \HLE %ADDED.v3
      = 0, 
      \label{eq:spe_Q} \\
  & & \hat{L}^{(s,s')} = \rmi \left. \frac{\partial \Omega}{\partial L^{(s,s')}} \right|_{\mathrm{saddle}}, 
      \label{eq:spe_Lhat} \\
  & & L^{(s,s')} = 
      \displaystyle \lim_{K \to \infty} \frac 1K\sum_{k=1}^K  \langle \tilde{b}^{(s)}\hat{u}^{(s')}\rangle_k, 
      \label{eq:spe_L}
\end{eqnarray}
for all $s$ and $s'$, respectively. 
Comparing them with (\ref{eq:def_average_overlap}) -- (\ref{eq:def_average_SU_response_function}), the following relationships are obtained: 
\begin{eqnarray}
  & & \eta^{(s)} = m^{(s)}, \label{eq:relatonship_m} \\
  & & q^{(s,s')} = C^{(s,s')}, \label{eq:relatonship_C}\\
  & & L^{(s,s')} = \rmi G^{(s,s')}. \label{eq:relatonship_G}
\end{eqnarray}
Therefore, we hereafter make use of $\{ m^{(s)}, C^{(s,s')}, \rmi G^{(s,s')} \}$ instead of $\{ \eta^{(s)}, q^{(s,s')}, L^{(s,s')} \}$. 
It should be noted that causality 
\begin{equation}
  \frac
  {\partial \langle \tilde{b}^{(s)} \rangle}
  {\partial \theta^{(s')}}
  =0, 
\end{equation}
should hold for $s \le s'$, therefore $L^{(s,s')}=G^{(s,s')}=0$ for $s \le s'$. 
\HLS In the next subsection, we calculate the remaining derivatives in the above saddle-point equations. \HLE %ADDED.v3

\subsection{Derivation of Saddle-Point Equations}
%~~~~~~~~~~~~~~~~~~~~~~~~~~~~~~~~~~~~~~~~~~~~~~~~~~~~~~~~~~~~~~~~~~~~~
\par
The integral in $\Omega$ with respect to $\hat{\vec{v}}$ and $\hat{\vec{w}}$ is calculated as 
\begin{eqnarray}
  \Omega
  &=& \frac 1\beta \int \frac{\rmd \hat{\vec{v}}}{(2\pi)^{(t+1)/2}} \frac{\rmd \hat{\vec{w}}}{(2\pi)^{(t+1)/2}} \nonumber \\
  & & \times e^{
      \rmi \hat{\vec{w}}^\top (\beta^{-1} \vec{1}) \hat{\vec{v}} 
      - \frac12 \hat{\vec{v}}^\top \vec{Q}       \hat{\vec{v}}
      - \frac12 \hat{\vec{v}}^\top \vec{B}^\top  \hat{\vec{w}}
      - \frac12 \hat{\vec{w}}^\top \vec{B}       \hat{\vec{v}}
      - \frac12 \hat{\vec{w}}^\top \hat{\vec{D}} \hat{\vec{w}} } \nonumber \\
  &=& \frac 1\beta \int \frac{\rmd \hat{\vec{v}}}{(2\pi)^{(t+1)/2}} \nonumber \\
  & & \times e^{
      - \frac12 \hat{\vec{v}}^\top \vec{Q} \hat{\vec{v}}} 
        |\hat{\vec{D}}|^{-1/2} 
        e^{ -\frac12 \hat{\vec{v}}^\top 
           (\beta^{-1} \vec{1} - \vec{B})^\top \hat{\vec{D}}^{-1} (\beta^{-1} \vec{1} - \vec{B}) \hat{\vec{v}}
        } \nonumber \\
  &=& -\frac 1{2\beta} \{ \ln |\hat{\vec{D}}| \nonumber \\
  & & + \ln |\vec{Q} + (\beta^{-1} \vec{1} - \vec{B})^\top \hat{\vec{D}}^{-1} (\beta^{-1} \vec{1} - \vec{B} )| \}, 
\end{eqnarray}
where $\vec{B}$ and $\hat{\vec{D}}$ are matrices whose elements are defined by 
\begin{eqnarray}
  & & B^{(s,s')} := -\rmi k^{(s')}-G^{(s,s')}, \\
  & & \hat{D}^{(s,s')} := \frac{\sigma_0^2}\beta + 1-m^{(s)}-m^{(s')}+C^{(s,s')}, 
\end{eqnarray}
respectively. 
The $\vec{1}$ denotes an identity matrix. 
\par
The saddle-point equations including $\Omega$ are evaluated as follows. 
We find 
\begin{align}
  \hat{\eta}^{(s)} 
  =& \rmi \left. \frac{\partial \Omega}{\partial m^{(s)}} \right|_{\mathrm{saddle}, \vec{k}=\vec{0}, \vec{Q}=\vec{O}} \nonumber \\
  =& \rmi \frac{\partial}{\partial m^{(s)}} \biggl( -\frac 1{2\beta} \{ \ln |\hat{\vec{D}}| \nonumber \\
   & + \ln |(\beta^{-1} \vec{1} + \vec{G})^\top \hat{\vec{D}}^{-1} (\beta^{-1} \vec{1} + \vec{G} )| \} \biggr) \nonumber \\
  =& \rmi \frac{\partial}{\partial m^{(s)}} \biggl( -\frac 1{\beta} \ln |\beta^{-1} \vec{1} + \vec{G}| \biggr) \nonumber \\
  =& 0, 
\end{align}
and find 
\begin{align}
  \hat{q}^{(s,s')} =0, 
\end{align}
in the same way. 
Similarly, the $\hat{L}^{(s,s')}$ becomes 
\begin{align}
  \hat{L}^{(s,s')} 
  =& \left. \frac{\partial \Omega}{\partial G^{(s,s')}} \right|_{\mathrm{saddle}, \vec{k}=\vec{0}, \vec{Q}=\vec{O}} \nonumber \\
  =& \frac\partial{\partial G^{(s,s')}} \biggl( -\frac 1\beta \ln |\vec{U}| \biggr) \nonumber \\
  \HLS = \HLE & \HLS -\frac1{\beta} \cdot \frac1{|U|} \frac{\partial |U|}{\partial G^{(s,s')}} \nonumber \\ \HLE %ADDED.v3
  \HLS \mathop{=}^{(a)} \HLE & \HLS -\frac1{\beta} \cdot \frac1{|U|} \frac{\partial}{\partial G^{(s,s')}} \sum_{\tau=-1}^{t-1} U^{(s,\tau)} \tilde{U}^{(s,\tau)} \nonumber \\ \HLE %ADDED.v3
  =& -\frac 1\beta \frac{\tilde{U}^{(s,s')}}{|\vec{U}|}, 
\end{align}
where we put $\vec{U} := \beta^{-1} \vec{1} + \vec{G}$, whose $(s,s')$ element is $U^{(s,s')} := \beta^{-1} \delta_{s,s'}+G^{(s,s')}$, and 
$\tilde{U}^{(s,s')}$ denotes a cofactor of the $(s,s')$ element of $\vec{U}$. % cofactor = YO-INSHI
\HLS We here use the cofactor expansion at $\mathop{=}^{(a)}$. \HLE %ADDED.v3
We therefore have 
\begin{align}
  \hat{\vec{L}} 
  =& -\frac 1\beta (\vec{U}^\top)^{-1} \nonumber \\
  =& -(\vec{1}+\beta \vec{G}^\top)^{-1}. 
\end{align}
Since $\vec{Q}=\vec{O}$, $\Omega$ can be expanded with respect to $\vec{Q}$ 
as $\ln |\vec{A} + \vec{Q}| = {\rm tr} \ln \vec{A} + {\rm tr}\, \vec{A}^{-1}\vec{Q}$. 
The $\hat{Q}^{(s,s')}$ can be evaluated as 
\begin{align}
  \hat{Q}^{(s,s')} 
  =& \rmi \lim_{\vec{Q} \to \vec{O}} \left. \frac{\partial \Omega}{\partial Q^{(s,s')}} \right|_{\mathrm{saddle}, \vec{k}=\vec{0}} \nonumber \\
  %\HLS = \HLE & \HLS - \frac{\rmi}{2\beta} \lim_{\vec{Q} \to \vec{O}} \frac{\partial}{\partial Q^{(s,s')}} \biggl\{ \ln |\hat{\vec{D}}| \nonumber \\ \HLE %ADDED.v3
  % & \HLS + \mathrm{tr}\, \ln (\beta^{-1} \vec{1} + \vec{G}) \hat{\vec{D}}^{-1} (\beta^{-1} \vec{1} + \vec{G}^\top) \nonumber \\ \HLE %ADDED.v3
  % & \HLS + \mathrm{tr}\, \ln [(\beta^{-1} \vec{1} + \vec{G}) \hat{\vec{D}}^{-1} (\beta^{-1} \vec{1} + \vec{G}^\top)]^{-1} \vec{Q} \biggr\} \nonumber \\ \HLE %ADDED.v3
  =& - \frac \rmi{2\beta} \lim_{\vec{Q} \to \vec{O}} \frac{\partial}{\partial Q^{(s,s')}} \mathrm{tr}\, \vec{M} \vec{Q} \nonumber \\
  %\HLS = \HLE & \HLS - \frac{\rmi}{2\beta} \lim_{\vec{Q} \to \vec{O}} \frac{\partial}{\partial Q^{(s,s')}} %ADDED.v3
  %\sum_{s=-1}^{t-1} \biggl( \sum_{\tau=-1}^{t-1} M^{(s,\tau)} Q^{(\tau,s)} \biggr) \nonumber \\ \HLE %ADDED.v3
  =& - \frac \rmi{2\beta} M^{(s',s)}, 
\end{align}
where we put 
\begin{eqnarray}
  & & \vec{M} := (\beta^{-1} \vec{1} + \vec{G})^{-1} \hat{\vec{D}} (\beta^{-1} \vec{1} + \vec{G}^\top)^{-1}, \\
  & & D^{(s,s')} := \beta \hat{D}^{(s,s')} \\
  & & \qquad\quad = \sigma_0^2 + \beta [ 1-m^{(s)}-m^{(s')}+C^{(s,s')} ]. 
\end{eqnarray}
We then have 
$\hat{\vec{Q}} = -\rmi \frac 1{2\beta} \vec{M}^\top 
= -\rmi \frac 12 (\vec{1}+\beta \vec{G})^{-1} \vec{D} (\vec{1}+\beta \vec{G}^\top)^{-1}$. 
Finally, we turn to $k^{(s)}$: 
%\HLS We here put $\vec{V}:=\vec{1}-\beta \vec{B}^\top$ whose $(s,s')$ element is %ADDED.v3
%$V^{(s,s')} = \delta_{s,s'} - \beta B^{(s',s)} = \delta_{s,s'} + \rmi \beta k^{(s)} + \beta G^{(s',s)}$, %ADDED.v3
%where $\delta_{s,s'}$ denotes Kronecker's delta taking 1 if $s=s'$, or 0 otherwise. \HLE %ADDED.v3
%We then have 
\begin{align}
  \hat{k}^{(s)} 
  =& \rmi \lim_{\vec{k}\to\vec{0}} \left. \frac{\partial \Omega}{\partial k^{(s,s')}} \right|_{\mathrm{saddle}, \vec{Q}=\vec{O}} \nonumber \\
  =& - \frac{\rmi}{\beta} \lim_{\vec{k}\to\vec{0}} \frac{\partial}{\partial k^{(s)}} \ln |\vec{1}-\beta \vec{B}^\top| \nonumber \\ 
  \HLS \mathop{=}^{(a)} \HLS & \HLS - \frac{\rmi}{\beta} \lim_{\vec{k}\to\vec{0}} \frac1{|\vec{1}-\beta \vec{B}^\top|} \frac{\partial}{\partial k^{(s)}} \nonumber \\ \HLE %ADDED.v3
  & \HLS 
    \left\{
    \left| 
      \begin{array}{ccc}
        \cdots & \delta_{-1,s'} + \rmi \beta k^{(-1)} + \beta G^{(s',-1)} & \cdots \\
        & \vdots & \\ 
        \cdots & \delta_{s,s'} + \beta G^{(s',s)} & \cdots \\
        & \vdots & \\ 
        \cdots & \delta_{t-1,s'} + \rmi \beta k^{(t-1)} + \beta G^{(s',t-1)} & \cdots \\
      \end{array}
    \right| 
    \right. 
    \nonumber \\ 
    \HLE %ADDED.v3
  & \HLS + \rmi \beta k^{(s)} 
    \left. 
    \left| 
      \begin{array}{ccc}
        \cdots & \delta_{-1,s'} + \rmi \beta k^{(-1)} + \beta G^{(s',-1)} & \cdots \\
        & \vdots & \\ 
        \cdots & 1 & \cdots \\
        & \vdots & \\ 
        \cdots & \delta_{t-1,s'} + \rmi \beta k^{(t-1)} + \beta G^{(s',t-1)} & \cdots \\
      \end{array}
    \right| 
    \right\} 
    \nonumber \\ 
    \HLE %ADDED.v3
  =& |\vec{\Lambda}_{[s]}|, 
  \label{eq:hatk}
\end{align}
where $\vec{\Lambda}_{[s]}$ denotes a $(s+2) \times (s+2)$ matrix whose $(s',s'')$ element is given by 
\begin{equation}
  \Lambda_{[s]}^{(s',s'')}=
  \left\{
  \begin{array}{ll}
    \delta_{s',s''} + \beta G^{(s'',s')}, & {\rm for} \; s'\ne s \\
    1, & {\rm for} \; s' = s \\
  \end{array}
  \right. . 
  \label{eq:def_Lambda}
\end{equation}
The $\vec{\Lambda}_{[s]}$ is a matrix whose elements in a row that represents values of stage $s$, 
i.e., the $(s+2)$th row, in $\vec{1}+\beta \vec{G}$ are replaced to $1$. 
Since $|\vec{1}-\beta \vec{B}^\top|$ contains $k^{(s)}$ only in a single row, 
$|\vec{1}-\beta \vec{B}^\top|$ is expanded with respect to the row at $\mathop{=}^{(a)}$ in (\ref{eq:hatk}). 
\begin{example}
  Parameter $\vec{\Lambda}_{[2]}$ has the following form: 
  \begin{equation}
    \vec{\Lambda}_{[2]} = 
    \left(
    \begin{array}{cccc}
      1 & \beta G^{(0,-1)} & \beta G^{(1,-1)} & \beta G^{(2,-1)} \\
      0 & 1                & \beta G^{(1, 0)} & \beta G^{(2, 0)} \\
      0 & 0                & 1                & \beta G^{(2, 1)} \\
      1 & 1                & 1                & 1                \\
    \end{array}
    \right) ,
  \end{equation}
  when $t=2$. 
  \qed
\end{example}

\subsection{Bit Error Rate}
%~~~~~~~~~~~~~~~~~~~~~~~~~~~~~~~~~~~~~~~~~~~~~~~~~~~~~~~~~~~~~~~~~~~~~
\par
One can obtain the polynomial expressions of soft-bits $\tilde{b}_k^{(s)}$, 
which are averaged over the path probability (\ref{eq:def_path_probability}), by GFA. 
Therefore, we can also evaluate the averaged value of analytic functions with respect to the soft-bits. 
The $n$-th order differentiation of the averaged generating functional, 
which has the forms of (\ref{eq:def_barZ}) and (\ref{eq:barZ}), with respect to $\psi_k^{(s)}$ gives 
\begin{align}
  \lim_{\vec{\psi}\to\vec{0}} \biggl( \frac{\partial}{\partial \psi_k^{(s)}} \biggr)^n \bar{Z}[\vec{\psi}]
  \mathop{=}^{(a)} & (- \rmi)^n \overline{\langle (\tilde{b}_k^{(s)})^n \rangle} \nonumber \\
  \mathop{=}^{(b)} & (- \rmi)^n \langle (\tilde{b}_k^{(s)})^n \rangle_k, 
  \label{eq:higherOrder}
\end{align}
where $\mathop{=}^{(a)}$ and $\mathop{=}^{(b)}$ are derived by differentiation 
of (\ref{eq:def_barZ}) and (\ref{eq:barZ}), repsectively. 
For arbitrary function $F(x)$ that can be expanded around $x=0$, we thus have the identity: 
\begin{align}
  \overline{\langle F(\tilde{b}_k^{(s)}) \rangle} 
  = \langle F(\tilde{b}_k^{(s)}) \rangle_k. 
  \label{eq:forArbitraryFunction}
\end{align}
\par
The Hamming distance between single letters is defined by 
\begin{equation}
  d(x,\hat{x}) := \biggl\{
  \begin{array}{ll}
    0, & \; {\rm if} \; x = \hat{x} \\
    1, & \; {\rm if} \; x \ne \hat{x} 
  \end{array} .
\end{equation}
The Hamming distance between $K$-bit sequences $\vec{b}$ and $\hat{\vec{b}}$ is measured 
by the averaged single-letter distortion as $d(\vec{x},\hat{\vec{x}}) := K^{-1} \sum_{k=1}^K d(b_k,\hat{b}_k)$. 
The hard-decision estimates can be represented by $\hat{b}_k^{(t)} = \sgn(\tilde{b}_k^{(t)})$ 
If we therefore choose $F(x)$ such as a function which approaches $\sgn(x)$, 
e.g., $\lim_{a\to\infty} \tanh(ax)$, 
the bit error rate $P_b^{(t)}$ can then be evaluated as 
\begin{align}
  P_b^{(t)} 
  :=& \overline{\langle d(\vec{b},\hat{\vec{b}}^{(t)} ) \rangle} \\
  =& \frac 12 \biggl(1 - \frac1K \sum_{k=1}^K \langle \sgn(\tilde{b}_k^{(t)}) \rangle_k \biggr), 
\end{align}
where $d(b,\hat{b})=(1-b\hat{b})/2$ for $b, \hat{b} \in \{ -1, 1 \}$. 
% Since 
% $\overline{\langle \lim_{\gamma' \to \infty} \sum_{n=0}^\infty a_n (\gamma' \tilde{b}_k^{(t)})^n \rangle}$ 
% $=\lim_{\gamma' \to \infty}$ $\sum_{n=0}^\infty$ $a_n \gamma'^n$ $\overline{\langle (\tilde{b}_k^{(t)})^n \rangle}$ 
% $=\lim_{\gamma' \to \infty}$ $\sum_{n=0}^\infty$ $a_n \gamma'^n$ $\langle (\tilde{b}_k^{(t)})^n \rangle_k$ 
% $=\langle \lim_{\gamma' \to \infty}$ $\tanh(\gamma' \tilde{b}_k^{(t)}) \rangle_k$ 
% by using (\ref{eq:higherOrder}) and the Taylor expansion of $\tanh(x) = \sum_{n=1}^\infty a_n x^{2n-1}$ 
% with $a_n := B_{2n} 4^n(4^n-1)/(2n)!$, 
% where $B_n := \lim_{t\to0} (\partial/\partial x)^n (x/(e^x-1))$ denotes the Bernoulli number. 

\subsection{Effective Single-User Problem and Analytical Result}
%~~~~~~~~~~~~~~~~~~~~~~~~~~~~~~~~~~~~~~~~~~~~~~~~~~~~~~~~~~~~~~~~~~~~~
\par
We here summarize our calculation. 
Some macroscopic parameters are found to vanish in the saddle-point: $k^{(s)}=Q^{(s,s')}=0$. 
The remaining macroscopic parameters can all be expressed in terms of three observables: 
the average overlap $m^{(s)}$, 
the average single-user correlation $C^{(s,s')}$, 
and the average single-user response function $G^{(s,s')}$, 
which are defined by (\ref{eq:def_average_overlap}) -- (\ref{eq:def_average_SU_response_function}). 
The averaged generating functional $\bar{Z}[\vec{\psi}]$ is dominated by a saddle-point for $K\to\infty$. 
We can thus simplify the saddle-point problem of Lemma \ref{lemma:DisorderAveragedGeneratingFunctional} in the large system limit. 
\par
We derive a single-user saddle-point problem. 
Note that we remained the user index $k$ to deduce the meaning of macroscopic parameters 
with respect to external messages $\{\theta_k^{(s)}\}$ and dummy functions $\{\psi_k^{(s)}\}$. 
We set these parameters as 
\begin{eqnarray}
  & & \psi_k^{(s)} \to 0, \label{eq:psi=0} \\
  & & \theta_k^{(s)} \to 0. \label{eq:theta=0}
\end{eqnarray}
%\begin{eqnarray}
%  & & \psi_k^{(s)} \to 0, \\
%  & & \theta_k^{(s)} \to \theta^{(s)}. 
%\end{eqnarray}
%We write the replacement of $\theta_k^{(s)}$ into a user-independent external message $\theta^{(s)}$ as 
%\begin{equation}
%  \vec{\theta}\to{\rm uim}. 
%\end{equation}
%In the end of analysis, the external message is set to $\theta^{(s)}=0$. 
Consequently, the single-user measure (\ref{eq:def_Single-UserMeasure}) becomes user independent. 
We then arrive at the following proposition. 
\par
%-------------------------------------------------PROPOSITION
\begin{proposition} \label{proposition:soft-PIC}
  Setting $\gamma\to\infty$ and $\theta^{(s)}=0$, 
  the dynamics of soft-PIC can be obtained as the following equations in the large system limit, i.e., $K\to\infty$.  
  \begin{eqnarray}
    m^{(s)}&=&\llangle \tilde{b}^{(s)} \rrangle, \label{eq:sp_m} \\
    C^{(s,s')}&=&\llangle \tilde{b}^{(s)} \tilde{b}^{(s')} \rrangle, \label{eq:sp_C} \\
    G^{(s,s')}&=&\llangle \tilde{b}^{(s)}(\vec{R}^{-1}\vec{v})^{(s')} \rrangle, \label{eq:sp_G} 
  \end{eqnarray}
  with the causality: $G^{(s,s')}=0 \; {\rm for} \; s \le s'$.
  The bit error rate of hard decisions $\{\sgn[\tilde{b}_k^{(s)}]\}$ at the $t^{\rm th}$ stage of soft-PIC (\ref{eq:deterministic_updating_rule}) is obtained by 
  \begin{equation}
    P_b^{(s)}= \frac 12 (1-\llangle \sgn ( \tilde{b}^{(s)} ) \rrangle ). 
    \label{eq:BER}
  \end{equation}
  The effective path measure is given by 
  \begin{align}
    & \llangle g(\tilde{\vec{b}},\vec{v}) \rrangle \nonumber \\ 
    & := \int {\cal D}\vec{v} \int_{\mathbb{R}^{t+2}} 
     \biggl( \prod_{s=-1}^{t-1} \rmd \tilde{b}^{(s)} \biggr) 
      g(\tilde{\vec{b}},\vec{v}) \; \delta [\tilde{b}^{(-1)}] \notag \\
    & \times \prod_{s=-1}^{t-1} \delta [ 
        \tilde{b}^{(s+1)}  - 
        f(\hat{k}^{(s)}+v^{(s)}+(\vec{\Gamma} \tilde{\vec{b}})^{(s)})
      ], 
    \label{eq:Dv}
  \end{align}
  where 
  \begin{eqnarray}
    && {\cal D}\vec{v} := \rmd \vec{v} |2\pi \vec{R}|^{-1/2} e^{-\frac 12\vec{v}\cdot \vec{R}^{-1}\vec{v}}, \\
    && \vec{R} = (\vec{1}+\beta \vec{G}^\top)^{-1} \vec{D} (\vec{1}+\beta \vec{G})^{-1}, \label{eq:R} \\
    && \vec{\Gamma} = (\vec{1}+\beta \vec{G})^{-1}\beta \vec{G}, \label{eq:Gamma_softPIC} \\
    && \hat{k}^{(s)} = |\vec{\Lambda}_{[s]}|. 
  \end{eqnarray}
  with 
  \begin{align}
    D^{(s,s')} :=& \sigma_0^2 + \beta [ 1-m^{(s)}-m^{(s')}+C^{(s,s')} ], \label{eq:elementOfD} \\
    \Lambda_{[s]}^{(s',s'')} =&
    \left\{
    \begin{array}{ll}
      \delta_{s',s''} + \beta G^{(s'',s')}, & {\rm for} \; s'\ne s \\
      1, & {\rm for} \; s' = s. \\
    \end{array}
    \right. \label{eq:sp_Lambda}
  \end{align}
  The terms $(\vec{R}^{-1}\vec{v})^{(s)}$ and $(\vec{\Gamma} \tilde{\vec{b}})^{(s)}$ denote 
  the $s^{\rm th}$ element of the vector $\vec{R}^{-1}\vec{v}$ and $\vec{\Gamma} \tilde{\vec{b}}$, respectively. 
  \qed
\end{proposition}
\par
\HLS Details of derivation is available in Appendix \ref{appendix:soft-PIC}. \HLE %ADDED.v3
Equations (\ref{eq:sp_m})-(\ref{eq:sp_Lambda}) entirely describe the dynamics of the system. 
Term $(\vec{\Gamma} \tilde{\vec{b}})^{(s)}$ in (\ref{eq:Dv}) is called the {\it Onsager reaction term}.

\subsection{Numerical Analysis and Experiments}
%~~~~~~~~~~~~~~~~~~~~~~~~~~~~~~~~~~~~~~~~~~~~~~~~~~~~~~~~~~~~~~~~~~~~~
\par
To validate the results obtained here, we performed numerical experiments in an $N=8,000$ system. 
%We chose $f(x)=\tanh(x/\sigma_0^2)$ as the transfer function of soft-PIC. 
The system of $N=8,000$ is too large for a practical system, 
but we are now concerned with the verification of the analytical result derived under the large system limit. 
\par
\begin{figure}[t]%[htbp]
  \begin{center}
    \includegraphics[width=.8\linewidth,keepaspectratio]{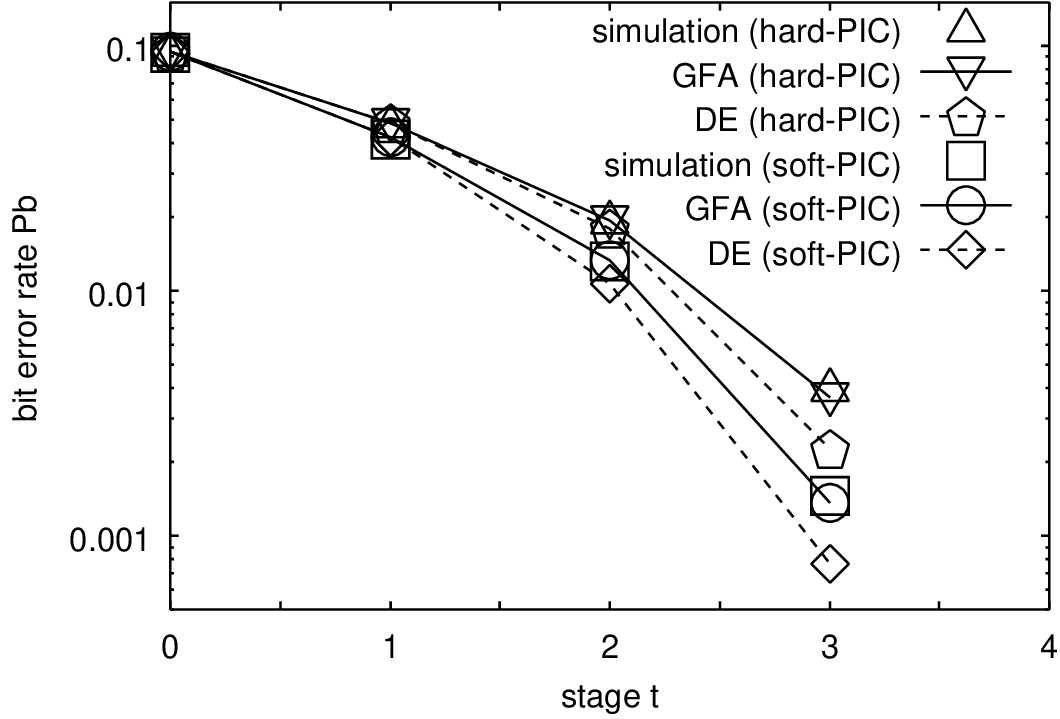} \\
    {\footnotesize \hspace*{7ex} (a)} \\[1em]
    \includegraphics[width=.8\linewidth,keepaspectratio]{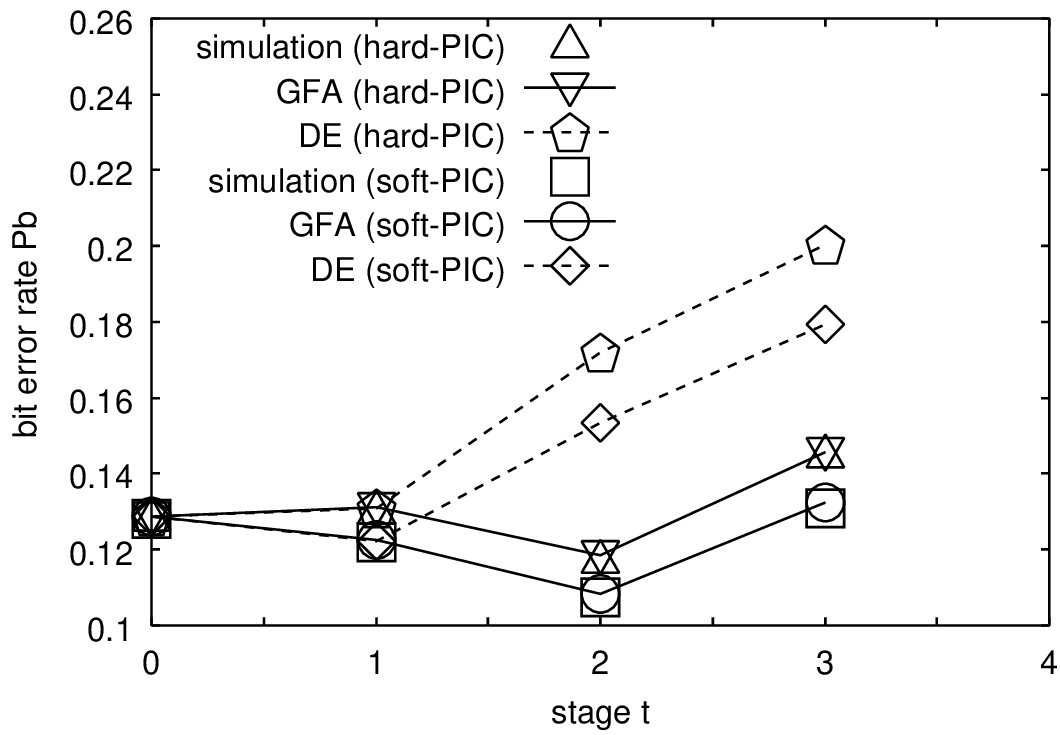} \\
    {\footnotesize \hspace*{7ex} (b)} \\
    \caption{
    First few stages of detection dynamics of the hard-PIC (upper) and soft-PIC 
    with $f(x)=\tanh(x/\sigma^2)$ with $\sigma=\sigma_0$ (lower) predicted by GFA (solid lines) and density evolution (dashed lines). 
    Computer simulations (triangles and squares) were evaluated with $N=8,000$ for $E_b/N_0=8$ [dB]. 
    The system loads were (a) $\beta=0.5<\beta_c$ and (b) $\beta=0.7>\beta_c$, respectively. 
    %from 100 experiments
    }
    \label{fig:soft-PIC}
  \end{center}
\end{figure}

Figure \ref{fig:soft-PIC}(a) plots the first few stages of the detection dynamics 
of the hard-PIC ($f(x)=\sgn(x)$) and soft-PIC ($f(x)=\tanh(x/\sigma^2)$ with $\sigma=\sigma_0$) 
predicted by GFA and density evolution \cite{Tanaka2005} for $E_b/N_0=8$ [dB]. 
%where $E_b/N_0$ [dB] denotes $10 \log_{10} E_b/N_0$. %DELETED
The detailed derivation of the density evolution analysis is available in the reference \cite{Tanaka2005}. 
The system load is $\beta=0.5<\beta_c$, where $\beta_c$ is the critical system load defined 
as the minimum system load at which the dynamics fail to converge to the replica-symmetric solution of the marginal-posterior-mode detector. 
The critical load $\beta_c$ of soft-PIC for $E_b/N_0=8$ [dB] is about $0.6$. 
The predictions of density evolution systematically deviate from computer simulation results at transients. 
\par
Figure \ref{fig:soft-PIC}(b) plots the first few stages of the detection dynamics of the hard-PIC 
and soft-PIC for $E_b/N_0=8$ [dB], predicted by GFA and density evolution with the system load of $\beta=0.7>\beta_c$. 
Oscillation of the detection dynamics was observed, when $\beta>\beta_c$. 
The density evolution results have residual deviations in Fig. \ref{fig:soft-PIC} 
due to the lack of the Onsager reaction term and the assumption that the summation over all messages follows a Gaussian distribution. 
In particular, the deviation between the density evolution predictions and the simulation results becomes large when $\beta>\beta_c$. 
In contrast, GFA exhibits good consistency with the simulation results. 

\subsection{Derivation of Existing Results by Generating Functional Analysis}
%~~~~~~~~~~~~~~~~~~~~~~~~~~~~~~~~~~~~~~~~~~~~~~~~~~~~~~~~~~~~~~~~~~~~~
\par
If we put $\vec{\Gamma} \tilde{\vec{b}} = \vec{0}$, viz., if we neglect the Onsager reaction term, 
the GFA recovers the density evolution framework based on the statistical neurodynamics \cite{Tanaka2005}. 
\par
The multiple integral including (\ref{eq:Dv}) can be partially performed, 
when we put $\vec{\Gamma} \tilde{\vec{b}} = \vec{0}$. 
Namely, $m^{(s)}$, $G^{(s')}$, and $P_b^{(s)}$ only depend on $v^{(s-1)}$ among $\vec{v}$; 
and $C^{(s,s')}$ only depends on $v^{(s-1)}$ and $v^{(s'-1)}$. 
We here separate $\vec{v}$ into two sub-vectors: $\vec{v}=(\vec{v}_1,\vec{v}_2)^\top$. 
The correlation matrix $\vec{R}$ can then be represented as 
\begin{align}
  \vec{R}=
  \left(
  \begin{array}{cc}
    \vec{R}_{11} & \vec{R}_{12} \\
    \vec{R}_{21} & \vec{R}_{22} \\
  \end{array}
  \right) . 
\end{align}
For example, $\vec{v}_1$ and $\vec{R}_{11}$ are chosen as $\vec{v}_1= (v^{(s-1)},v^{(s'-1)})^\top$ and 
\begin{align}
  \vec{R}_{11}=
  \left(
  \begin{array}{cc}
    R^{(s -1,s-1)} & R^{(s -1,s'-1)} \\
    R^{(s'-1,s-1)} & R^{(s'-1,s'-1)} \\
  \end{array}
  \right) , 
\end{align}
to calculate (\ref{eq:sp_C}). 
For arbitrary function $f(\vec{v}_1)$, the following identity holds: 
\begin{align}
  \int \mathcal{D}\vec{v} f(\vec{v}_1) 
  = \frac 1{\sqrt{|2\pi\vec{R}_{11}|}} 
    \int \rmd \vec{v}_1 e^{-\frac 12 \vec{v}_1 \cdot \vec{R}_{11}^{-1} \vec{v}_1} f(\vec{v}_1). 
\end{align}
Since it turns out that the response function becomes zero except for $G^{(s,s-1)}$, 
the $\hat{k}^{(s)}$ has a simple form. 
\par
We then obtain the following result. 
The bit error rate $P_b(t)$ of hard decisions $\hat{b}_k(t)=\sgn[u_k(t-1)]$ at the $t$th stage is given by 
\begin{equation}
  P_b(t) = \mathcal{Q} \biggl( \frac {\hat{k}^{(s-1)}} {\sqrt{R^{(s-1,s'-1)}}} \biggr), 
\end{equation}
where $\mathcal{Q}(z) := \int_z^\infty \rmD x$ denotes the error function 
and we here put $\rmD z := \rmd z (2\pi)^{-1/2} e^{-z^2/2}$ to simplify the notation. 
The $m^{(s)}$ are to be evaluated with the following set of recursive equations: 
\begin{align}
  \hat{k}^{(s)} =& 1-\beta G^{(s,s-1)} \hat{k}^{(s-1)}, 
\end{align}
\begin{align}
  R^{(s,s')} =& D^{(s,s')} + \beta^2 G^{(s,s-1)} G^{(s',s'-1)} R^{(s-1,s'-1)} \nonumber \\
              & + \sum_{\lambda=-1}^{s -1} D^{(\lambda,s')} \prod_{\tau=\lambda+1}^{s } (-\beta G^{(\tau,\tau-1)}) \nonumber \\
              & + \sum_{\lambda=-1}^{s'-1} D^{(\lambda,s )} \prod_{\tau=\lambda+1}^{s'} (-\beta G^{(\tau,\tau-1)}), 
\end{align}
\begin{align}
  D^{(s,s')} =& \sigma_0^2 + \beta ( 1 - m^{(s)} - m^{(s')} + C^{(s,s')} ), 
\end{align}
\begin{align}
  m^{(s+1)}  =& \int_{\mathbb{R}} \rmD z f(\hat{k}^{(s-1)} + z \sqrt{R^{(s-1,s-1)}} ), 
\end{align}
\begin{align}
  G^{(s,s-1)}=& \frac1{\sqrt{R^{(s-1,s-1)}}} \int_{\mathbb{R}} \rmD z z f( \hat{k}^{(s-1)} + z \sqrt{R^{(s-1,s-1)}} ), 
\end{align}
\begin{align}
  C^{(s,s')} =& \int_{\mathbb{R}^3} \rmD z \rmD u \rmD v \nonumber \\
              & \times f (\hat{k}^{(s -1)} + z \sqrt{R^{(s-1,s'-1)}} \nonumber \\
              & \qquad + u \sqrt{R^{(s -1,s -1)}-R^{(s-1,s'-1)}} ) \nonumber \\
              & \times f (\hat{k}^{(s'-1)} + z \sqrt{R^{(s-1,s'-1)}} \nonumber \\
              & \qquad + u \sqrt{R^{(s'-1,s'-1)}-R^{(s-1,s'-1)}} ). 
\end{align}
The initialization condition is 
$R^{(-1,-1)}=D^{(-1,-1)}=\sigma_0^2+\beta$, $\hat{k}^{(-1)}=1$
and $m^{(-1)}=G^{(-1,-1)}=C^{(-1,-1)}=0$, 
\par
This result is identical to that of density evolution \cite{Tanaka2005}. 
In the derivation by means of density evolution, 
it is assumed that the local field $u_k(t)$ follows the Gaussian distribution with mean $B_t$ and covariance $C_{t,\tau}$. 
Furthermore, the Onsager reaction term is ignored. 
The GFA, on the other hand, can treat the Onsager reaction term correctly.

\section{Generating Functional Analysis for ORC-PIC \label{section:GFA_ORC-PIC}}
%~~~~~~~~~~~~~~~~~~~~~~~~~~~~~~~~~~~~~~~~~~~~~~~~~~~~~~~~~~~~~~~~~~~~~
\par
This section discusses the dynamics of ORC-PIC. 
%-------------------------------------------------PROPOSITION
\begin{proposition} \label{proposition:ORC-PIC}
  The dynamics of ORC-PIC is given by 
  the following average over the effective path measure: 
  \begin{align}
    & \llangle g(\tilde{\vec{b}},\vec{v}) \rrangle \nonumber \\
    & := \int {\cal D}\vec{v} \int_{\mathbb{R}^{t+2}} 
      \biggl( \prod_{s=-1}^{t-1} \rmd \tilde{b}^{(s)} \biggr) 
      g(\tilde{\vec{b}},\vec{v}) \; \delta [ \tilde{b}^{(-1)} ] \notag \\
    & \times \prod_{s=-1}^{t-1} \delta [ 
        \tilde{b}^{(s+1)}-f(\hat{k}^{(s)}+v^{(s)}+(\vec{\Gamma} \tilde{\vec{b}})^{(s)}-(\hat{\vec{\Gamma}} \tilde{\vec{b}})^{(s)})
      ]. 
    \label{eq:Dv-ORC-PIC}
  \end{align}
  All other parameters are identical to those in Proposition \ref{proposition:soft-PIC}. 
  \qed
\end{proposition}
\par
For ORC-PIC, the summation over all messages, 
which is the argument of $f(\cdots)$ in (\ref{eq:def_ORC-PIC}) becomes 
\begin{align}
  u_k^{(t)} =
  & \sum_{k=1}^K W_{kk'} (b_{k'} -\tilde{b}_{k'}^{(t)}) 
    + \tilde{b}_k^{(t)}
    \nonumber \\
  & - \sum_{s'=-1}^s \hat{\Gamma}_k^{(s,s')} \tilde{b}_k^{(s')}
    + \frac{\sigma_0}{\sqrt{N}} \sum_{\mu=1}^N s_k^\mu n^\mu 
    + \theta_k^{(t)}, 
  \label{eq:u_k_ORC-PIC}
\end{align}
with external messages $\{\theta_k^{(t)}\}$. 
The averaged generating functional $\bar{Z}[\vec{\psi}]$ is represented as 
\begin{align}
  \bar{Z}[\vec{\psi}] 
  =& \mathbb{E}_{\vec{s}_1,\cdots,\vec{s}_K,\vec{n}} \biggl[ \int_{\mathbb{R}^{(t+1)K}} \underline{\rmd}\vec{u} \nonumber \\
  & \times 
    p[\tilde{\vec{b}}^{(-1)}] 
    \biggl( \prod_{s=-1}^{t-1} \frac{\gamma}{\sqrt{2\pi}} e^{-\frac{\gamma^2}2 \{ \tilde{b}_k^{(s+1)}-f(u_k^{(s)}) \}^2} \biggr) 
    \nonumber \\
  & \times \exp \biggl[ -\rmi\sum_{s=-1}^t\tilde{\vec{b}}^{(s)} \cdot \vec{\psi}^{(s)} \biggr]
    \nonumber \\
  & \times \prod_{s=0}^{t-1} \prod_{k=1}^K 
      \delta \biggl( u_k^{(s)} - \sum_{k=1}^K W_{kk'} (b_{k'}-\tilde{b}_{k'}^{(s)}) - \tilde{b}_k^{(s)}  \nonumber \\
  & + \sum_{s'=-1}^s \hat{\Gamma}_k^{(s,s')} \tilde{b}_k^{(s')}
    - \frac{\sigma_0}{\sqrt{N}} \sum_{\mu=1}^N s_k^\mu n^\mu 
    - \theta_k^{(s)} \biggr) \biggr] . 
  \label{eq:def_barZ_ORC-PIC}
\end{align}
The difference between (\ref{eq:def_barZ}) and (\ref{eq:def_barZ_ORC-PIC}) is 
only in the fourth term $\sum_{s'=-1}^s \hat{\Gamma}_k^{(s,s')} \tilde{b}_k^{(s')}$ 
in $\delta ( \cdots )$ in (\ref{eq:def_barZ_ORC-PIC}). 
One can straightforwardly show Proposition \ref{proposition:ORC-PIC} 
in the same manner as the derivation of Proposition \ref{proposition:soft-PIC}. 
The difference in the analytical results only appears in (\ref{eq:Dv}) of Proposition \ref{proposition:soft-PIC}. 
\par
We will now consider how to choose matrix $\hat{\vec{\Gamma}}=(\hat{\Gamma}^{(s,s')})$ 
in the Onsager reaction canceling term to cancel the Onsager reaction term. 
%-------------------------------------------------PROPOSITION
\begin{proposition}
  \label{proposition:HowToChooseGammaHat}
  When the Onsager reaction term $(\hat{\vec{\Gamma}} \hat{\vec{b}})^{(t)}$ is chosen as 
  \begin{align}
    (\hat{\vec{\Gamma}} \hat{\vec{b}})^{(t)} 
     = \sum_{s=0}^{t-1}(-1)^{t-s-1}\beta^{t-s}\biggl( \prod_{\tau=s}^{t-1} G^{(\tau+1,\tau)} \biggr) \tilde{b}^{(s)}, 
    \label{eq:HowToChooseGammaHat}
  \end{align}
  for $t>0$ and $(\hat{\vec{\Gamma}} \hat{\vec{b}})^{(-1)}=(\hat{\vec{\Gamma}} \hat{\vec{b}})^{(0)}=0$, 
  then the Onsager reaction term $\vec{\Gamma} \hat{\vec{b}}$ is canceled, 
  viz., $\vec{\Gamma} \hat{\vec{b}} - \hat{\vec{\Gamma}} \hat{\vec{b}} =O$, 
  and the response functions become $G^{(t,t')}=0$ except for $t=t'+1$. 
  \qed
\end{proposition}
\par
Details of derivation is available in Appendix \ref{appendix:HowToChooseGammaHat}. 
Note that (\ref{eq:GammaHatRecursive}) is a recursive formula of (\ref{eq:HowToChooseGammaHat}). 
% Using Proposition \ref{proposition:ORC-PIC}, we arrive the following PIC algorithm. 
One can confirm the ORC-PIC algorithm can correctly cancel the Onsager reaction term by GFA. 
The density evolution results completely agree with the GFA results \cite{Tanaka2005}. 
The density evolution curves of Fig. \ref{fig:soft-PIC} are equivalent to the 
%proformance %DELETED.v3
\HLS performance \HLE %ADDED.v3
of ORC-PIC. 
\par
\begin{figure}[t]%[htbp]
  \begin{center}
    \includegraphics[width=.8\linewidth,keepaspectratio]{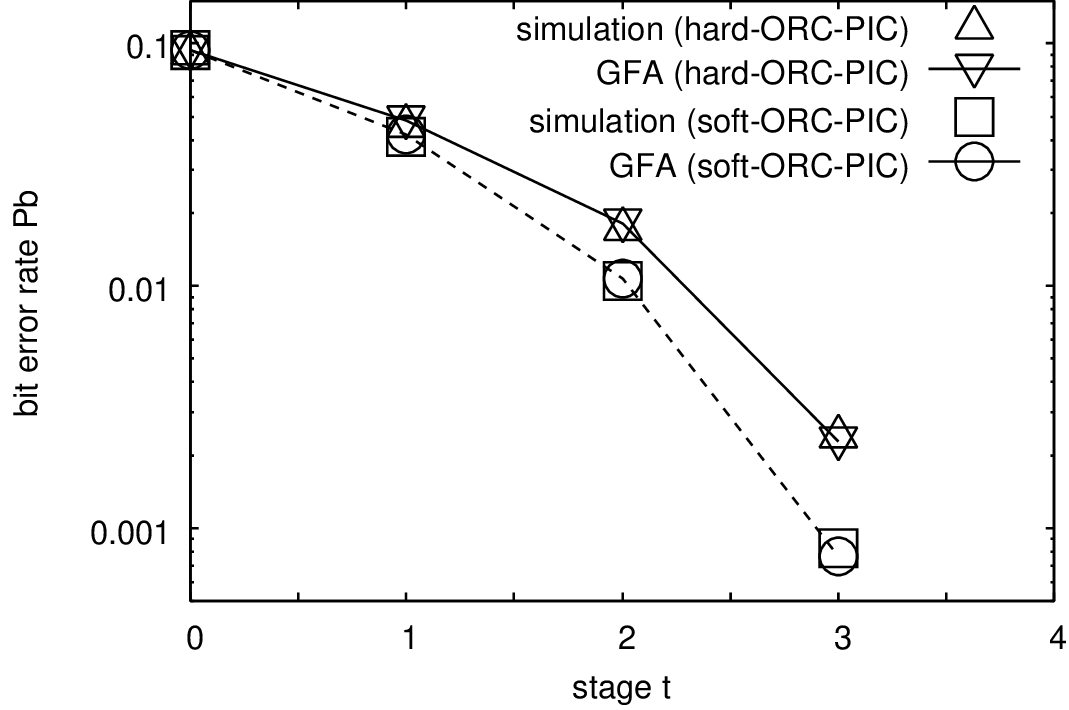} \\
    {\footnotesize \hspace*{7ex} (a)} \\[1em]
    \includegraphics[width=.8\linewidth,keepaspectratio]{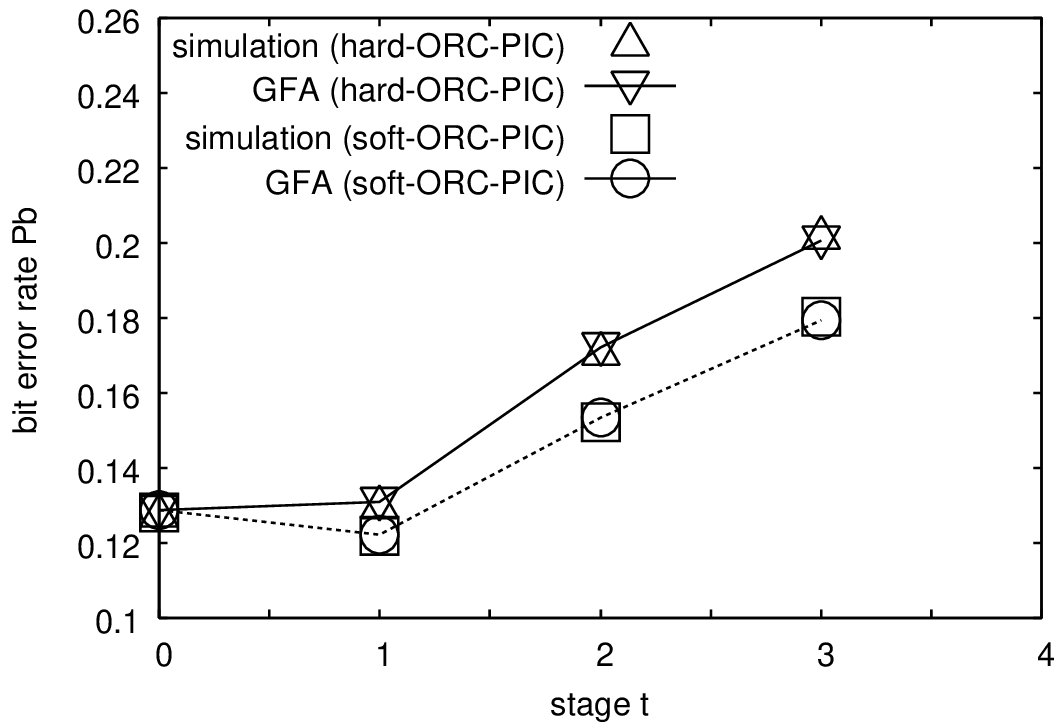} \\
    {\footnotesize \hspace*{7ex} (b)} \\
    \caption{
    First few stages of detection dynamics of the hard-ORC-PIC (upper) and soft-ORC-PIC 
    with $f(x)=\tanh(x/\sigma^2)$ with $\sigma=\sigma_0$ (lower) predicted by GFA (solid lines). 
    In this case, the density evolution result is identical to the GFA result. 
    Computer simulations (triangles and squares) were evaluated with $N=8,000$ for $E_b/N_0=8$ [dB]. 
    The system loads were (a) $\beta=0.5<\beta_c$ and (b) $\beta=0.7>\beta_c$, respectively. 
    %from 100 experiments
    }
    \label{fig:ORC-PIC}
  \end{center}
\end{figure}

To validate the results obtained here, we performed numerical experiments in an $N=8,000$ system. 
%We chose $f(x)=\tanh(x/\sigma_0^2)$ as the transfer function of soft-ORC-PIC. 
Figure \ref{fig:ORC-PIC}(a) plots the first few stages of the detection dynamics 
of the hard-ORC-PIC ($f(x)=\sgn(x)$) and soft-ORC-PIC ($f(x)=\tanh(x/\sigma^2)$ with $\sigma=\sigma_0$) predicted by GFA \cite{Tanaka2005} for $E_b/N_0=8$ [dB]. 
Figure \ref{fig:ORC-PIC}(b) plots the first few stages of the detection dynamics of the hard-ORC-PIC 
and soft-ORC-PIC for $E_b/N_0=8$ [dB], predicted by GFA and density evolution with the system load of $\beta=0.7>\beta_c$. 
Oscillation of the detection dynamics was observed, when $\beta>\beta_c$. 

\section{Generating Functional Analysis for BP-based detector \label{section:GFA_BP}}
%~~~~~~~~~~~~~~~~~~~~~~~~~~~~~~~~~~~~~~~~~~~~~~~~~~~~~~~~~~~~~~~~~~~~~
\par
We next apply this scheme to the BP-based detector. 
One can obtain the following result in the same manner as soft-PIC. 
%-------------------------------------------------PROPOSITION
\begin{proposition} \label{proposition:BP-based_detector}
  The dynamics of the BP-based detector represented by (\ref{eq:BP-basedDetector}) 
  is described by equations (\ref{eq:sp_m})-(\ref{eq:Dv}) and the following equations: 
  $\vec{R} =$ $(\vec{1} - \beta \vec{B})^{-1} \vec{D} (\vec{1} - \beta \vec{B}^\top)^{-1}$, 
  $\vec{\Gamma} =$ ${\rm diag} (A^{(-1)},$ $\cdots, A^{(t-1)})$ 
  $- [ \sum_{s=-1}^{t-1}$ $\vec{J}_{s+1} \otimes \vec{\Delta}^{s+1} (\vec{U}^\top)^{-1} ]^\top$ and 
  $\hat{k}^{(s)} = |\vec{\Lambda}_{[s]}|$ 
  where $\vec{B}$, $\vec{U}$, and $\vec{D}$ are matrices having matrix elements 
  \begin{align}
    B^{(s,s')} = - \sum_{\tau=-1}^s J^{(s,\tau)} G^{(\tau,s')}, 
  \end{align}
  \begin{align}
    U^{(s,s')} = \delta_{s,s'} + \beta \sum_{\tau=-1}^s J^{(s,\tau)} G^{(\tau,s')}, 
    \label{eq:def_U}
  \end{align}
  \begin{align}
    D^{(s,s')} =& \; \sigma_0^2 R^{(s)} R^{(s')} + \beta \biggl[ R^{(s)} R^{(s')} \notag \\ 
    & - R^{(s)} \!\!\! \sum_{\tau'=-1}^{s'} \!\!\! J^{(s',\tau')} m^{(\tau')} - R^{(s')} \!\!\! 
      \sum_{\tau=-1}^{s} \!\!\! J^{(s,\tau)} m^{(\tau)} \notag \\
    & + \sum_{\tau=-1}^{s} \sum_{\tau'=-1}^{s'} J^{(s,\tau)} J^{(s',\tau')} C^{(\tau,\tau')} \biggr], 
  \end{align}
  and 
  \begin{align}
    \Lambda_s^{(s',s'')} =
    \left\{ \!\!\!
    \begin{array}{ll}
      \displaystyle \delta_{s',s''} + \beta \sum_{\tau'=-1}^{s'} J^{(s',\tau')} G^{(\tau',s')}, & \!\! s'\ne s \\
      R^{(s')}, & \!\! s' = s. \\
    \end{array}
    \right. , 
    \label{eq:def_Lambda}
  \end{align}
  and matrix $\vec{J}_s$ denotes 
  \begin{align}
    \vec{J}_s := 
    \left(
    \begin{array}{ccc}
      J^{(s-1,-1)} & \cdots & J^{(s-1,-1)} \\
      \vdots & & \vdots \\
      J^{(t-1,t-s-1)} & \cdots & J^{(t-1,t-s-1)} \\
      0 & \cdots & 0 \\
      \vdots & & \vdots \\
      0 & \cdots & 0 \\
    \end{array}
    \right). 
    \label{eq:def_Js}
  \end{align}
  Here, $\vec{\Delta} = (\delta_{s+1,s'})$ denotes a $(t+1)\times(t+1)$ matrix 
  whose $(s,s')$ element is given by $\delta_{s+1,s'}$, 
  and the operator $\otimes$ denotes the Hadamard product 
  $A \otimes B=(a_{ij} b_{ij})$ for $A=(a_{ij})$ and $B=(b_{ij})$. 
  \qed
\end{proposition}
\par
\begin{figure}[t]%[htbp]
  \begin{center}
    \includegraphics[width=.8\linewidth,keepaspectratio]{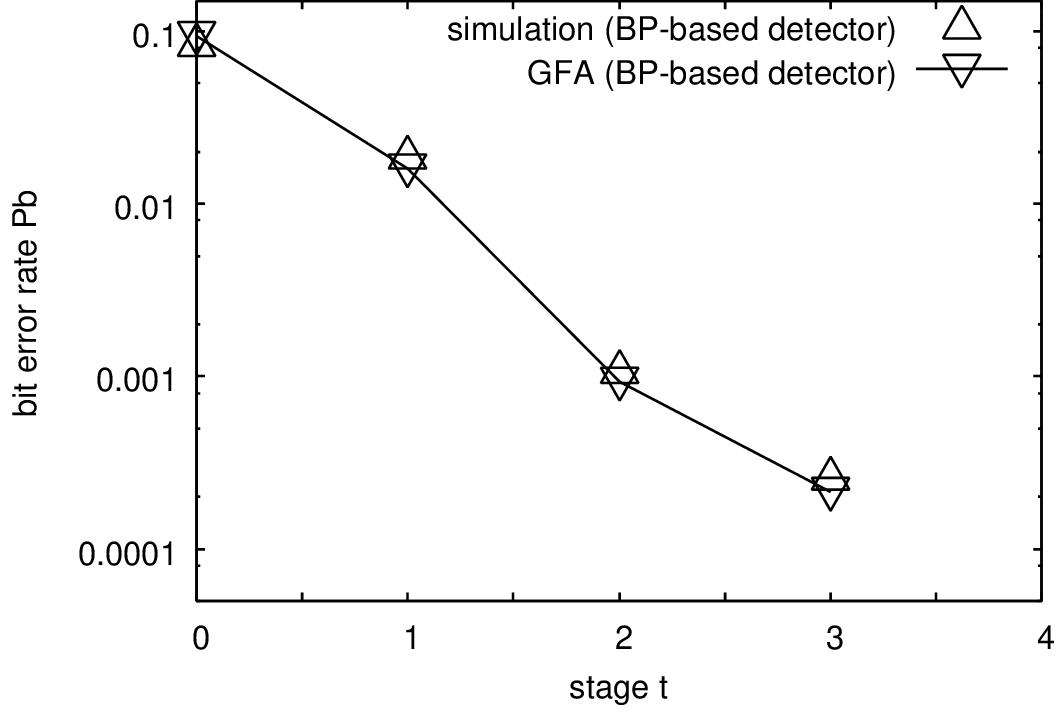}
    \caption{
    First few stages of detection dynamics of the BP-based detector predicted by GFA (solid lines). 
    Computer simulations (triangles and squares) were evaluated with $N=8,000$ for $E_b/N_0=8$ [dB]. 
    The control parameter and the system load were set as $\sigma=\sigma_0$ and $\beta=0.5$. 
    %from 100 experiments
    }
    \label{fig:BP-based}
  \end{center}
\end{figure}

For the BP-based detector, the summation over all messages, 
which is the argument of $\tanh(\cdots)$ of (\ref{eq:BP-basedDetector}) becomes 
\begin{align}
  u_k^{(t)} =
  R^{(t)} h_k  + A^{(t)} \tilde{b}_k^{(t)} 
  - \sum_{s=-1}^t J^{(t,s)} \sum_{k'=1}^K W_{kk'} \tilde{b}_k^{(s)}. 
  \label{eq:u_k_BP-basedDetector}
\end{align}
The averaged generating functional $\bar{Z}[\vec{\psi}]$ is represented as 
\begin{align}
  \bar{Z}[\vec{\psi}] 
  =& \mathbb{E}_{\vec{s}_1,\cdots,\vec{s}_K,\vec{n}} \biggl[ \int_{\mathbb{R}^{(t+1)K}} \underline{\rmd}\vec{u} \nonumber \\
  & \times 
    p[\tilde{\vec{b}}^{(-1)}] 
    \biggl( \prod_{s=-1}^{t-1} \frac{\gamma}{\sqrt{2\pi}} e^{-\frac{\gamma^2}2 \{ \tilde{b}_k^{(s+1)}-f(u_k^{(s)}) \}^2} \biggr) 
    \nonumber \\
  & \times \exp \biggl[ -\rmi\sum_{s=-1}^t \tilde{\vec{b}}^{(s)} \cdot \vec{\psi}^{(s)} \biggr]
    \nonumber \\
  & \times \prod_{s=0}^{t-1} \prod_{k=1}^K 
      \delta \biggl( u_k^{(s)} - R^{(t)} h_k - A^{(t)} \tilde{b}_k^{(t)} \nonumber \\
  & + \sum_{s=-1}^t J^{(t,s)} \sum_{k'=1}^K W_{kk'} \tilde{b}_k^{(s)}  \biggr) \biggr] . 
  \label{eq:def_barZ_BP-basedDetector}
\end{align}
Details on the derivation are given in Appendix \ref{app:GFA_BP-basedDetector}. 
For the first few time-steps, 
we confirmed explicit expressions for the solutions to our dynamic equations as 
\begin{eqnarray}
  && m^{(t)} = \int Dz \tanh (\sqrt{R^{(t,t)}}z+\hat{k}^{(t)}), \\
  && q^{(t,t)} = \int Dz \tanh^2 (\sqrt{R^{(t,t)}}z+\hat{k}^{(t)}), \\
  && \hat{k}^{(t+1)} = \frac 1{ \sigma^2+\beta[1-q^{(t,t)}] }, \\
  && R^{(t+1,t+1)} = \frac {\beta[1-2m^{(t)}+q^{(t,t)}])+\sigma_0^2}{(\sigma^2+\beta[1-q^{(t,t)}])^2},  
\end{eqnarray}
where the BER of hard decisions at the $t^{\rm th}$ stage 
is $P_b^t = H ( \hat{k}^{(t)} / \sqrt{R^{(t,t)}} )$ with $H(x)=\int_x^\infty Dz$. 
In the BP-based detector, the Onsager reaction term vanished. 
These are identical to the results from density evolution \cite{Kabashima2003}. 
Figure \ref{fig:BP-based} plots the first few stages of the detection dynamics %ADDED
of the BP-based detector predicted by GFA \cite{Tanaka2005} for $E_b/N_0=8$ [dB]. %ADDED
The control parameter correct was set as $\sigma = \sigma_0$. %ADDED
\par
On the other hand, Kabashima analyzed the stability of the density evolution results 
and obtained an unstable condition for the fixed point solution \cite{Kabashima2003}. 
This unstable condition coincided with the Almeida-Thouless (AT) instability condition \cite{Almeida1978} 
of replica analysis obtained by Tanaka \cite{Tanaka2002}. 
The AT instability is the local instability of the RS saddle-point solution in the replica analysis. %ADDED
In GFA, it is known that the AT instability condition is derived 
by a condition that the response function diverges \cite{Sompolinsky1982}. %ADDED(reference)
When the AT instability condition is satisfied, 
the relaxation times of GFA dynamic equations are expected to extend to infinity. 
In the system addressed here, this condition must be equivalent to the AT instability condition. 
In Section \ref{subsection:instability}, we will discuss the stability analysis that corresponds to the AT instability. 

\section{Stationary Estimates \label{section:stationary_estimates}}
%~~~~~~~~~~~~~~~~~~~~~~~~~~~~~~~~~~~~~~~~~~~~~~~~~~~~~~~~~~~~~~~~~~~~~
\par
This section explains how we approximately extract the stationary estimate, 
which is a stationary state of the iterative algorithm, under some assumptions. 
To simplify the problem, we restrict ourselves to the soft-PIC algorithm. 
We follow the method of Ref. \cite{Coolen2005}.

\subsection{Analysis of Stationary Estimates}
%~~~~~~~~~~~~~~~~~~~~~~~~~~~~~~~~~~~~~~~~~~~~~~~~~~~~~~~~~~~~~~~~~~~~~
\par
We recognize that the representation of the effective path measure given by (\ref{eq:Dv}) 
is fully equivalent to the measure corresponding to a single-user process of the form: 
\begin{align}
  \tilde{b}^{(t+1)} = f( \hat{k}^{(t)}+v^{(t)}+\theta^{(t)}+(\vec{\Gamma} \tilde{\vec{b}})^{(t)} ). 
  \label{eq:def_Single-userProcess_prepare}
\end{align}
Variable $\vec{v}=(v^{(t)})$ can be regarded 
as a random variable that obeys Normal distribution $\mathcal{N}(\vec{0},\vec{R})$, e.g., 
$\langle v^{(t)} \rangle_{\vec{v}} = 0$ and $\langle v^{(t)} v^{(t')} \rangle_{\vec{v}} = R^{(t,t')}$, from (\ref{eq:R}), 
where $\langle \cdots \rangle_{\vec{v}}$ denotes the average over random variable $\vec{v}$. 
We here put 
\begin{align}
  \phi^{(t+1)} := \hat{k}^{(t)}+v^{(t)}+\theta^{(t)}+(\vec{\Gamma} \tilde{\vec{b}})^{(t)}, 
\end{align}
then the single-user process can be rewritten as $\tilde{b}^{(t)} = f( \phi^{(t)} )$. 
We therefore have the following form. 
\begin{align}
  \phi^{(t+1)} = \hat{k}^{(t)}+v^{(t)}+\theta^{(t)} + \sum_{t'=-1}^t \Gamma^{(t,t')} f(\phi^{(t')}), 
  \label{eq:def_Single-userProcess}
\end{align}
The relationship of (\ref{eq:def_Single-userProcess}) is referred as a {\it single-user process}. 
Using (\ref{eq:def_Single-userProcess}), 
the overlap (\ref{eq:sp_m}), 
the correlation funciton (\ref{eq:sp_G}),
the response function (\ref{eq:sp_G}), 
and the bit error rate (\ref{eq:BER}) are obtained as 
\begin{align}
  & m^{(t)} = \langle f(\phi^{(t)}) \rangle_{\vec{v}}, \\
  & C^{(t,t')} = \langle f(\phi^{(t)}) f(\phi^{(t')}) \rangle_{\vec{v}}, \\
  & G^{(t,t')} = \langle f(\phi^{(t)}) (\vec{R}^{-1} \vec{v})^{(t')} \rangle_{\vec{v}}, \\
  & P_b^{(t)} = \frac 12 (1 - \langle \sgn(f(\phi^{(t)})) \rangle_{\vec{v}}) . 
\end{align}
\par
We make the following assumptions to evaluate the stationary estimates. 
\begin{assumption} %USED
  \label{assumption:TTI}
  (Time-translation invariance: TTI) 
  The dynamics reaches a time-translation invariant estimate: 
  \begin{align}
    & \lim_{t\to\infty} m^{(t)} = m , \label{eq:TTI_m} \\
    & \lim_{t\to\infty} C^{(t+\tau,\tau)} = C^{(\tau)} , \label{eq:TTI_C} \\
    & \lim_{t\to\infty} G^{(t+\tau,\tau)} = G^{(\tau)} . \label{eq:TTI_G}
  \end{align}
  If this property holds, the dynamics reaches stationary estimates. 
  \qed
\end{assumption}
\par
\begin{assumption}%USED
  \label{assumption:FIR}
  (Finite integrated response: FIR) 
  The {\it integrated response} 
  \begin{align}
    \chi := \lim_{t\to\infty} \sum_{t' \le t} G^{(t,t')} 
    \label{eq:integratedResponse} 
  \end{align}
  is a finite non-negative number, i.e., $\chi < \infty$. 
  \qed
\end{assumption}
\par
The integrated response is also called as a {\it susceptibility}. 
\par
\begin{assumption}%USED
  \label{assumption:WLTM}
  (Weak long-term memory: WLTM) 
  \begin{align}
    \lim_{t\to\infty} G^{(t,t')} = 0 
  \end{align}
  for any finite $t'$ \cite{Cugliandolo1993, Cugliandolo1994}. 
  \qed
\end{assumption}
\par
Since the response function represents the memory which means what happened to the system, 
the weakness of the long-term memory implies that the system responds to its past in an averaged way. 
The details of what takes place during finite stages tend to be washed away. 
\par
We assume that the stationary estimate is unique 
and Assumptions \ref{assumption:TTI} -- \ref{assumption:WLTM} hold. 
Note that violation of WLTM immediately implies $\chi = \infty$. 
Under Assumption \ref{assumption:TTI}, 
the matrices $\vec{C}=(C^{(s,s')}$ and $\vec{G}=(G^{(s,s')})$ can be regarded as symmetric Toeplitz matrices, 
since we consider that the size of square matrices $\vec{C}$, $\vec{G}$, and $\vec{D}$ are sufficiently large, 
viz., $t+2$ is sufficiently large. 
The matrix $\vec{D}=(D^{(s,s')})$ whose element is given by (\ref{eq:elementOfD}) 
is also regarded as a symmetric Toeplitz matrix from 
$
  D^{(s',s)} 
  = \sigma_0^2 + \beta [ 1-m^{(s')}-m^{(s)}+C^{(s',s)} ] 
  = \sigma_0^2 + \beta [ 1-m^{(s')}-m^{(s)}+C^{(s,s')} ] 
  = D^{(s,s')}
$ and 
$\lim_{t\to\infty} D^{(t+\tau,t)} = \sigma_0^2 + \beta [1-2m+C(\tau)] =: D^{(\tau)}$, 
where we again use Assumption \ref{assumption:TTI}. 
Therefore, we hereafter make the following assumption. 
\begin{assumption}%USED
  \label{assumption:commute}
  The matrices $\{ \vec{C}, \vec{G}, \vec{D} \}$ and their powers are Toeplitz matirices. 
  %All of them commute each other. %DELETED.v3
  \HLS All pairs of matrices commute. \HLE %ADDED.v3
  \qed
\end{assumption}
\par
In this subsection, we derive the following proposition under some assumptions. 
\begin{proposition}
  \label{proposition:stationaryEstimates}
  Let Assumptions \ref{assumption:TTI} -- \ref{assumption:commute} hold, 
  then soft-PIC converges to an unique stationary estimate whose BER is given by 
  \begin{align}
    P_b = \frac 12 \biggl( 1- \int_{\mathbb{R}} \rmD z \sgn(g(z)) \biggr), 
    \label{eq:BER_at_stationaryEstimate}
  \end{align}
  where $g(z)$ is to be determined by equations that describe the stationary estimate: 
  \begin{align}
    & m = \int_{\mathbb{R}} \rmD z g(z), \label{eq:stationaryEstimates_m} \\
    & c = \int_{\mathbb{R}} \rmD z g(z)^2, \\
    & \chi = \frac 1{\sqrt{\mathsf{F}}} \int_{\mathbb{R}} \rmD z z g(z), \\
    & g(z) = f(\mathsf{E} + z \sqrt{\mathsf{F}} + \mathsf{G} g(z) + \bar{\theta} ), 
      \label{eq:stationaryEstimates_g(z)} \\
    & \mathsf{E} = \frac 1{1+\beta \chi} , \\
    & \mathsf{F} = \frac{\sigma_0^2+\beta(1-2m+c)}{(1+\beta \chi)^2} , \\
    & \mathsf{G} = \frac{\beta \chi}{1 + \beta \chi}. \label{eq:stationaryEstimates_G}
  \end{align}
  %$Q(z) := \int_z^\infty \rmD x$ denotes the error function. 
  \qed
\end{proposition}
\par
This result is identical to that from the statistical-mechanical analysis \cite{Tanaka200x}, 
which is called the naive mean-field theory \cite{Bray1986}. 
If we put $f(x)=\tanh(x/\sigma^2)$, $\bar{\theta}=0$ and $\mathsf{G}=0$, 
the BER of Proposition \ref{proposition:stationaryEstimates} can then be rewritten as 
\begin{align}
  P_b = \mathcal{Q} \biggl( \frac {\tilde{\mathsf{E}}} {\sqrt{\tilde{\mathsf{F}}}} \biggr), 
\end{align}
with 
\begin{align}
  & m = \int_{\mathbb{R}} \rmD z \tanh   ( \tilde{\mathsf{E}} + z \sqrt{\tilde{\mathsf{F}}} ), \\
  & c = \int_{\mathbb{R}} \rmD z \tanh^2 ( \tilde{\mathsf{E}} + z \sqrt{\tilde{\mathsf{F}}} ), \\
  & \tilde{\mathsf{E}} = \frac 1{\sigma^2 + \beta (1-c)} , \label{eq:stationaryEstimates_MPMdetector_E} \\
  & \tilde{\mathsf{F}} = \frac{\sigma_0^2+\beta(1-2m+c)}{[\sigma^2 + \beta (1-c)]^2}, \label{eq:stationaryEstimates_MPMdetector_F}
\end{align}
where the identity, 
\begin{align}
  \chi = \frac{1-c}{\sigma^2}, 
  \label{eq:stationaryEstimates_MPMdetector_chi}
\end{align}
holds in this parameter settings. 
This result recovers that of the marginal-posterior-mode detectors \cite{Tanaka2002}. 
%It is derived under some assumptions, %DELETED
%but it may mean that $\mathsf{G}=0$, viz., the self-coupling term is vanished, %DELETED
%is required to achieve individually optimal. %DELETED
It may mean that the relationship $\mathsf{G}=0$, 
\HLS which corresponds to the case where the self-coupling term vanishes, \HLE %ADDED.v3
%which corresponds to the case where the self-coupling term is vanished, %DELETED.v3
is required to achieve the individually optimal performance. %ADDED

\subsection{Derivation of Proposition \ref{proposition:stationaryEstimates}}
%~~~~~~~~~~~~~~~~~~~~~~~~~~~~~~~~~~~~~~~~~~~~~~~~~~~~~~~~~~~~~~~~~~~~~
\par
Since the covariance matrix $\vec{R}$ of (\ref{eq:R}) becomes 
\begin{align}
  \vec{R} 
  =& \biggl( \sum_{m=0}^\infty (-\beta \vec{G}^\top)^m \biggr)
     \vec{D}
     \biggl( \sum_{n=0}^\infty (-\beta \vec{G})^n \biggr) \\
  =& \vec{D}(\vec{1} + \beta \vec{G}^\top + \beta \vec{G} + \beta^2 \vec{G}^\top \vec{G})^{-1}, 
\end{align}
by using Assumption \ref{assumption:commute}, 
we then have $\vec{R}^\top = \vec{R}$. 
Similarly, the reaction matrix $\vec{\Gamma}$ is represented as 
\begin{align}
  \vec{\Gamma}
  =& (\vec{1}+\beta \vec{G})^{-1} \beta \vec{G} \\
  =& - \sum_{n=1}^\infty (-\beta)^n \vec{G}^n. 
\end{align}
Since $\vec{G}$ is a lower triangular Toeplitz matrix, 
the $\vec{\Gamma}$ is also a lower triangular Toeplitz matrix, 
i.e., $\Gamma^{(s,s')} := \Gamma^{(s-s')}$ for $s>s'$ and $\Gamma^{(s,s')}=0$ for $s \le s'$. 
\par
Under Assumption \ref{assumption:WLTM}, 
it can be considered that the system responds to its past in an averaged way. 
%Let Assumptions \ref{assumption:TTI} -- \ref{assumption:commute} hold, 
%we can calculate the long time average for the effective single user equation. %Heimel p62-63
We therefore consider the average of $\phi^{(t)}$ instead of $\lim_{t\to\infty} \phi^{(t+1)}$, which gives 
\begin{align}
  \bar{\phi}
  :=& \frac 1{t+2} \sum_{t'=-1}^t \phi^{(t'+1)} \\
  =& \bar{\hat{k}} + \bar{v} + \bar{\theta} 
     + \frac 1{t+2} \sum_{t'=-1}^t \sum_{s=-1}^{t'} \Gamma^{(t',s)} f(\phi^{(s)}), 
  \label{eq:average_phi}
\end{align}
where we put 
$\bar{\hat{k}} := \frac 1{t+2} \sum_{t'=-1}^t \hat{k}^{(s)}$, 
$\bar{v} := \frac 1{t+2} \sum_{t'=-1}^t v^{(s)}$ 
and $\bar{\theta} := \frac 1{t+2} \sum_{t'=-1}^t \theta^{(s)}$. 
Under Assumptions \ref{assumption:TTI} and \ref{assumption:FIR}, $\sum_{t'=-1}^\infty (G^{(t,t')})^n$ is written by 
\begin{align}
   & \lim_{t\to\infty} \sum_{t'=-1}^t (G^{(t,t')})^n \\
  =& \lim_{t\to\infty} \sum_{t'=-1}^t \biggl( \sum_{t_1=-1}^t \cdots \sum_{t_{n-1}=-1}^t G^{(t,t_1)} G^{(t,t_1)} \times \nonumber \\
   & \cdots \times G^{(t_{n-2},t_{n-1})} G^{(t_{n-1},t')} \biggr) \\
  =& \chi \lim_{t\to\infty} \sum_{t_1=-1}^t \cdots \sum_{t_{n-1}=-1}^t G^{(t,t_1)} G^{(t,t_1)} \times \nonumber \\
   & \cdots \times G^{(t_{n-2},t_{n-1})} \\
  =& \chi^n. 
\end{align}
Therefore, the last term on the right-hand side of (\ref{eq:average_phi}) 
in the $t\to\infty$ limit is evaluated as 
\begin{align}
   & \lim_{t\to\infty} \frac 1{t+2} \sum_{t'=-1}^t \sum_{s=-1}^{t'} \Gamma^{(t',s)} f(\phi^{(s)}) \\
  %=& \lim_{t\to\infty} \sum_{s=0}^t \Gamma^{(s)} \biggl( \frac 1{t+2} \sum_{t'=s-1}^{t } f(\phi^{(t'-s)}) \biggr) \\
  %=& \bar{f} \lim_{t\to\infty} \sum_{s=-1}^t \Gamma^{(t,s)} \\
  =& \bar{f} \sum_{n=1}^\infty (-\beta)^n \biggl(  \lim_{t\to\infty} \sum_{s=-1}^t (G^{(t,s)})^n \biggr) \\
  =& \frac{\beta \chi}{1+\beta \chi} \bar{f}, 
  \label{eq:LimitOfOnsagerTerm}
\end{align}
where 
\begin{align}
  \bar{f} := \frac 1{t+2} \sum_{t'=-1}^t f( \phi^{(s)} ). 
\end{align}
Taking the $t\to\infty$ limit and using (\ref{eq:LimitOfOnsagerTerm}), 
the average single-user process $\bar{\phi}$ of (\ref{eq:average_phi}) becomes 
\begin{align}
  \bar{\phi} = \bar{\hat{k}} + \bar{v} + \bar{\theta} + \frac{\beta \chi}{1+\beta \chi} \bar{f}. 
  \label{eq:bar_phi}
\end{align}
\par
First, we evaluate $\bar{\hat{k}}$, which corresponds to the signal part. 
If series $\hat{k}^{(-1)}, \hat{k}^{(0)}, \cdots$ converges, then $\bar{\hat{k}}=\hat{k}^{(\infty)}$. 
Since $\bar{\hat{k}}$ is obtained as 
$\bar{\hat{k}} 
= 1 - \beta \hat{k}^{(\infty)} \sum_{t'=-1}^\infty G^{(t')} 
= 1 - \beta \chi \bar{\hat{k}}$, 
we then have 
\begin{align}
  \bar{\hat{k}}
  = \frac 1{1+\beta \chi} =: \mathsf{E}. 
\end{align}
%Under Assumptions \ref{assumption:FIR}, the signal does not vanish, i.e., $0<\mathsf{E}\le1$. 
\par
Next, we consider the variance of Gaussian random variable $\bar{v}$. 
The average of $\bar{v}$ is obviously $\langle \bar{v} \rangle_{\vec{v}}$ 
$= \lim_{t\to\infty} \frac 1{t+2} \sum_{t'=-1}^t$ $\langle v^{(t')} \rangle_{\vec{v}}$ $=0$. 
Using the definition of the persistent correlation $c := \lim_{t\to\infty} C(t)$, 
the variance of $\bar{v}$ is given by 
\begin{align}
  \langle \bar{v}^2 \rangle_{\vec{v}} 
  =& \lim_{\tau\to\infty} \frac 1{(\tau+2)^2} \sum_{t=-1}^\tau \sum_{t'=-1}^\tau \nonumber \\
   & ( (\vec{1}+\beta \vec{G}^\top)^{-1} \vec{D} (\vec{1}+\beta \vec{G})^{-1} )^{(t,t')} \\
  =& \sum_{s=0}^\infty \sum_{s'=0}^\infty 
     ( (\vec{1}+\beta \vec{G}^\top)^{-1} )^{(s )}
     ( (\vec{1}+\beta \vec{G}^\top)^{-1} )^{(s')} \nonumber \\
   & \times \lim_{\tau\to\infty} \frac 1{(\tau+2)^2} \sum_{t=s-1}^\tau \sum_{t'=s'-1}^\tau D^{(t-s-t'+s')} \\
  =& \frac{\sigma_0^2+\beta(1-2m+c)}{(1+\beta \chi)^2}, \\
  =:& \mathsf{F}, 
\end{align}
where we here write $((\vec{1}+\beta \vec{G}^\top)^{-1})^{(t+\tau,t)}$ 
as $((\vec{1}+\beta \vec{G}^\top)^{-1})^{(\tau)}$, 
since $(\vec{1}+\beta \vec{G}^\top)^{-1}$ is a Toeplitz matrix. 
Namely, $\bar{v}$ obeys a Gaussian distribution with mean zero and variance $\mathsf{F}$: $\bar{v} \sim \mathcal{N}(0,\mathsf{F})$. 
The Gaussian random variable $\bar{v}$ can be represented as $\bar{v}=\sqrt{\mathsf{F}}z$ 
by using standard Gaussian random variable $z \sim \mathcal{N}(0,1)$. 
%\vec{R}=(\vec{1}+\beta \vec{G}^\top)^{-1} \vec{D} (\vec{1}+\beta \vec{G})^{-1}, (\ref{eq:R})
\par
Limit $f(\phi^{(t)})$ as $t\to\infty$ converges 
by using Assumptions \ref{assumption:TTI}, \ref{assumption:FIR} and \ref{assumption:WLTM}. 
We therefore have $\bar{\phi} = \lim_{t\to\infty} \frac 1{t+2} \sum_{t'=-1}^t \phi^{(t')} = \phi^{(\infty)}$ 
and $\bar{f} = \lim_{t\to\infty} \frac 1{t+2} \sum_{t'=-1}^t f(\phi^{(t')}) = f(\phi^{(\infty)})$. 
These relationships give $\bar{f}=f(\bar{\phi})$. 
Applying function $f$ to both sides of (\ref{eq:bar_phi}) and letting $g(z) := \bar{f}$, we obtain 
\begin{align}
  g(z) = f(\mathsf{E} + z \sqrt{\mathsf{F}} + \bar{\theta} + \mathsf{G} g(z)), 
\end{align}
where $\mathsf{G} := \beta \chi / ( 1 + \beta \chi )$. 
The persistent overlap $m := \lim_{t\to\infty} m^{(t)}$, which is identical to that of (\ref{eq:TTI_m}), then becomes 
\begin{align}
  m
  =& \lim_{t\to\infty} \langle f(\phi^{(t)}) \rangle_{\vec{v}} \\
  =& \langle \bar{f} \rangle_{\bar{v}} 
  = \int_{\mathbb{R}} \rmD z g(z). 
\end{align}
We then have the persistent correlation $c$: 
\begin{align}
  c 
  =& \lim_{\tau\to\infty} \lim_{t\to\infty} \langle f(\phi^{(t+\tau)}) f(\phi^{(t)}) \rangle_{\vec{v}} \\
  =& \langle \bar{f}^2 \rangle_{\bar{v}} 
  =  \int_{\mathbb{R}} \rmD z g(z)^2, 
\end{align}
in a similar way. 
Since we assume Assumption \ref{assumption:FIR}, 
the integrated response $\chi$ of (\ref{eq:integratedResponse}) is a finite non-negative number and is then given by 
\begin{align}
  \chi 
  =& \lim_{t\to\infty} \sum_{t'=-1}^t \frac{\partial \langle f(\phi^{(t)}) \rangle_{\vec{v}}}{\partial \theta^{(t')}} \\
  =& \langle \frac{\partial \bar{f}}{\partial \bar{\theta}} \rangle_{\bar{v}} 
  =  \int_{\mathbb{R}} \rmD z \frac{\partial g(z)}{\partial \bar{\theta}} 
  =  \frac 1{\sqrt{\mathsf{F}}} \int_{\mathbb{R}} \rmD z z g(z). 
\end{align}
For arbitrary function $F(x)$ that can be expanded around $x=0$, 
the bit error rate $P_b := \lim_{t\to\infty} P_b^{(t)}$ can be evaluated as 
\begin{align}
  \lim_{t\to\infty} \langle F(f(\phi^{(t)})) \rangle_{\vec{v}}
  = \langle F(\bar{f}) \rangle_{\bar{v}} 
  = \int_{\mathbb{R}} \rmD z F(g(z)). 
\end{align}
Therefore, the bit error rate $P_b$ can be evaluated as (\ref{eq:BER_at_stationaryEstimate}). 
\par
Setting $\theta^{(s)}=0$, i.e., $\bar{\theta}=0$, 
we have then arrived at Proposition \ref{proposition:stationaryEstimates}.

\subsection{Stability Analysis for Stationary Estimates \label{subsection:instability}}
%~~~~~~~~~~~~~~~~~~~~~~~~~~~~~~~~~~~~~~~~~~~~~~~~~~~~~~~~~~~~~~~~~~~~~
\par
If Assumption \ref{assumption:WLTM} (WLTM) is violated, 
the dynamics do not achieve the stationary estimate. 
For example, oscillatory behavior in the dynamics corresponds to this situation. 
In this subsection, we investigate where Assumption \ref{assumption:WLTM} is violated 
and obtain the following proposition. 
\begin{proposition}
  \label{proposition:ATinstability}
  The condition where WLTM is violated is given by 
  \begin{align}
    \frac{\beta}{(1 + \beta \chi)^2} 
    \int_{\mathbb{R}} \rmD z \biggl( \frac{\partial g(z)}{\partial \bar{\theta}} \biggr)^2 
    > 1, 
    \label{eq:ATinstability}
  \end{align}
  where $g(z)$ and all parameters are identical 
  to the set of (\ref{eq:stationaryEstimates_m}) -- (\ref{eq:stationaryEstimates_G}). 
  \qed
\end{proposition}
If we put $f(x)=\tanh(x/\sigma^2)$ and $\bar{\theta}=0$ and $\mathsf{G}=0$, 
condition (\ref{eq:ATinstability}) can be simplified as 
\begin{align}
  \frac{\beta}{[\sigma^2 + \beta (1-c)]^2} 
  \int_{\mathbb{R}} \rmD z [ 1 - \tanh^2 (\tilde{\mathsf{E}} + z \sqrt{\tilde{\mathsf{F}}} ) ]^2 
  > 1, 
  \label{eq:ATinstability_of_MPMdetector}
\end{align}
where $\tilde{\mathsf{E}}$, $\tilde{\mathsf{F}}$, and $c$ are the same as 
(\ref{eq:stationaryEstimates_MPMdetector_E}), (\ref{eq:stationaryEstimates_MPMdetector_F}), 
and (\ref{eq:stationaryEstimates_MPMdetector_chi}), respectively. 
Equation (\ref{eq:ATinstability_of_MPMdetector}) coincides with 
the AT instability \cite{Almeida1978} of the marginal-posterior-mode detector \cite{Tanaka2002}. 
This condition (\ref{eq:ATinstability_of_MPMdetector}) is also identical 
to the microscopic instability, which is derived by the belief update 
of the BP-based detector \cite{Kabashima2003}. 
\par
In the rest of this subsection, we derive Proposition \ref{proposition:ATinstability}. 
Let us separate the response function $G^{(t,t')}$ and the Onsager reaction term $\Gamma^{(t,t')}$ explicitly 
into TTI parts and further small contributions: 
\begin{align}
  G^{(t,t')} =& \tilde{G}^{(t-t')} + \hat{G}^{(t,t')}, \label{eq:Separation_of_G} \\
  \Gamma^{(t,t')} =& \tilde{\Gamma}^{(t-t')} + \hat{\Gamma}^{(t,t')}, \label{eq:Separation_of_Gamma}
\end{align}
with $\lim_{t\to\infty} G^{(t+\tau,t)} = \tilde{G}^{(\tau)}$, 
and $\lim_{t\to\infty} \Gamma^{(t+\tau,t)} = \tilde{\Gamma}^{(\tau)}$, 
where $\tilde{G}^{(t-t')}$ and $\tilde{\Gamma}^{(t-t')}$ denote TTI parts that are referred to as {\it persistent parts}. 
The $\hat{G}^{(t,t')}$ and $\hat{\Gamma}^{(t,t')}$ denote small contribution terms that are referred to as {\it non-persistent parts}. 
If weak long-term memory holds, then $\hat{G}^{(t,t')}$ and $\hat{\Gamma}^{(t,t')}$ vanish for $t\to\infty$. 
We here assume that some small long-term memory, 
but such that the limits 
$\hat{G}^{(t')} := \lim_{t\to\infty} \hat{G}^{(t,t')}$ 
and $\hat{\Gamma}^{(t')} := \lim_{t\to\infty} \hat{\Gamma}^{(t,t')}$ exist and also both 
$\hat{\chi} := \lim_{t\to\infty} \sum_{t'=-1}^t \hat{G}(t')$, 
$\hat{\gamma} := \lim_{t\to\infty} \sum_{t'=-1}^t \hat{\Gamma}(t')$, 
$\tilde{\chi} := \lim_{t\to\infty} \sum_{t'=-1}^t \tilde{G}(t')$ 
and $\tilde{\gamma} := \lim_{t\to\infty} \sum_{t'=-1}^t \tilde{\Gamma}(t')$ exist. 
When Assumption \ref{assumption:WLTM} holds, 
the identities $\lim_{t\to\infty} \hat{G}^{(t,t')} = 0$ and 
$\lim_{t\to\infty} \hat{\Gamma}^{(t,t')} = 0$ hold. % small long-term memory (small) <> weak long-term memory (zero)
\par
We expand (\ref{eq:Gamma_softPIC}) for small $\hat{\vec{G}}$, and find 
\begin{align}
  \vec{\Gamma}
  =& \sum_{n=1}^\infty (-\beta)^n \sum_{m=0}^n \left(
     \begin{array}{c}
       n \\
       m
     \end{array}
     \right) \tilde{\vec{G}}^{n-m} \hat{\vec{G}}^m \\
  =& (\vec{1}+\beta \tilde{\vec{G}})^{-1} \beta \tilde{\vec{G}} \nonumber \\
   & + \sum_{n=1}^\infty \sum_{m=0}^{n-1} 
       (-\beta \tilde{\vec{G}})^m
       (\beta \hat{\vec{G}})
       (-\beta \tilde{\vec{G}})^{n-1-m} + O(\tilde{\vec{G}}^2), 
  \label{eq:gammaHat_chiHat_chiTilde}
\end{align}
where the $(t,t')$ elements of the $(t+2) \times (t+2)$ matrices 
$\tilde{\vec{G}}$, $\hat{\vec{G}}$, $\tilde{\vec{\Gamma}}$, and $\hat{\vec{\Gamma}}$ 
are $\tilde{G}^{(t,t')}$, $\hat{G}^{(t,t')}$, $\tilde{\Gamma}^{(t,t')}$, and $\hat{\Gamma}^{(t,t')}$, respectively. 
From (\ref{eq:gammaHat_chiHat_chiTilde}), the persistent and the non-parsistent parts of $\vec{\Gamma}$ is given by 
$\tilde{\vec{\Gamma}}=(\vec{1}+\beta \tilde{\vec{G}})^{-1} \beta \tilde{\vec{G}}$ 
and $\hat{\vec{\Gamma}}$. 
Then, $\hat{\gamma}$ becomes 
\begin{align}
  \hat{\gamma}
  =& \beta \lim_{t\to\infty} \sum_{t'=-1}^t \sum_{s=-1}^t \sum_{n=0}^\infty 
     (n+1) (-\beta \tilde{G}^{(t,s)})^n \hat{G}^{(s,t')} \\
  =& \frac{\beta \hat{\chi}}{(1+\beta \tilde{\chi})^2}. 
  \label{eq:gammaHat}
\end{align}
We next substitute (\ref{eq:Separation_of_G}) into (\ref{eq:def_Single-userProcess}) 
and take the average with respect to iterative steps $t$, 
i.e., $\frac 1{t+2}\sum_{t'=-1}^t$, to both-sides of the substituted equation, which gives 
\begin{align}
  \bar{\phi} 
  = \bar{\hat{k}} + \bar{v} + \bar{\theta} + \frac{\beta \tilde{\chi}}{1+\beta\tilde{\chi}} 
  + \frac 1{t+2} \sum_{t'=-1}^t \sum_{s=-1}^t \Gamma^{(t',s)} f(\phi^{(s)}). 
  \label{eq:def_Separeted_Single-userProcess}
\end{align}
\par
As $\lim_{t\to\infty} \tilde{G}^{(t,t')}$, 
since we assume Assumption \ref{assumption:FIR} (finite integrated response), we find 
\begin{align}
  \hat{G}^{(t)}
  =& \lim_{t\to\infty} G^{(t,t')} \\
  =& \lim_{t\to\infty} \frac{\partial \langle f(\phi^{(t)}) \rangle_{\vec{v}}}{\partial \theta^{(t')}} \\
  =& \lim_{t\to\infty} \biggl\langle 
     \biggl( \frac{\partial \bar{f}}{\partial \bar{\theta}} \biggr)^2 
     \frac1{t+2} \sum_{s=-1}^t \hat{\Gamma}^{(s)} 
     \biggr\rangle_{\bar{v}}. 
\end{align}
Substituting this equation into the definition of $\hat{\chi}$, the following identity is found: 
\begin{align}
  \hat{\chi}=\hat{\gamma}
  \biggl\langle 
  \biggl( \frac{\partial \bar{f}}{\partial \bar{\theta}} \biggr)^2 
  \biggr\rangle_{\bar{v}}. 
  \label{eq:chiHat}
\end{align}
Although $\hat{\chi}=\hat{\gamma}=0$ always solves this equation, 
one finds another solution when 
$(\hat{\gamma}/\hat{\chi}) 
\langle ( \partial \bar{f} / \partial \bar{\theta} )^2 \rangle_{\bar{v}} =1$ holds, 
which is equivalent to the AT line \cite{Almeida1978}. 
Since $\hat{\chi}=0$ is required to hold Assumption \ref{assumption:FIR} and (\ref{eq:chiHat}) 
represents an evaluation of $\hat{\chi}$, 
the evaluated value should not be greater than $\hat{\chi}$, 
i.e., $\hat{\chi} < \hat{\gamma} \langle ( \partial \bar{f}/\partial \bar{\theta} )^2 \rangle_{\bar{v}}$, 
to diverge stationary estimates. 
We therefore have an instability condition of Assumption \ref{assumption:FIR} as 
\begin{align}
  \frac{\beta}{(1+\beta \tilde{\chi})^2} 
  \biggl\langle 
  \biggl( \frac{\partial \bar{f}}{\partial \bar{\theta}} \biggr)^2 
  \biggr\rangle_{\bar{v}} > 1, 
\end{align}
by substituting (\ref{eq:gammaHat}). 
Using $\chi = \tilde{\chi} + \hat{\chi} \approx \tilde{\chi}$ and (\ref{eq:stationaryEstimates_g(z)}), 
we have then arrived at Proposition \ref{proposition:ATinstability}.

\section{Decoupling Principle \label{section:decouple}}
%~~~~~~~~~~~~~~~~~~~~~~~~~~~~~~~~~~~~~~~~~~~~~~~~~~~~~~~~~~~~~~~~~~~~~
\par
% With the statistical-mechanical approach \cite{Tanaka2002, Guo2005, Guo2007}, %DELETED
% a complex {\it many-body problem} such as an inference problem %DELETED
% where many users' transmitted signals in the CDMA system %DELETED
% is reduced to an equivalent {\it sone-body problem}. %DELETED
With the statistical-mechanical approach \cite{Tanaka2002, Guo2005, Guo2007}, %ADDED
a complex {\it many-user problem} such as an inference problem %ADDED
where many users' transmitted signals in the CDMA system %ADDED
is reduced to an equivalent {\it single-user problem}. %ADDED
This approach brings a significant interpretation in the communication theory. 
This is known as the {\it decoupling principle} found by Guo and Verd\'u \cite{Guo2005}. 
It claims that the vector channel concatenated with optimal detection is equivalent 
to a bank of independent single-user additive white Gaussian noise channels, 
whose signal-to-noise ratio degrades due to the multiple-access interference, 
in the large-system limit under a certain randomness assumption of the channel. 
\par
The decoupling principle has recently attracted a great deal of attention 
and has been investigated in detail \cite{Ikehara2007, Nakamura2008}. 
Especially, by applying density evolution, 
Ikehara and Tanaka have found that the decoupling principle holds not only at equilibrium 
but also at each stage of the BP-based detector \cite{Ikehara2007}. 
Their analysis was however based on an assumption of independence of messages. 
The GFA allows us to study algorithms without assuming the independence of messages 
\cite{Coolen2000, Coolen2005, Mimura2005, Mimura2006, Mimura2007}. 
In this section, we investigate the decoupling principle via detection dynamics 
using GFA by considering an arbitrary fading model.

\subsection{Generic Detection Algorithm}
%~~~~~~~~~~~~~~~~~~~~~~~~~~~~~~~~~~~~~~~~~~~~~~~~~~~~~~~~~~~~~~~~~~~~~
\par
In this section, we focus on a generic system model 
and a generic iterative multiuser detection algorithm. 
\par
We now consider a $K$-input $N$-output vector channel, 
which can be regarded as a fully-synchronous $K$-user CDMA system with spreading factor $N$ 
or a multiple-input multiple-output (MIMO) system. 
Let $\vec{b}=(b_1,\cdots,b_K)^\top \in \mathbb{R}^K$ denote the input vector, 
and $\vec{y}=(y_1,\cdots,y_N)^\top \in \mathbb{R}^N$ denote the output vector of the channel. 
%where operator ${}^\top$ denotes the transpose. %DELETED
The output $\vec{y}$ is given by a linear transform as 
\begin{equation}
\vec{y} = \vec{S} \vec{b} + \vec{n}, 
\end{equation}
where $\vec{S}=(a_1 \vec{s}_1,\cdots, a_K \vec{s}_K)$ denotes an $N \times K$ channel state matrix, 
which includes the received amplitude $a_k$. 
$\vec{s}_k = \frac1{\sqrt{N}} (s_{1k}, \cdots, s_{Nk})^\top$ denote the channel parameters 
for the spreading code sequence of user $k$, 
and $\vec{n}$ is a vector consisting of i.i.d. zero-mean Gaussian random variables with variance $\sigma_0^2$. 
We use control parameter $\sigma^2$ as the estimate of the channel noise variance instead of true noise variance. 
Each element $s_{\mu k} \in \{-1,1\}$ is an i.i.d. random variable with equal probability. 
Also, let $p_{\vec{b}}(\vec{b})=\prod_{k=1}^K p_b(b_k)$ denote the joint prior distribution of input $\vec{b} \in \{ -1,1 \}^K$. 
Input distribution $p_b(b)$ is arbitrary. 
The fading model is represented by arbitrary probability distribution $p_{\vec{a}}(\vec{a})=\prod_{k=1}^K p_a(a_k)$. 
The distribution $p_a(a)$ is assumed to have finite moments. 
Especially it has the second-order moment $\gamma$. 
We analyze the system in the large-system limit where $K,N\to\infty$, while the ratio $\beta = K/N$ is kept finite.

\begin{definition}
  (Generic iterative algorithm) 
  The updating rule for tentative decision $\tilde{b}_k^{(t)}\in\mathbb{R}$ 
  of bit signal $b_k$ at stage $t$ is 
  \begin{align}
    \tilde{b}_k^{(t+1)} = f (u_k^{(t)} ), \label{eq:GenericAlgorithm} 
  \end{align}
  with 
  \begin{align}
    u_k^{(t)} 
    =& \; R^{(t)} h_k - \sum_{s=-1}^t J^{(t,s)} \sum_{k=1}^K W_{k k'} \tilde{b}_{k'} (s) \nonumber \\
     & \; - \sum_{s=-1}^t \hat{\Gamma}_k^{(t,s)} \tilde{b}_k^{(s)} + \theta_k^{(t)}, 
  \end{align}
  where $\vec{h}=(h_1,\cdots,h_K)^\top$, $\vec{W}=(W_{kk'})$, 
  $R^{(t)}$, $J^{(t,s)}$, $A^{(s)}$, $\hat{\Gamma}_k^{(t,s)}$, and $\theta^{(t)}$ are parameters. 
  Transfer function $f:\mathbb{R}\to\mathbb{R}$ is arbitrary and applied componentwise. 
  \qed
\end{definition}
\par
The $u_k^{(t)}$ is a summation of messages from tentative decisions. 
The $\theta_k^{(t)}$ is again introduced to define a response function. 
Let $h_k$ be an output of the matched filter for user $k$, i.e., $h_k = \vec{s}_k \cdot \vec{y}$, 
and let $\vec{W}$ be a $K \times K$ correlation matrix 
which is defined by $\vec{W} = \tilde{\vec{S}}^\top \tilde{\vec{S}}$ with $\tilde{\vec{S}}=(\vec{s}_1,\cdots, \vec{s}_K)$. 
For initialization, we use $\tilde{b}_k^{(-1)} = 0$. %where $\delta(x)$ denotes the Dirac delta function. 
This generic algorithm includes various types of iterative multiuser detectors as in the following examples. 
\par
\begin{example}
  If we set the transfer function $f$ and the parameters as 
  \begin{align}
    f(x) =& \tanh(x), \\
    R^{(s)} =& \sigma^{-2}, \\
    J^{(s,s')} =& \sigma^{-2} \delta_{s,s'}, \\
    A^{(s)} =& \theta_k^{(s)}=\hat{\Gamma}_k^{(s,s')}=0, 
  \end{align}
  the algorithm of (\ref{eq:GenericAlgorithm}) is identical to soft-PIC \cite{Kaiser1997}: 
  $\tilde{b}_k^{(t+1)} = \tanh [ \frac 1{\sigma^2} ( h_k - \sum_{k'=1}^K W_{kk'} \tilde{b}_k^{(t)} ) ]$. 
  %Here, $\delta_{s,s'}$ denotes Kronecker's delta taking 1 if $s=s'$, or 0 otherwise. %DELETED.v3
  \qed
\end{example}
\par
\begin{example}
  All parameters except $\hat{\Gamma}_k^{(s,s')}$ use those of soft-PIC. 
  If the $\hat{\Gamma}_k^{(s,s')}$ are chosen to cancel the Onsager reaction, 
  then algorithm (\ref{eq:GenericAlgorithm}) gives ORC-PIC \cite{Tanaka2005}. 
  \qed
\end{example}
\par
\begin{example}
  We next consider the following parameters and function: 
  \begin{align}
    f(x) =& \tanh(x), \\
    A^{(t)} =& \{ \sigma^2 + \beta [1-Q^{(t)}] \}^{-1}, \\
    R^{(t)} =& A^{(t)} + A^{(t)} \beta [1-Q^{(t)}] R^{(t-1)}, \\
    \theta_k^{(s)} =& 0, \\
    \hat{\Gamma}_k^{(s,s')} =& - \delta_{s,s'} A^{(s)}, \\
    J^{(t,s)} =& \mathbb{I}(0 \le s < t) A^{(s)} \prod_{s'=s+1}^t A^{(s')} \beta [1-q^{(s')}] \nonumber \\
               & + \mathbb{I}(s=t) A^{(t)}, 
  \end{align}
  with $Q^{(t)} = K^{-1} \sum_{k=1}^K [\tilde{b}_k^{(t)}]^2$. 
  Here, $\mathbb{I}(\mathcal{P})$ denotes an indicator function, which takes value 1 if $\mathcal{P}$ is true, or 0 otherwise. 
  We assume that $Q(t)$ can be regarded as a constant with respect to $\tilde{\vec{b}}$ by using the central limit theorem. 
  Then (\ref{eq:GenericAlgorithm}) is equivalent to the BP-based detector \cite{Kabashima2003}. 
  \qed
\end{example}
\par
As previously mentioned, the iterative detection algorithm (\ref{eq:GenericAlgorithm}) includes various types of iterative algorithms. 
It should be noted that in all these detectors, $\tilde{b}_k^{(t)}$ gives 
an approximate value for the posterior-mean at the $t^{\rm th}$ stage \cite{Tanaka2005, Kabashima2003}. 
%
%From the posterior average $\tilde{b}_k(t)$, the tentative decision at the $t^{\rm th}$ update is evaluated as $\hat{b}_k(t) = \sgn [\tilde{b}_k(t)]$. 
%Therefore we can derive the following proposition. 
%
%The widely used measure of the performance of a demodulator is the bit error rate (BER). 
%The BER $P_b(t)$ of hard decisions $\hat{b}_k(t)=\sgn[\tilse{b}_k(t)]$ at the $t^{\rm th}$ stage of soft PIC is given by $P_b(t)=[1-m_h(t)]/2$, 
%where $m_h(t)=\frac1K\sum_{k=1}^K b_k \hat{b}_k(t)$ is the overlap between the information bits vector $\vec{b}(t)=(b_1,\cdots,b_K)^\top$ and 
%the tentative hard-decision vector $(\hat{b}_1(t),\cdots,\hat{b}_K(t))^\top$. 
%The $\theta_k(t)$ are set to zero in the end of the analysis. 
We obtain the analytical result, which involves the following measure: 
\begin{align}
  \langle g(\hat{\vec{b}}_k,\vec{v}_k) \rangle_{*_k} = 
  & \int \mathcal{D} \vec{v}_k 
    \int_{\mathbb{R}^{t+2}} \biggl( \prod_{s=-1}^{t-1} \rmd \hat{b}_k^{(s)} \biggr) 
    g(\hat{\vec{b}}_k,\vec{v}_k) \notag \\
  & \times \delta \bigl[ \tilde{b}_k^{(-1)} \bigr] 
    \prod_{s=-1}^{t-1} \delta \bigl[ \hat{b}_k^{(s+1)}-f(u_k^{(s)}) \bigr] , 
  \label{eq:E}
\end{align}
with 
\begin{align}
  u_k^{(s)} = a_k \hat{k}^{(s)} b_k + v_k^{(s)} + \sum_{s'=-1}^s [\Gamma^{(s,s')}-\hat{\Gamma}_k^{(s,s')}] \tilde{b}_k^{(s')}, 
  \label{eq:u}
\end{align}
where $g(\hat{\vec{b}}_k,\vec{v}_k)$ denotes an arbitrary function of 
$\tilde{\vec{b}}_k = (\hat{b}_k^{(-1)},\cdots,\tilde{b}_k^{(t)} )^\top$ and 
$\vec{v}_k = (v_k^{(-1)},\cdots,v_k^{(t)} )^\top$. 
The parameters are as follows: 
\begin{align}
  & \vec{R}=(\vec{1} - \beta \vec{B})^{-1} \vec{D} (\vec{1} - \beta \vec{B}^\top)^{-1}, \\
  & \vec{\Gamma}=- [ \sum_{s=-1}^{t-1} \vec{J}_{s+1} \otimes \vec{\Delta}^{s+1} (\vec{U}^\top)^{-1} ]^\top \\
  & \hat{k}^{(s)}=|\vec{\Lambda}_s|, 
\end{align}
where $\vec{B}$, $\vec{U}$, and $\vec{D}$ are matrices with elements 
\begin{align}
  B^{(s,s')} =& - \sum_{\tau=-1}^s J^{(s,\tau)} G^{(\tau,s')}, \\
  U^{(s,s')} =& \delta_{s,s'} + \beta \sum_{\tau=-1}^s J^{(s,\tau)} G^{(\tau,s')} \\
  D^{(s,s')} =& \sigma_0^2 R^{(s)} R^{(s')} + \beta [ R^{(s)} R^{(s')} \nonumber \\
              & - R^{(s)}  \sum_{\tau'=-1}^{s'} J^{(s',\tau')} m^{(\tau')} \nonumber \\
              & - R^{(s')} \sum_{\tau =-1}^{s} J^{(s,\tau)} m^{(\tau)} \nonumber \\
              & + \sum_{\tau=-1}^{s} \sum_{\tau'=-1}^{s'} J^{(s,\tau)} J^{(s',\tau')} C^{(\tau,\tau')} ], \\
  \Lambda_s^{(s',s'')} =& \mathbb{I}(s' \ne s) \biggl( \delta_{s',s''} + \beta \sum_{\tau'=-1}^{s'} J^{(s',\tau')} G^{(\tau',s')} \biggr) \nonumber \\
                        &  + \mathbb{I}(s' = s) R^{(s')}. 
\end{align}
with  $\mathcal{D}\vec{v}=|2\pi \vec{R}|^{-1/2} d\vec{v} \exp [ -\frac 12\vec{v}\cdot \vec{R}^{-1}\vec{v}]$. 
All these parameters mentioned above are obtained from the following three kinds of quantities: 
\begin{align}
  m^{(s)}
  =& \lim_{K \to \infty} \frac 1K \sum_{k=1}^K \langle a_k b_k \tilde{b}_k^{(s)} \rangle_{*_k} \nonumber \\
  =& \langle a_k b_k \tilde{b}_k^{(s)} \rangle_{*_k,b_k,a_k}, \\
  C^{(s,s')}
  =& \lim_{K \to \infty} \frac 1K \sum_{k=1}^K \langle \tilde{b}_k^{(s)} \tilde{b}_k^{(s')} \rangle_{*_k} \nonumber \\
  =& \langle \tilde{b}_k^{(s)} \tilde{b}_k^{(s')} \rangle_{*_k,b_k,a_k}, \\
  G^{(s,s')}
  =& \mathbb{I}(s \le s') \lim_{K \to \infty} \frac 1K \sum_{k=1}^K \langle \tilde{b}_k^{(s)} (\vec{R}^{-1}\vec{v}_k)^{(s')} \rangle_{*_k} \nonumber \\
  =& \mathbb{I}(s \le s') \langle \tilde{b}_k^{(s)} (\vec{R}^{-1}\vec{v}_k)^{(s')} \rangle_{*_k,b_k,a_k}, 
\end{align}
where $G^{(s,s')}=0$ for $s<s'$ due to causality. 
Here, $\vec{\Delta} = (\delta_{s+1,s'})$ is a $(t+1)\times(t+1)$ matrix. 
Operator $\langle \, \cdot \, \rangle_U$ denotes the expectation with respect to random variable $U$ 
%$\langle \, \cdot \, \rangle_{b_k} = \int db p_b(b_k) \, \cdot$ and $\langle \, \cdot \, \rangle_{a_k} = \int db p_a(a_k) \, \cdot$
and operator $\otimes$ denotes the Hadamard product, i.e., $A \otimes B=(a_{ij} b_{ij})$ for $A=(a_{ij})$ and $B=(b_{ij})$. 
Terms $(\vec{R}^{-1}\vec{v})^{(s)}$ denote the $s^{\rm th}$ element of the vector $\vec{R}^{-1}\vec{v}$. 
Term $\sum_{s'=-1}^s \Gamma^{(s,s')} \tilde{b}^{(s')}$ in (\ref{eq:u}) represents 
the retarded self-interaction, which is called the Onsager reaction. 
%This result entirely describe the dynamics of the system. 
%
\begin{figure}[t]%[htbp]
\begin{center}
\includegraphics[width=.96\linewidth,keepaspectratio]{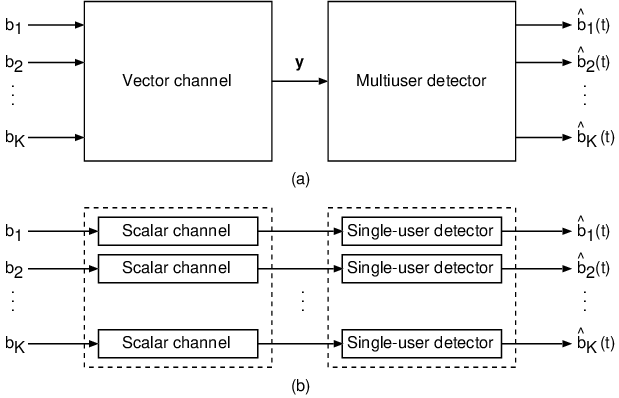}
\caption{
Schematic of decoupling principle. 
(a) Vector channel followed by multiuser detection and estimates at stage $t$. 
(b) Equivalent scalar channels followed by single-user detectors and estimates at stage $t$. 
}
\label{fig:SchematicFigure}
\end{center}
\end{figure}
\begin{figure}[t]%[htbp]
\begin{center}
\includegraphics[width=.96\linewidth,keepaspectratio]{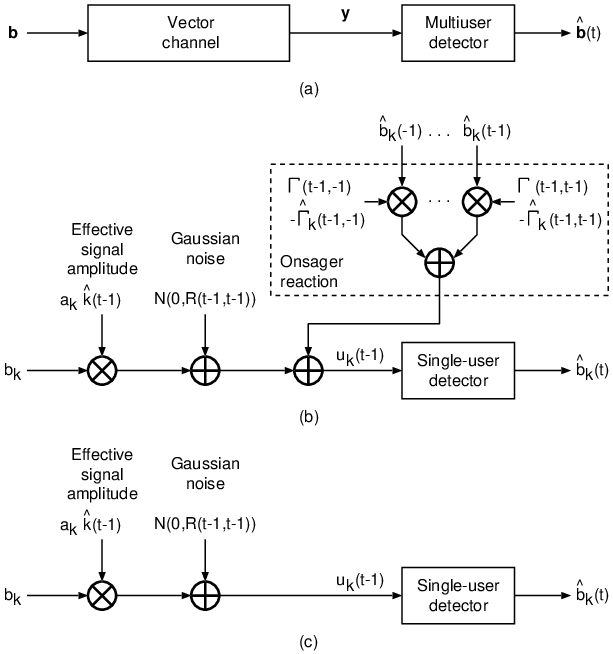}
\caption{
Vector channnel and equivalent single-user channels. 
(a) Vector channel, multiuser detection, and tentative decision. 
(b) Equivalent single-user non-Gaussian channel, single-user detection, and tentative decision. 
(c) Equivalent single-user Gaussian channel, single-user detection, and tentative decision. 
}
\label{fig:DecouplingPrinciple}
\end{center}
\end{figure}

\subsection{Decoupling Principle}
%~~~~~~~~~~~~~~~~~~~~~~~~~~~~~~~~~~~~~~~~~~~~~~~~~~~~~~~~~~~~~~~~~~~~~

From the average of (\ref{eq:E}) we find the effective single-user process: 
$\tilde{b}_k^{(s+1)}=f(u_k^{(s)})$ and $\tilde{b}_k^{(-1)}=0$. 
Variable $\vec{v}_k$ can be regarded as the $(t+1)$-dimensional Gaussian variable 
with mean vector $\vec{0}$ and covariance matrix $\vec{R}$. 
Using (\ref{eq:u}), the effective single-user iterative process can be written in the following simple form. 
\begin{align}
  u_k^{(s)} = 
  &  a_k \hat{k}^{(s)} b_k  + v_k^{(s)} \notag \\
  & + \sum_{s'=-1}^s [ \Gamma^{(s,s')} - \hat{\Gamma}_k^{(s,s')} ] f(u_k^{(s'-1)}), \label{eq:esp}
\end{align}
with $v_k^{(s)} \sim \mathcal{N}(0,R^{(s-1,s-1)})$, 
where $\tilde{b}_k^{(s)}=f(u_k^{(s-1)})$, $\tilde{b}_k^{(-1)}=f(u_k^{(-2)})=0$. 
Introducing the average over $\vec{v}_k$ as 
$\langle \, \cdot \, \rangle_{\vec{v}_k} = \int \mathcal{D}\vec{v}_k \, \cdot$ and using (\ref{eq:esp}), 
then $m^{(s)}$, $C^{(s,s')}$, and $G^{(s,s')}$ are obtained as 
$m^{(s)} = \langle b_k \tilde{b}_k^{(s)} \rangle_{\vec{v},b_k,a_k}$, 
$C^{(s,s')} = \langle \tilde{b}_k^{(s)} \tilde{b}_k^{(s')} \rangle_{\vec{v},b_k,a_k}$, and 
$G^{(s,s')} = \mathbb{I}(s \le s') \langle \tilde{b}_k^{(s)} (\vec{R}^{-1}\vec{v}_k)^{(s')} \rangle_{\vec{v}_k,b_k,a_k}$, respectively. 
The macroscopic parameters contain the effects of all users as averages; 
therefore, these can be represented by each user's distribution. 
Hence, (\ref{eq:esp}) holds for any user $k$. 
All the parameters in (\ref{eq:esp}) can only be represented by each user's distribution. 
\par
This result (\ref{eq:esp}) means that each user experiences an equivalent single-user additive noise channel 
whose signal-to-noise ratio degrades due to both the multiple-access interference and the Onsager reaction, 
at each stage of the algorithm. 
This property is called the decoupling principle \cite{Guo2005}. 
The additive channel noise generally becomes non-Gaussian 
due to the existence of the Onsager reaction $\sum_{t'=-1}^{t-1} \Gamma^{(t-1,t')} f(u_k^{(t'-1)})$. 
Figure \ref{fig:SchematicFigure} has a schematic of the decoupling principle. 
\par
Figure \ref{fig:DecouplingPrinciple} shows the results of the analysis. 
Approximate {\it posterior-mean estimator (PME)} $\tilde{b}_k^{(t)} = f(u_k^{(t-1)})$ is used as the single-user detector. 
It should be noted that (\ref{eq:esp}) implies that equivalent channel noise becomes Gaussian 
if the Onsager reaction is cancelled. 
However, for soft-PIC, the Onsager reaction is not cancelled (Fig. \ref{fig:DecouplingPrinciple} (b)). 
If the Onsager reaction does not vanish, the noise distribution is not centered at zero. 
Therefore, soft-PIC must be suboptimal. 
\par
Under the assumption of the large-system limit, the performance of algorithm (\ref{eq:GenericAlgorithm}) at stage $t$ 
can be evaluated by using the equivalent scalar channel determined 
from three types of macroscopic parameters $(\vec{m},\vec{C},\vec{G})$, where $\vec{m}$, $\vec{C}$, and $\vec{G}$ are 
a $(t+1)$-dimensional vector $(m^{(-1)},\cdots,m^{(t-1)})^\top$, 
a $(t+1) \times (t+1)$ matrix $(C^{(s,s')})$ and 
a $(t+1) \times (t+1)$ matrix $(G^{(s,s')})$, respectively. 
The bit error rate $P_b^{(t)}$ of hard decisions $\sgn [\tilde{b}_k^{(t)}]$ at the $t^{\rm th}$ stage 
is given by $P_b^{(t)}=\frac 12 (1- \langle b_k \sgn [ \tilde{b}_k^{(t)}] \rangle_{\vec{v},b_k,a_k})$. 
%where function $\sgn(x)$ denotes the sign function taking 1 for $x \ge 0$ and -1 for $x<0$. %DELETED
\par
We next consider the case where we set the parameters in (\ref{eq:GenericAlgorithm}) to those of ORC-PIC. 
The reaction term in the effective single-user process is determined 
by two matrices $\vec{\Gamma}$ and $\hat{\vec{\Gamma}}_k$. 
It is considered that the parameter $\vec{\Gamma}$ represents a retarded self-interaction caused 
by iterative calculation. 
The parameter $\hat{\vec{\Gamma}}_k$, on the other hand, is arbitrary; 
therefore we can choose $\hat{\vec{\Gamma}}_k$ that cancel the Onsager reaction. 
Using the inductive method, we can show that the Onsager reaction is entirely cancelled at each stage $t$, 
when the parameter $\hat{\Gamma}^{(s,s')}$ is set to 
$(\hat{\vec{\Gamma}} \tilde{\vec{b}})^{(t)} = \beta G^{(t,t-1)} [ \tilde{b}^{(t-1)} - (\hat{\vec{\Gamma}} \tilde{\vec{b}})^{(t-1)} ]$ and 
$(\hat{\vec{\Gamma}} \tilde{\vec{b}})^{(-1)} = (\hat{\vec{\Gamma}} \tilde{\vec{b}})^{(0)} = 0$. 
This parameter $\hat{\vec{\Gamma}}$ coincides with the parameter derived by density evolution \cite{Tanaka2005}. 
%This implies that both the result from GFA and density evolution if the Onsager reaction is cancelled. %DELETED
Density evolution cannot treat the Onsager reaction, however density evolution can be applied to derive an algorithm that can correctly cancel the Onsager reaction. %ADDED
\par
In the case where we set the parameters in (\ref{eq:GenericAlgorithm}) to those of the BP-based detector, 
we confirmed the Onsager reaction is cancelled at least at stage 8 through straightforward calculations. 
In this case, $\vec{\Gamma}$ becomes ${\rm diag} (A^{(-1)}, 0, \cdots, 0)$. 
Therefore, for both ORC-PIC and the BP-based detector (at least at stage 8), 
the equivalent single-user channels are given as an additive white Gaussian noise channel (Fig. \ref{fig:DecouplingPrinciple} (c)).

\section{Conclusion}
%~~~~~~~~~~~~~~~~~~~~~~~~~~~~~~~~~~~~~~~~~~~~~~~~~~~~~~~~~~~~~~~~~~~~~
\par
We evaluated the detection dynamics 
of soft-PIC, ORC-PIC and the BP-based detector by applying GFA 
in the large-system limit. 
We also showed that GFA could treat the dynamics of belief propagation. 
\par
From the practical point of view, 
iterative algorithms are generally utilized as suboptimal methods. 
It is important to understand the detection dynamics in detail to improve detectors. 
We confirmed that the results from the density evolution analysis could be obtained 
from those of the GFA analysis by omitting the Onsager reaction term. 
The GFA could correctly treat the Onsager reaction, 
which is a retarded self-interaction due to iterations, 
and gives density evolution a basic theory. 
%To validate the results, we performed numerical experiments in an $N=8,000$ system. 
%The system of $N=8,000$ is too large for a practical system, 
%but we are now concerned with the verification of the analytical result derived under the large system limit. 
\par
We also studied the decoupling principle in iterative multiuser detection algorithms using GFA. 
and found that the decoupling principle holds. 
%The equivalent additive noise is generally non-Gaussian. 
%However, if algorithms can cancel the Onsager reaction, 
%such as ORC-PIC and the BP-based detector, 
%equivalent single-user channels coincide with the additive white Gaussian noise channel at each stage. 
The reconstruction algorithm of compressed sensing could also be analyzed by GFA \cite{Mimura2011}.

%=====================================================================
% if have a single appendix:
%\appendix[Proof of the Zonklar Equations]
% or
%\appendix  % for no appendix heading
% do not use \section anymore after \appendix, only \section*
% is possibly needed

% use appendices with more than one appendix
% then use \section to start each appendix
% you must declare a \section before using any
% \subsection or using \label (\appendices by itself
% starts a section numbered zero.)
%

\appendices

\section{Notations} \label{app:notations}
%~~~~~~~~~~~~~~~~~~~~~~~~~~~~~~~~~~~~~~~~~~~~~~~~~~~~~~~~~~~~~~~~~~~~~
\par
We here summarize notations, which are used in this paper, in Table \ref{table:notations}. 
It should be noted that notations to describe the GFA result don't contain the user number $k$, 
since a many-user problem is reduced to an equivalent single-user problem in GFA.  
In Table \ref{table:notations} 'BP' denotes BP-based detector. 

\newpage

\begin{table}[b!]
  \begin{center}
    \caption{Summary of notations}
    \label{table:notations}
    \begin{tabular}{lll}
      \hline
               & the first   &             \\
      variable & or top-left & description \\
               & element     &             \\
      \hline \\
      (notations for models) & & \\
      $K \in \mathbb{N}^+$ & -- & the number of users \\
      $N \in \mathbb{N}^+$ & -- & spreading code length \\
      $k, k' \in \{1 ,\cdots, K\}$ & -- & user number \\
      $\mu, \mu' \in \{1 ,\cdots, N\}$ & -- & chip interval index \\
      $t \in \mathbb{N}^+$ & -- & stage \\
      $s, s', t'\in \{-1, 0, \cdots, t\}$ & -- & stage \\
      $\vec{s}_k =(s_k^\mu) \in \{ \pm 1 \}^N$ & $s_k^1$ & spreading code of user $k$ \\
      $\vec{b} =(b_k) \in \{ \pm 1 \}^K$ & $b_1$ & information bits \\
      $\tilde{\vec{b}}^{(t)} =(\tilde{b}_k^{(t)}) \in \mathbb{R}^K$ & $\tilde{b}_1^{(t)}$ & tentative soft decisions \\
      $\hat{\vec{b}}^{(t)} =(\hat{b}_k^{(t)}) \in \{\pm 1\}^K$ & $\hat{b}_1^{(t)}$ & tentative hard decisions \\
      $\vec{y} =(y^\mu) \in \mathbb{R}^N$ & $y^1$ & received signals \\
      $\vec{n} =(n^\mu) \in \mathbb{R}^N$ & $n^1$ & noise vector \\
      $\sigma_0\in\mathbb{R}^+$ & -- & true noise variance \\
      $\sigma\in\mathbb{R}^+$ & -- & control parameter \\
      %& & (estimated variance) \\
      $\vec{W}=(W_{kk'}) \in \mathbb{R}^{K \times K}$ & $W_{11}$ & correlation matrix \\
      $\vec{\psi}^{(t)}=(\psi_k^{(t)}) \in \mathbb{R}^K$ & $\psi_1^{(t)}$ & dummy functions \\
      $\vec{\theta}^{(t)}=(\theta_k^{(t)})\in\mathbb{R}^K$ & $\theta_1^{(t)}$ & external message \\
      $\vec{h}=(h_k)\in\mathbb{R}^K$ & $h_1$ & matched filter outputs \\
      $f:\mathbb{R}\to\mathbb{R}$ & -- & transfer function \\
      $\hat{\vec{\Gamma}}_k=(\hat{\Gamma}_k^{(s,s')})\in\mathbb{R}^{(\!t+1\!) \! \times \! (\!t+1\!)}$ & $\hat{\Gamma}_k^{(\!-1,-1\!)}$ & parameter for ORC-PIC \\
      $R^{(t)}\in\mathbb{R}^+$ & -- & parameter for BP \\
      $A^{(t)}\in\mathbb{R}^+$ & -- & parameter for BP \\
      $Q^{(t)}\in\mathbb{R}^+$ & -- & parameter for BP \\
      $J^{(t,s)}\in\mathbb{R}^+$ & -- & parameter for BP \\
      $\gamma \in \mathbb{R}^+$ & -- & inverse temperature \\
      $\vec{S}=(a_k \vec{s}_k) \in \mathbb{R}^{N \times K}$ & $a_1 s_1^1$ & channel state matrix \\
      $a_k \in \mathbb{R}^+$ & -- & received amplitude \\
      $P_b^{(t)}\in\mathbb{R}^+$ & -- & bit error rate \\ \\
      %$u_k^{(t)}\in\mathbb{R}$ & -- & summation over all messages \\
      (notations for GFA) & & \\
      $\tilde{\vec{b}} =(\tilde{b}^{(t)}) \in \mathbb{R}^{t+1}$ & $\tilde{b}^{(-1)}$ & tentative soft decisions \\
      $\hat{\vec{b}} =(\hat{b}^{(t)}) \in \{\pm 1\}^{t+1}$ & $\hat{b}^{(-1)}$ & tentative hard decisions \\
      $\vec{\psi}=(\psi^{(t)}) \in \mathbb{R}^{t+1}$ & $\psi^{(-1)}$ & dummy functions \\
      $\vec{\theta}=(\theta^{(t)})\in\mathbb{R}^{t+1}$ & $\theta^{(-1)}$ & external message \\
      $\vec{\eta}=(\eta^{(s)})\in\mathbb{R}^{t+1}$ & $\eta^{(-1)}$ & macroscopic parameter \\
      $\hat{\vec{\eta}}=(\hat{\eta}^{(s)})\in\mathbb{R}^{t+1}$ & $\hat{\eta}^{(-1)}$ & macroscopic parameter \\
      $\vec{k}=(k^{(s)})\in\mathbb{R}^{t+1}$ & $k^{(-1)}$ & macroscopic parameter \\
      $\hat{\vec{k}}=(\hat{k}^{(s)})\in\mathbb{R}^{t+1}$ & $\hat{k}^{(-1)}$ & macroscopic parameter \\
      $\vec{q}=(q^{(s,s')})\in\mathbb{R}^{(\!t+1\!) \! \times \! (\!t+1\!)}$ & $q^{(\!-1,-1\!)}$ & macroscopic parameter \\
      $\hat{\vec{q}}=(\hat{q}^{(s,s')})\in\mathbb{R}^{(\!t+1\!) \! \times \! (\!t+1\!)}$ & $\hat{q}^{(\!-1,-1\!)}$ & macroscopic parameter \\
      $\vec{Q}=(Q^{(s,s')})\in\mathbb{R}^{(\!t+1\!) \! \times \! (\!t+1\!)}$ & $Q^{(\!-1,-1\!)}$ & macroscopic parameter \\
      $\hat{\vec{Q}}=(\hat{Q}^{(s,s')})\in\mathbb{R}^{(\!t+1\!) \! \times \! (\!t+1\!)}$ & $\hat{Q}^{(\!-1,-1\!)}$ & macroscopic parameter \\
      $\vec{L}=(L^{(s,s')})\in\mathbb{R}^{(\!t+1\!) \! \times \! (\!t+1\!)}$ & $L^{(\!-1,-1\!)}$ & macroscopic parameter \\
      $\hat{\vec{L}}=(\hat{L}^{(s,s')})\in\mathbb{R}^{(\!t+1\!) \! \times \! (\!t+1\!)}$ & $\hat{L}^{(\!-1,-1\!)}$ & macroscopic parameter \\
      $\vec{m}=(m^{(s)})\in\mathbb{R}^{t+1}$ & $m^{(-1)}$ & overlap \\
      $\vec{C}=(C^{(s,s')})\in\mathbb{R}^{(\!t+1\!) \! \times \! (\!t+1\!)}$ & $C^{(\!-1,-1\!)}$ & correlation function \\
      $\vec{G}=(G^{(s,s')})\in\mathbb{R}^{(\!t+1\!) \! \times \! (\!t+1\!)}$ & $G^{(\!-1,-1\!)}$ & response function \\ \\
      \hline
    \end{tabular}
  \end{center}
\end{table}

\section{Derivation of Lemma \ref{lemma:DisorderAveragedGeneratingFunctional}} \label{app:Z_soft-PIC}
%~~~~~~~~~~~~~~~~~~~~~~~~~~~~~~~~~~~~~~~~~~~~~~~~~~~~~~~~~~~~~~~~~~~~~
\par
Evaluating the averaged generating functional straightforwardly, 
one can obtain Lemma \ref{lemma:DisorderAveragedGeneratingFunctional}. 
Substituting (\ref{eq:SystemModel}) and (\ref{eq:h_k}) into (\ref{eq:def_of_SumOverAllMessage}), 
the summation over all messages becomes 
\begin{align}
  u_k^{(t)}
  =& \sum_{k'=1}^K W_{kk'} (b_{k'} -\tilde{b}_{k'}^{(t)}) 
  + \tilde{b}_k^{(t)}
  + \frac{\sigma_0}{\sqrt{N}} \sum_{\mu=1}^N s_k^\mu n^\mu 
  + \theta_k^{(t)}. 
  \label{eq:u_k}
\end{align}
The average generating functional $\bar{Z}[\vec{\psi}]$ is represented 
using $\vec{u}^{(s)}=(u_1^{(s)},\cdots,u_K^{(s)})^\top$ whose elements are defined by (\ref{eq:u_k}). 
Using Dirac's delta function $\delta$, we first introduce the definition of $\vec{u}^{(-1)}, \cdots, \vec{u}^{(t-1)}$ 
into $Z[\vec{\psi}]$, which is written in terms of them, as 
\begin{align}
  \bar{Z}[\vec{\psi}] 
  =& \mathbb{E}_{\vec{s}_1,\cdots,\vec{s}_K,\vec{n}} \biggl[ \int_{\mathbb{R}^{(t+1)K}} \underline{\rmd}\vec{u} \nonumber \\
  & \times 
    p[\tilde{\vec{b}}^{(-1)}] 
    \biggl( \prod_{s=-1}^{t-1} \frac{\gamma}{\sqrt{2\pi}} e^{-\frac{\gamma^2}2 \{ \tilde{b}_k^{(s+1)}-f(u_k^{(s)}) \}^2} \biggr) 
    \nonumber \\
  & \times \exp \biggl[ -\rmi\sum_{s=-1}^t\tilde{\vec{b}}^{(s)} \cdot \vec{\psi}^{(s)} \biggr]
    \nonumber \\
  & \times \prod_{s=-1}^{t-1} \prod_{k=1}^K 
      \delta \biggl( u_k^{(s)} - \sum_{k'=1}^K W_{kk'} (b_{k'}-\tilde{b}_{k'}^{(s)}) - \tilde{b}_k^{(s)} \nonumber \\
  & - \frac{\sigma_0}{\sqrt{N}} \sum_{\mu=1}^N s_k^\mu n^\mu 
    - \theta_k^{(s)} \biggr) \biggr], 
\end{align}
where $\underline{\rmd}\vec{u}$$:=$$\prod_{s=-1}^{t-1} \prod_{k=1}^K \frac{\rmd u_k^{(s)}}{\sqrt{2\pi}}$. 
The arguments of Dirac's delta functions $\delta( \cdot )$ represent the definition of $\{ u_k^{(t)} \}$. 
The term of $\delta$ is rewritten by applying the Fourier integral form 
of Dirac's delta function $\delta(x)=\int_{\mathbb{R}} \rmd \hat{x} e^{\rmi \hat{x} x}$ as 
$\delta (u_k^{(s)} - [\cdots]) = \int_{\mathbb{R}} \rmd \hat{u}_k^{(s)} \exp\{ \rmi \hat{u}_k^{(s)} (u_k^{(s)} - [\cdots]) \}$. 
We then have 
\begin{eqnarray}
  \bar{Z}[\vec{\psi}]
  &=& \sum_{\tilde{\vec{b}}^{(-1)},\cdots,\tilde{\vec{b}}^{(t)}} p[\tilde{\vec{b}}^{(-1)}] 
      \int_{\mathbb{R}^{2(t+1)K}} \underline{\rmd}\vec{u}\underline{\rmd}\hat{\vec{u}} \nonumber \\
  & & \times \exp \biggl[ \rmi \sum_{s=-1}^{t-1} \sum_{k=1}^K 
      \hat{u}_k^{(s)} \{ u_k^{(s)} - \tilde{b}_k^{(s)} - \theta_k^{(s)} \} \nonumber \\
  & & - \rmi \sum_{s=-1}^t \sum_{k=1}^K \tilde{b}_k^{(s)} \psi_k^{(s)} \nonumber \\
  & & + \sum_{s=-1}^t \sum_{k=1}^K \{ \ln \frac{\gamma}{\sqrt{2\pi}} 
      - \frac{\gamma^2}2 [ \tilde{b}_k^{(t+1)}-f(u_k^{(s)}) ]^2 \} \biggr] \nonumber \\
  & & \times \mathbb{E}_{ \vec{s}_1, \cdots, \vec{s}_K, \vec{n} } \biggl\{ \exp \biggl[ \nonumber \\
  & & - \rmi \sqrt{\beta} \sigma_0 \sum_{\mu=1}^N \sum_{s=-1}^{t-1} 
      \biggl( \frac1{\sqrt{K}} \sum_{k=1}^K s_k^\mu  \hat{u}_k^{(s)} \biggr) n^\mu \nonumber \\
  & & - \rmi \beta \sum_{\mu=1}^N \sum_{s=-1}^{t-1} 
      \biggl( \frac1{\sqrt{K}} \sum_{k=1}^K s_k^\mu  \hat{u}_k^{(s)} \biggr) \nonumber \\
  & & \times \biggl( \frac1{\sqrt{K}} \sum_{k'=1}^K s_{k'}^\mu \{ b_{k'} - \tilde{b}_{k'}^{(s)} \} \biggr) \biggr] \biggr\} , 
  \label{eq:barZ_calc}
\end{eqnarray}
where $\underline{\rmd}\hat{\vec{u}}$$:=$$\prod_{s=-1}^{t-1} \prod_{k=1}^K \frac{\rmd \hat{u}_k^{(s)}}{\sqrt{2\pi}}$. 
Without loss of generality, we can set to $b_k=1 \; (\forall k)$. 
\par
To take the average of $\vec{s}_1, \cdots, \vec{s}_K$, 
we introduce variables $v_\mu^{(s)}$ and $w_\mu^{(s)}$ which are defined as: 
\begin{eqnarray}
  & & v_k^{(s)} := \frac1{\sqrt{K}} \sum_{k=1}^K s_k^\mu \hat{u}_k^{(s)}, \\
  & & w_k^{(s)} := \frac1{\sqrt{K}} \sum_{k=1}^K s_k^\mu \{ 1-\tilde{b}_k^{(s)} \}. 
\end{eqnarray}
It should be noted that the averaged generating functional $\bar{Z}[\vec{\psi}]$ 
of (\ref{eq:barZ_calc}) includes $\vec{s}_1, \cdots, \vec{s}_K$ 
only in terms of these variables $v_\mu^{(s)}$ and $w_\mu^{(s)}$. 
Due to this, random variables $\vec{s}_1, \cdots, \vec{s}_K$ 
are isolated and are able to be averaged. 
Introducing $v_\mu^{(s)}$ and $w_\mu^{(s)}$ 
into term $\mathbb{E}_{ \vec{s}_1, \cdots, \vec{s}_K, \vec{n} } \{\cdots\}$ in (\ref{eq:barZ_calc}), 
one obtains 
\begin{eqnarray}
  & & \int_{\mathbb{R}^{2tN}} \underline{\rmd}\vec{v}\underline{\rmd}\vec{w} 
      \mathbb{E}_{ \vec{s}_1, \cdots, \vec{s}_K, \vec{n} } \biggl\{ \nonumber \\
  & & \exp \biggl[ - \rmi \sqrt{\beta} \sigma_0 \sum_{\mu=1}^N \sum_{s=-1}^{t-1} v_\mu^{(s)} n^\mu 
      - \rmi \beta \sum_{\mu=1}^N \sum_{s=-1}^{t-1} v_\mu^{(s)} w_\mu^{(s)} \biggr] \biggr\} \nonumber \\
  & & \times \delta \biggl( v_k^{(s)} - \frac1{\sqrt{K}} \sum_{k=1}^K s_k^\mu \hat{u}_k^{(s)} \biggr) \nonumber \\
  & & \times \delta \biggl( w_k^{(s)} - \frac1{\sqrt{K}} \sum_{k=1}^K s_k^\mu \{ 1-\tilde{b}_k^{(s)} \} \biggr) \nonumber \\
  &=& \int_{\mathbb{R}^{4tN}} \underline{\rmd}\vec{v}\underline{\rmd}\hat{\vec{v}}\underline{\rmd}\vec{w}\underline{\rmd}\hat{\vec{w}} \nonumber \\
  & & \times \exp \biggl[ \rmi \sum_{\mu=1}^N \sum_{s=-1}^{t-1} 
     \{ \hat{v}_\mu^{(s)} v_\mu^{(s)} + \hat{w}_\mu^{(s)} w_\mu^{(s)} - \beta v_\mu^{(s)} w_\mu^{(s)} \} \biggr] \nonumber \\
  & & \times \mathbb{E}_{ \vec{n} } \biggl\{ \exp \biggl[ 
      - \rmi \sqrt{\beta} \sigma_0 \sum_{\mu=1}^N \sum_{s=-1}^{t-1} v_\mu^{(s)} n^\mu \biggr] \biggr\} \nonumber \\
  & & \times \mathbb{E}_{ \vec{s}_1, \cdots, \vec{s}_K } \biggl\{ \exp \biggl[ 
      - \rmi \frac1{\sqrt{K}} \sum_{\mu=1}^N \sum_{s=-1}^{t-1} 
      \{ \hat{v}_\mu^{(s)} \sum_{k=1}^K s_k^\mu \hat{u}_k^{(s)} \nonumber \\
  & & + \hat{w}_\mu^{(s)} \sum_{k=1}^K s_k^\mu (1-\tilde{b}_k^{(s)}) \} \biggr] \biggr\}. 
  \label{eq:DisorderedAverage}
\end{eqnarray}
We here again used the Fourier integral form of Dirac's delta. 
The term $\mathbb{E}_{ \vec{n} } \{\cdots\}$ in (\ref{eq:DisorderedAverage}) becomes 
\begin{eqnarray}
  & & \mathbb{E}_{ \vec{n} } \biggl\{ \exp \biggl[ - \rmi \sqrt{\beta} \sigma_0 \sum_{\mu=1}^N \sum_{s=-1}^{t-1} v_\mu^{(s)} n^\mu \biggr] \biggr\} \nonumber \\
  % = \prod_{\mu=1}^N \biggl< \exp \biggl[ - \rmi \sqrt{\beta} \sigma_0 \sum_{s=-1}^{t-1} v_\mu^{(s)} n^\mu \biggr] \biggr>_n \nonumber \\
  % = \prod_{\mu=1}^N \int Dn^\mu \exp \biggl[ - \rmi \sqrt{\beta} \sigma_0 \sum_{s=-1}^{t-1} v_\mu^{(s)} n^\mu \biggr] 
  %   \quad \because Dn^\mu := (2\pi)^{-1/2}e^{-(n^\mu)^2/2}dn^\mu \nonumber \\
  % = \prod_{\mu=1}^N \exp \biggl[ \frac12 \biggl( - \rmi \sqrt{\beta} \sigma_0 \sum_{s=-1}^{t-1} v_\mu^{(s)} \biggr)^2 \biggr] \nonumber \\
  & & = \prod_{\mu=1}^N \exp \biggl[ -\frac12 \beta \sigma_0^2 \sum_{s=-1}^{t-1} \sum_{s'=-1}^{t-1} v_\mu^{(s)} v_\mu^{(s')} \biggr]. 
\end{eqnarray}
Since $t$ is finite, term $\mathbb{E}_{ \vec{s}_1, \cdots, \vec{s}_K } \{\cdots\}$ in (\ref{eq:DisorderedAverage}) is given by 
\begin{eqnarray}
  & & \mathbb{E}_{ \vec{s}_1, \cdots, \vec{s}_K } \biggl\{ \exp \biggl[ - \rmi \frac1{\sqrt{K}} \sum_{\mu=1}^N \sum_{s=-1}^{t-1} \{ \hat{v}_\mu^{(s)} \sum_{k=1}^K s_k^\mu \hat{u}_k^{(s)} \nonumber \\
  & & + \hat{w}_\mu^{(s)} \sum_{k=1}^K s_k^\mu (1-\tilde{b}_k^{(s)}) \} \biggr] \biggr\} \nonumber \\
  &=& \prod_{\mu=1}^N \prod_{k=1}^K \cos \biggl[ - \frac1{\sqrt{K}} \sum_{s=-1}^{t-1} \{ \hat{v}_\mu^{(s)} \hat{u}_k^{(s)} + \hat{w}_\mu^{(s)} (1-\tilde{b}_k^{(s)}) \} \biggr] \nonumber \\
  &=& \prod_{\mu=1}^N \prod_{k=1}^K \exp \biggl[ - \frac12 \biggl( \frac1{\sqrt{K}} \sum_{s=-1}^{t-1} \{ \hat{v}_\mu^{(s)} \hat{u}_k^{(s)} \nonumber \\
  & & + \hat{w}_\mu^{(s)} (1-\tilde{b}_k^{(s)}) \} \biggr)^2 +O(K^{-2}) \biggr] \nonumber \\
  % &=& \prod_{\mu=1}^N \prod_{k=1}^K \exp \biggl[ - \frac1{2K} \sum_{s=-1}^{t-1} \sum_{s'=-1}^{t-1} \{ \hat{v}_\mu^{(s)} \hat{u}_k^{(s)} 
  %     + \hat{w}_\mu^{(s)} (1-\tilde{b}_k^{(s)}) \} \{ \hat{v}_\mu(s') \hat{u}_k^{(s')} + \hat{w}_\mu(s') (1-\tilde{b}_k^{(s')}) \} \biggr] \nonumber \\
  % &=& \prod_{\mu=1}^N \exp \biggl[ - \frac12 \sum_{s=-1}^{t-1} \sum_{s'=-1}^{t-1} \biggl\{ \hat{v}_\mu^{(s)} 
  %     \biggl( \frac 1K \sum_{k=1}^K \hat{u}_k^{(s)} \hat{u}_k^{(s')} \biggr) \hat{v}_\mu(s') 
  %     + \hat{v}_\mu^{(s)} \biggl( \frac 1K \sum_{k=1}^K \hat{u}_k^{(s)} (1-\tilde{b}_k^{(s')}) \biggr) \hat{w}_\mu(s') \nonumber \\
  % & & \quad + \hat{w}_\mu^{(s)} \biggl( \frac 1K \sum_{k=1}^K (1-\tilde{b}_k^{(s)}) \hat{u}_k^{(s')} \biggr) \hat{v}_\mu(s') 
  %     + \hat{w}_\mu^{(s)} \biggl( \frac 1K \sum_{k=1}^K (1-\tilde{b}_k^{(s)}) (1-\tilde{b}_k^{(s')}) \biggr) \hat{v}_\mu(s') \biggr\} \biggr] \nonumber \\
  &=& \prod_{\mu=1}^N \exp \biggl[ - \frac12 \sum_{s=-1}^{t-1} \sum_{s'=-1}^{t-1} \biggl\{ \hat{v}_\mu^{(s)} \biggl( \frac 1K \sum_{k=1}^K \hat{u}_k^{(s)} \hat{u}_k^{(s')} \biggr) \hat{v}_\mu^{(s')} \nonumber \\
  & & + \hat{v}_\mu^{(s)} \biggl( \frac 1K \sum_{k=1}^K \hat{u}_k^{(s)} - \frac 1K \sum_{k=1}^K \tilde{b}_k^{(s')} \hat{u}_k^{(s)} \biggr) \hat{w}_\mu^{(s')} \nonumber \\
  & & + \hat{w}_\mu^{(s)} \biggl( \frac 1K \sum_{k=1}^K \hat{u}_k^{(s')} - \frac 1K \sum_{k=1}^K \tilde{b}_k^{(s)} \hat{u}_k^{(s')} \biggr) \hat{v}_\mu^{(s')} \nonumber \\
  & & + \hat{w}_\mu^{(s)} \biggl( 1 - \frac 1K \sum_{k=1}^K \tilde{b}_k^{(s)} - \frac 1K \sum_{k=1}^K \tilde{b}_k^{(s')} \nonumber \\
  & & + \frac 1K \sum_{k=1}^K \tilde{b}_k^{(s)} \tilde{b}_k^{(s')} \biggr) \hat{v}_\mu^{(s')} \biggr\} +O(K^{-1}) \biggr]. 
  \label{eq:SpreadSequenceAverage}
\end{eqnarray}
We next separate the relevant one-stage and two-stage values which are appeared in (\ref{eq:SpreadSequenceAverage}): 
\begin{eqnarray}
  & & \eta^{(s)} := \frac1{\sqrt{K}} \sum_{k=1}^K \tilde{b}_k^{(s)}, \label{eq:def_mu} \\
  & & k^{(s)} := \frac1{\sqrt{K}} \sum_{k=1}^K \hat{u}_k^{(s)}, \\
  & & q^{(s,s')} := \frac1{\sqrt{K}} \sum_{k=1}^K \tilde{b}_k^{(s)} \tilde{b}_k^{(s')}, \\
  & & Q^{(s,s')} := \frac1{\sqrt{K}} \sum_{k=1}^K \hat{u}_k^{(s)} \hat{u}_k^{(s')}, \\
  & & L^{(s,s')} := \frac1{\sqrt{K}} \sum_{k=1}^K \tilde{b}_k^{(s)} \hat{u}_k^{(s')}. \label{eq:def_L}
\end{eqnarray}
Equation (\ref{eq:SpreadSequenceAverage}) is only written in terms of these variables 
$\eta^{(s)}$, $k^{(s)}$, $q^{(s,s')}$, $Q^{(s,s')}$ and $L^{(s,s')}$. 
We hereafter call these variables {\it macroscopic parameters}, which describe the nature of the system 
and are also often called the {\it order parameters}. 
Similar to the way to derive (\ref{eq:DisorderedAverage}), 
one can introduce $m^{(s)}$, $k^{(s)}$, $q^{(s,s')}$, $Q^{(s,s')}$ and $L^{(s,s')}$ into (\ref{eq:SpreadSequenceAverage}). 
The term of $\delta$ in (\ref{eq:SpreadSequenceAverage}) is again rewritten by applying the Fourier integral form 
of Dirac's delta such as 
$\delta(\eta^{(s)}$ $-[\cdots])$ $=\int_{\mathbb{R}}$ $\rmd \hat{\eta}^{(s)}$ $\exp\{ \rmi \hat{\eta}^{(s)}(\eta^{(s)}$ $-[\cdots])\}$, 
$\delta(k^{(s)}$ $-[\cdots])$ $=\int_{\mathbb{R}}$ $\rmd \hat{k}^{(s)}$ $\exp\{ \rmi \hat{k}^{(s)}(k^{(s)}$ $-[\cdots])\}$, 
$\delta(q^{(s,s')}$ $-[\cdots])$ $=\int_{\mathbb{R}}$ $\rmd \hat{q}^{(s,s')}$ $\exp\{ \rmi \hat{q}^{(s,s')}(q^{(s,s')}$ $-[\cdots])\}$, 
$\delta(Q^{(s,s')}$ $-[\cdots])$ $=\int_{\mathbb{R}}$ $\rmd \hat{Q}^{(s,s')}$ $\exp\{ \rmi \hat{Q}^{(s,s')}(Q^{(s,s')}$ $-[\cdots])\}$ and 
$\delta(L^{(s,s')}$ $-[\cdots])$ $=\int_{\mathbb{R}}$ $\rmd \hat{L}^{(s,s')}$ $\exp\{ \rmi \hat{L}^{(s,s')}(L^{(s,s')}$ $-[\cdots])\}$. 
We here introduce the notations $\underline{\rmd}u := \prod_{s=-1}^{t-1} \frac{\rmd u^{(s)}}{\sqrt{2\pi}}$ 
$\underline{\rmd} \hat{u} := \prod_{s=-1}^{t-1} \frac{\rmd \hat{u}^{(s)}}{\sqrt{2\pi}}$, 
$\rmd\vec{\eta} := \prod_{s=-1}^{t-1} \rmd \eta^{(s)}$, 
$\rmd\hat{\vec{\eta}} := \prod_{s=-1}^{t-1} \rmd \hat{\eta}^{(s)}$, 
$\rmd\vec{k} := \prod_{s=-1}^{t-1} \rmd k^{(s)}$, 
$\rmd\hat{\vec{k}} := \prod_{s=-1}^{t-1} \rmd \hat{k}^{(s)}$, 
$\rmd\vec{q} := \prod_{s=-1}^{t-1} \prod_{s'=-1}^{t-1} \rmd q^{(s,s')}$, 
$\rmd\hat{\vec{q}} := \prod_{s=-1}^{t-1} \prod_{s'=-1}^{t-1} \rmd \hat{q}^{(s,s')}$, 
$\rmd\vec{Q} := \prod_{s=-1}^{t-1} \prod_{s'=-1}^{t-1} \rmd Q^{(s,s')}$, 
$\rmd\hat{\vec{Q}} := \prod_{s=-1}^{t-1} \prod_{s'=-1}^{t-1} \rmd \hat{Q}^{(s,s')}$, 
$\rmd\vec{L} := \prod_{s=-1}^{t-1} \prod_{s'=-1}^{t-1} \rmd L^{(s,s')}$ 
$\rmd\hat{\vec{L}} := \prod_{s=-1}^{t-1} \prod_{s'=-1}^{t-1} \rmd \hat{L}^{(s,s')}$. 
$\underline{\rmd}\vec{v} := \prod_{\mu=1}^{N} \prod_{s=-1}^{t-1} \frac{\rmd v_\mu^{(s)}}{\sqrt{2\pi}}$, 
$\underline{\rmd}\hat{\vec{v}} := \prod_{\mu=1}^{N} \prod_{s=-1}^{t-1} \frac{\rmd \hat{v}_\mu^{(s)}}{\sqrt{2\pi}}$, 
$\underline{\rmd}\vec{w} := \prod_{\mu=1}^{N} \prod_{s=-1}^{t-1} \frac{\rmd w_\mu^{(s)}}{\sqrt{2\pi}}$, 
and $\underline{\rmd}\hat{\vec{w}} := \prod_{\mu=1}^{N} \prod_{s=-1}^{t-1} \frac{\rmd \hat{w}_\mu^{(s)}}{\sqrt{2\pi}}$. 
Since the initial probability, which is given as $p[\hat{\vec{b}}^{(-1)}] := \prod_{k=1}^K \delta(\tilde{b}_k^{(-1)})$, is factorizable, 
the averaged generating functional $\bar{Z}[\vec{\psi}]$ factorizes into single-user contributions, i.e., with respect to user index $k$. 
We have then arrived at Lemma \ref{lemma:DisorderAveragedGeneratingFunctional}.

\HLS %ADDED.v3(FROM HERE)
\section{Derivation of Proposition \ref{proposition:soft-PIC}} \label{appendix:soft-PIC}
%~~~~~~~~~~~~~~~~~~~~~~~~~~~~~~~~~~~~~~~~~~~~~~~~~~~~~~~~~~~~~~~~~~~~~
\par
Taking the limit of (\ref{eq:psi=0}) and (\ref{eq:theta=0}) for all $k$ and $s$, the single-user measure (\ref{eq:average_over_singleUserMeasure}) becomes user independent. 
It can be, therefore, represented without the user index $k$. 
Equation (\ref{eq:average_over_singleUserMeasure}) becomes 
\begin{align}
  & \lim_{\{\psi_k^{(s)} \to 0\}} \lim_{\{\theta_k^{(s)} \to 0\}} \langle f(\tilde{\vec{b}},\vec{u},\hat{\vec{u}}) \rangle_k \nonumber \\
  & = \frac
    {
      \displaystyle
      \int \underline{\rmd}u \underline{\rmd} \hat{u} 
      \int_{\mathbb{R}^{t+2}} \biggl( \prod_{s=-1}^{t} \rmd \tilde{b}^{(s)} \biggr) 
      w_*(\tilde{\vec{b}},\vec{u},\hat{\vec{u}}) f(\tilde{\vec{b}},\vec{u},\hat{\vec{u}}) 
    }
    {
      \displaystyle
      \int \underline{\rmd}u \underline{\rmd} \hat{u} 
      \int_{\mathbb{R}^{t+2}} \biggl( \prod_{s=-1}^{t} \rmd \tilde{b}^{(s)} \biggr) 
      w_*(\tilde{\vec{b}},\vec{u},\hat{\vec{u}}) 
    }
    \label{eq:singleUserMeasure_theta=0} \\
  & =: \langle f(\tilde{\vec{b}},\vec{u},\hat{\vec{u}}) \rangle_*, 
\end{align}
where 
\begin{align}
  & w_*(\tilde{\vec{b}},\vec{u},\hat{\vec{u}}) \nonumber \\
  & := \delta [\tilde{b}^{(-1)}] \exp \biggl[ 
      \sum_{s=-1}^{t-1} \{ \ln \frac{\gamma}{\sqrt{2\pi}} - \frac{\gamma^2}2 [\tilde{b}^{(s+1)}-f(u^{(s)})]^2 \} \nonumber \\
  & \quad - \rmi \sum_{s=0}^{t-1} \sum_{s'=0}^{t-1} \{ \hat{Q}^{(s,s')} \hat{u}^{(s)} \hat{u}^{(s')} + \hat{L}^{(s,s')} \tilde{b}^{(s)} \hat{u}^{(s')} \} \nonumber \\
  & \quad + \rmi \sum_{s=0}^{t-1} \hat{u}^{(s)} \{ u^{(s)} - \tilde{b}^{(s)} - \hat{k}^{(s)} \} \biggr] \biggl. \biggr|_{\mathrm{saddle}}. 
\end{align}
\par
We consider two kind of functions as $f(\tilde{\vec{b}},\vec{u},\hat{\vec{u}})$. 
First, we treat an arbitrary function $f(\tilde{\vec{b}})$ that does not include $\vec{u}$ and $\hat{\vec{u}}$. 
In this case, one can, therefore, perform the integral with respect to $\vec{u}$ and $\hat{\vec{u}}$. 
The numerator of (\ref{eq:singleUserMeasure_theta=0}) becomes 
\begin{align}
  & \quad \int \underline{\rmd}u \underline{\rmd} \hat{u} 
      \int_{\mathbb{R}^{t+2}} \biggl( \prod_{s=-1}^{t} \rmd \tilde{b}^{(s)} \biggr) 
      w_*(\tilde{\vec{b}},\vec{u},\hat{\vec{u}}) f(\tilde{\vec{b}}) \nonumber \\
  & = \int \underline{\rmd}u \underline{\rmd} \hat{u}
      \int_{\mathbb{R}^{t+2}} \biggl( \prod_{s=-1}^{t} \rmd \tilde{b}^{(s)} \biggr) 
      \delta [\tilde{b}^{(-1)}] \nonumber \\
  & \quad \times \biggl( \prod_{s=-1}^{t-1} \frac{\gamma}{\sqrt{2\pi}} \rme^{- \frac{\gamma^2}2[\tilde{b}^{(s+1)}-f(u^{(s)})]} \biggr) \nonumber \\
  & \quad \times \rme^{ - \frac 12 \hat{\vec{u}}^\top (\vec{1}+\beta\vec{G})^{-1}\vec{D}(\vec{1}+\beta\vec{G})^{-1} \hat{\vec{u}} } \nonumber \\
  & \quad \times \rme^{ 
      \rmi \tilde{\vec{b}}^\top (\vec{1}+\beta \vec{G}^\top)^{-1} \hat{\vec{u}} 
      + \rmi \hat{\vec{u}} \cdot (\vec{u}-\tilde{\vec{b}}-\hat{\vec{k}}) 
    } f(\tilde{\vec{b}}) \nonumber \\
  & = \int \underline{\rmd}u \underline{\rmd} \hat{u} 
      \int_{\mathbb{R}^{t+2}} \biggl( \prod_{s=-1}^{t} \rmd \tilde{b}^{(s)} \biggr) 
      \delta [\tilde{b}^{(-1)}] \nonumber \\
  & \quad \times \biggl( \prod_{s=-1}^{t-1} \frac{\gamma}{\sqrt{2\pi}} \rme^{- \frac{\gamma^2}2[\tilde{b}^{(s+1)}-f(u^{(s)})]} \biggr) \nonumber \\
  & \quad \times \rme^{ - \frac 12 \hat{\vec{u}}^\top \vec{R} \hat{\vec{u}} 
      + \rmi \hat{\vec{u}} \cdot (\vec{u}-\hat{\vec{k}}-\vec{\Gamma}\tilde{\vec{b}}) 
    } f(\tilde{\vec{b}}), 
\end{align}
where 
$\vec{R} := (\vec{1}+\beta\vec{G})^{-1}\vec{D}(\vec{1}+\beta\vec{G})^{-1}$ and 
$\vec{\Gamma} := (\vec{1}+\beta\vec{G})^{-1} \beta\vec{G}$. 
We use variable transformations 
$\vec{v}=\vec{u}-\hat{\vec{k}}-\vec{\Gamma}\tilde{\vec{b}}$ and $\vec{w}=\hat{\vec{u}}$, which gives 
\begin{align}
  & \quad \int \biggl( \prod_{s=-1}^{t-1} \frac{\partial(u^{(s)},\hat{u}^{(s)})}{\partial(v^{(s)},w^{(s)})} \frac{\rmd v^{(s)} \rmd w^{(s)}}{2\pi} \biggr) 
    \rme^{\rmi \vec{v} \cdot \vec{w} - \frac 12 \vec{w} \cdot \vec{R} \vec{w}} \nonumber \\
  & \quad \times \int_{\mathbb{R}^{t+2}} \biggl( \prod_{s=-1}^{t} \rmd \tilde{b}^{(s)} \biggr) f(\tilde{\vec{b}}) \delta [\tilde{b}^{(-1)}] \nonumber \\
  & \quad \times \biggl( \prod_{s=-1}^{t-1} \frac{\gamma}{\sqrt{2\pi}} 
    \rme^{- \frac{\gamma^2}2[\tilde{b}^{(s+1)}-f(\hat{k}^{(s)}+v^{(s)}+(\vec{\Gamma}\tilde{\vec{b}})^{(s)})]^2} \biggr) \nonumber \\
  & = \int \biggl( \prod_{s=-1}^{t-1} \frac{\rmd v^{(s)}}{\sqrt{2\pi}} \biggr) \sqrt{|\vec{R}^{-1}|} \rme^{- \frac 12 \vec{v} \cdot \vec{R} \vec{v}} \nonumber \\
  & \quad \times \int_{\mathbb{R}^{t+2}} \biggl( \prod_{s=-1}^{t} \rmd \tilde{b}^{(s)} \biggr) f(\tilde{\vec{b}}) \delta [\tilde{b}^{(-1)}] \nonumber \\
  & \quad \times \biggl( \prod_{s=-1}^{t-1} \frac{\gamma}{\sqrt{2\pi}} 
    \rme^{- \frac{\gamma^2}2[\tilde{b}^{(s+1)}-f(\hat{k}^{(s)}+v^{(s)}+(\vec{\Gamma}\tilde{\vec{b}})^{(s)})]^2} \biggr). \label{eq:app_numerator}
\end{align}
Taking the limit $\gamma \to \infty$, the numerator of (\ref{eq:average_over_singleUserMeasure}), finally, arrives at 
\begin{align}
  & \int {\cal D}\vec{v} \int_{\mathbb{R}^{t+2}} 
    \biggl( \prod_{s=-1}^{t-1} \rmd \tilde{b}^{(s)} \biggr) 
    f(\tilde{\vec{b}}) \; \delta [\tilde{b}^{(-1)}] \notag \\
  & \times \prod_{s=-1}^{t-1} \frac{\gamma}{\sqrt{2\pi}} \rme^{-\frac{\gamma^2}{2}( \tilde{b}^{(s+1)} - f(\hat{k}^{(s)}+v^{(s)}+(\vec{\Gamma} \tilde{\vec{b}})^{(s)}) )^2} 
    =: \llangle f(\tilde{\vec{b}}) \rrangle, \label{eq:average_over_singleUserMeasure_final}
\end{align}
where $\mathcal{D}\vec{v} := \rmd \vec{v} |2\pi\vec{R}|^{-1/2}e^{-\frac12\vec{v}\cdot\vec{R}^{-1}\vec{v}}$. 
Since $\llangle 1 \rrangle =1$, the effective path measure (\ref{eq:average_over_singleUserMeasure}) in the limit $\gamma \to \infty$ is given by 
\begin{align}
  \lim_{\gamma \to \infty} \langle f(\tilde{\vec{b}}) \rangle_*
  & = \frac{\llangle f(\tilde{\vec{b}}) \rrangle}{\llangle 1 \rrangle} \nonumber \\ 
  & = \llangle f(\tilde{\vec{b}}) \rrangle. \label{eq:finalform_of_effective_path_measure}
\end{align}
Equation (\ref{eq:sp_m}) can be obtained from (\ref{eq:finalform_of_effective_path_measure}), (\ref{eq:spe_m}) and (\ref{eq:relatonship_m}). 
Equation (\ref{eq:sp_C}) is given by (\ref{eq:finalform_of_effective_path_measure}), (\ref{eq:spe_q}) and (\ref{eq:relatonship_C}). 
\par
Next, we consider a function $\tilde{b}^{(s)}\hat{\vec{u}}$ that corresponds to (\ref{eq:spe_L}). 
In a similar way to have (\ref{eq:average_over_singleUserMeasure_final}), one obtain 
\begin{align}
  & \quad \int \underline{\rmd}u \underline{\rmd} \hat{u} 
      \int_{\mathbb{R}^{t+2}} \biggl( \prod_{s=-1}^{t} \rmd \tilde{b}^{(s)} \biggr) 
      w_*(\tilde{\vec{b}},\vec{u},\hat{\vec{u}}) \tilde{b}^{(s)}\hat{\vec{u}} \nonumber \\
  & = \int \biggl( \prod_{s=-1}^{t-1} \frac{\partial(u^{(s)},\hat{u}^{(s)})}{\partial(v^{(s)},w^{(s)})} \frac{\rmd v^{(s)} \rmd w^{(s)}}{2\pi} \biggr) 
    \vec{w} \rme^{\rmi \vec{v} \cdot \vec{w} - \frac 12 \vec{w} \cdot \vec{R} \vec{w}} \nonumber \\
  & \quad \times \int_{\mathbb{R}^{t+2}} \biggl( \prod_{s=-1}^{t} \rmd \tilde{b}^{(s)} \biggr) \tilde{b}^{(s)} \delta [\tilde{b}^{(-1)}] \nonumber \\
  & \quad \times \biggl( \prod_{s=-1}^{t-1} \frac{\gamma}{\sqrt{2\pi}} 
    \rme^{- \frac{\gamma^2}2[\tilde{b}^{(s+1)}-f(\hat{k}^{(s)}+v^{(s)}+(\vec{\Gamma}\tilde{\vec{b}})^{(s)})]^2} \biggr) \nonumber \\
  & = \int {\cal D}\vec{v} \int_{\mathbb{R}^{t+2}} 
    \biggl( \prod_{s=-1}^{t-1} \rmd \tilde{b}^{(s)} \biggr) 
    \rmi \tilde{b}^{(s)} \vec{R}^{-1} \vec{v} \delta [\tilde{b}^{(-1)}] \notag \\
  & \quad \times \prod_{s=-1}^{t-1} \frac{\gamma}{\sqrt{2\pi}} \rme^{-\frac{\gamma^2}{2}( \tilde{b}^{(s+1)} - f(\hat{k}^{(s)}+v^{(s)}+(\vec{\Gamma} \tilde{\vec{b}})^{(s)}) )^2} \nonumber \\
  & = \llangle \rmi \tilde{b}^{(s)} \vec{R}^{-1} \vec{v} \rrangle, 
\end{align}
by using the notation of (\ref{eq:average_over_singleUserMeasure_final}). 
We therefore have 
\begin{align}
  & \quad \int \underline{\rmd}u \underline{\rmd} \hat{u} 
      \int_{\mathbb{R}^{t+2}} \biggl( \prod_{s=-1}^{t} \rmd \tilde{b}^{(s)} \biggr) 
      w_*(\tilde{\vec{b}},\vec{u},\hat{\vec{u}}) \tilde{b}^{(s)}\hat{u}^{(s')} \nonumber \\
  & = \llangle \rmi \tilde{b}^{(s)} (\vec{R}^{-1} \vec{v})^{(s')} \rrangle. \label{eq:special_case}
\end{align}
Equation (\ref{eq:sp_G}) can be obtained from (\ref{eq:special_case}), (\ref{eq:finalform_of_effective_path_measure}), (\ref{eq:spe_L}) and (\ref{eq:relatonship_G}). 
We have then arrived at Proposition \ref{proposition:soft-PIC}. 
\HLE %ADDED.v3(TO HERE)

\section{Derivation of Proposition \ref{proposition:HowToChooseGammaHat}} \label{appendix:HowToChooseGammaHat}
%~~~~~~~~~~~~~~~~~~~~~~~~~~~~~~~~~~~~~~~~~~~~~~~~~~~~~~~~~~~~~~~~~~~~~
\par
The inductive method is applied. 
\par
(i) The case of $t=-1$. 
This is an initial stage, therefore the Onsager reaction term does not exist. 
To cancel the Onsager reaction term, the matrix $\hat{\Gamma} \hat{\vec{b}}$ 
is simply chosen as $(\hat{\Gamma} \hat{\vec{b}})^{(-1)}=0$. 
This is automatically held by the initializaiton $\tilde{b}^{(-1)}=0$. 
The matrix $G$ is $G=0$ because of causality. 
\par
(ii) The case of $t=0$. 
In this case, $G$ is a $1 \times 1$ zero matrix $O$ and $\Gamma$ becomes $\Gamma=(\vec{1}+\beta G)^{-1} \beta G = O$. 
The Onsager reaction term is therefore $\Gamma \hat{\vec{b}}=O$. 
To cancel the Onsager reaction term, we only have to choose matrix $\hat{\Gamma}$ as $(\hat{\Gamma}\hat{\vec{b}})^{(-1)}=0$. 
Matrix $G$ is 
\begin{equation}
  G=\biggl( 
  \begin{array}{cc}
    0 & 0 \\
    G^{(0,-1)} & 0 \\
  \end{array}
  \biggr). 
\end{equation}
until here. 
\par
(iii) The case of stage $t=1$. 
Since $G \in \mathbb{R}^{2 \times 2}$ is a nilpotent matrix, i.e., $G^2=O$, 
then the Onsager reaction term $\Gamma \hat{\vec{b}}$ becomes 
\begin{eqnarray}
  \Gamma \hat{\vec{b}}
  &=& [(\vec{1}+\beta G)^{-1} \beta G] \hat{\vec{b}} \nonumber \\
  &=& \beta G \hat{\vec{b}} \nonumber \\
  &=& \biggl( 
  \begin{array}{cc}
   0 & 0 \\
   \beta G^{(0,-1)} & 0 \\
  \end{array}
  \biggr)
  \left( 
  \begin{array}{c}
   0 \\
   \tilde{b}^{(0)} \\
  \end{array}
  \right)  \nonumber \\
  &=&
  \left( 
  \begin{array}{c}
   0 \\
   0 \\
  \end{array}
  \right). 
\end{eqnarray}
To cancel the Onsager reaction term, one needs $(\hat{\Gamma}\hat{\vec{b}})^{(0)}=0$. 
If $(\hat{\Gamma}\hat{\vec{b}})^{(0)}$ is chosen as this, 
the average $\llangle \tilde{b}^{(1)} \rrangle$ 
$= \int \mathcal{D} \vec{v}$
$\sgn [\hat{k}^{(0)}+v^{(0)}+\theta^{(0)}]$ 
only includes $\theta^{(0)}$ and does not depend on $\theta^{(-1)}$. 
Therefore, $G^{(1,-1)}$ $= \partial \llangle \tilde{b}^{(1)} \rrangle$$/\partial \theta^{(-1)}$ $= 0$ holds. 
The matrix $G$ is 
\begin{equation}
  G=\left( 
  \begin{array}{ccc}
    0 & 0 & 0 \\
    G^{(0,-1)} & 0 & 0\\
    0 & G^{(1,0)} & 0\\
  \end{array}
  \right) 
\end{equation}
until here. 
\par
(iv) The case of stage $t=2$. 
Since $G \in \mathbb{R}^{3 \times 3}$ is a nilpotent matrix, i.e., $G^3=O$, 
then the ORCT $\Gamma \hat{\vec{b}}$ is 
\begin{eqnarray}
  \Gamma \hat{\vec{b}}
  &=& [(\vec{1}+\beta G)^{-1} \beta G] \hat{\vec{b}}\nonumber \\
  &=& [\beta G - (\beta G)^2] \hat{\vec{b}} \nonumber \\
  &=&
  \left( 
  \begin{array}{ccc}
    0 & 0 & 0 \\
    \beta G^{(0,-1)} & 0 & 0\\
    -\beta^2 G^{(1,0)}G^{(0,-1)} & \beta G^{(1,0)} & 0\\
  \end{array}
  \right)
  \left( 
  \begin{array}{c}
    0 \\
    \tilde{b}^{(0)} \\
    \tilde{b}^{(1)} \\
  \end{array}
  \right) \nonumber \\
  &=&
  \left( 
  \begin{array}{c}
    0 \\
    0 \\
    \beta G^{(1,0)} \tilde{b}^{(0)} \\
  \end{array}
  \right). 
\end{eqnarray}
Therefore, if $\hat{\Gamma}\hat{\vec{b}}$ is chosen as 
\begin{equation}
  (\hat{\Gamma}\hat{\vec{b}})^{(1)}=\beta G^{(1,0)} \tilde{b}^{(0)}, 
\end{equation}
then it cancels the Onsager reaction term. 
When we choose $(\hat{\Gamma} \hat{\vec{b}})^{(1)}$ as this, 
the average $\llangle \tilde{b}^{(2)} \rrangle$ 
$= \int \mathcal{D} \vec{v}$
$\sgn [\hat{k}^{(1)}+v^{(1)}+\theta^{(1)}]$ 
only includes $\theta(1)$ and does not depend on $\theta^{(-1)}$ and $\theta(0)$. 
Therefore, $G^{(2,s)}$ $= \partial \llangle \tilde{b}^{(2)} \rrangle$$/\partial \theta^{(-1)}$ $= 0$ holds 
for $s\in\{-1,0\}$. 
The matrix $G$ is 
\begin{equation}
  G=\left( 
  \begin{array}{cccc}
    0          & 0         & 0         & 0 \\
    G^{(0,-1)} & 0         & 0         & 0 \\
    0          & G^{(1,0)} & 0         & 0 \\
    0          & 0         & G^{(2,1)} & 0 \\
  \end{array}
  \right)
\end{equation}
until here. 
\par
(v) The case of stage $t=3$. 
Since $G \in \mathbb{R}^{4 \times 4}$ is a nilpotent matrix, i.e., $G^4=O$, 
then the ORCT $\Gamma \hat{\vec{b}}$ is 
\begin{eqnarray}
  \Gamma \hat{\vec{b}}
  &=&[(\vec{1}+\beta G)^{-1} \beta G] \hat{\vec{b}} \nonumber \\
  &=&[\beta G - (\beta G)^2 + (\beta G)^3] \hat{\vec{b}} \nonumber \\
  &=&
  \left( 
  \begin{array}{c}
    0 \\
    0 \\
    \beta G^{(1,0)} \tilde{b}^{(0)} \\
    \beta G^{(2,1)} \tilde{b}^{(1)} - \beta^2 G^{(2,1)}G^{(1,0)} \tilde{b}^{(0)} \\
  \end{array}
  \right). \nonumber \\
\end{eqnarray}
Therefore, if $\hat{\Gamma}\hat{\vec{b}}$ is chosen as 
\begin{equation}
  (\hat{\Gamma}\hat{\vec{b}})^{(2)}=\beta G^{(2,1)} \tilde{b}^{(1)} - \beta^2 G^{(2,1)}G^{(1,0)} \tilde{b}^{(0)}, 
  \label{app:ORCT_typical_form}
\end{equation}
then it cancels the Onsager reaction term. 
When we choose $(\hat{\Gamma} \hat{\vec{b}})^{(2)}$ as this, 
the average $\llangle \tilde{b}^{(3)} \rrangle$ 
$= \int \mathcal{D} \vec{v}$
$\sgn [\hat{k}^{(2)}+v^{(2)}+\theta^{(2)}]$ 
only includes $\theta(1)$ and does not depend on $\theta^{(-1)}, \cdots, \theta(1)$. 
Therefore, $G^{(3,s)}$ $= \partial \llangle \tilde{b}^{(3)} \rrangle$$/\partial \theta^{(-1)}$ $= 0$ holds for $s\in\{-1,0,1\}$. 
The matrix $G$ is 
\begin{equation}
  G=\left( 
  \begin{array}{ccccc}
    0 & 0 & 0 & 0 & 0 \\
    G^{(0,-1)} & 0 & 0 & 0 & 0 \\
    0 & G^{(1,0)} & 0 & 0 & 0 \\
    0 & 0 & G^{(2,1)} & 0 & 0 \\
    0 & 0 & 0 & G^{(3,2)} & 0 \\
  \end{array}
  \right)
\end{equation}
until here. 
Note that it is expected that 
\begin{eqnarray}
  (\hat{\Gamma}\hat{\vec{b}})^{(t)} 
  &=& \beta G^{(t,t-1)} [ \tilde{b}^{(t-1)} - (\hat{\Gamma}\hat{\vec{b}})^{(t-1)} ] \nonumber \\
  &=& - \sum_{s=0}^{t-1} (-\beta)^{t-s} \biggl( \prod_{\tau=s}^{t-1} G^{(\tau+1,\tau)} \biggr) \tilde{b}^{(s)}
\end{eqnarray}
will holds for any $t \ge 1$ in an analogy with (\ref{app:ORCT_typical_form}). 
\par
(vi) Stage $t \ge 3$ 
\par
We here assume that the Theorem holds for stage $t \in \{-1,0,1,\cdots,t\}$. 
Namely, we assume that $(\hat{\Gamma}\hat{\vec{b}})^{(-1)},\cdots,(\hat{\Gamma}\hat{\vec{b}})^{(t)}$ are given by 
\begin{eqnarray}
  && (\hat{\Gamma}\hat{\vec{b}})^{(-1)} = 0 \nonumber \\
  && (\hat{\Gamma}\hat{\vec{b}})^{(0)} = 0 \nonumber \\
  && (\hat{\Gamma}\hat{\vec{b}})^{(1)} = \displaystyle \beta G^{(1,0)} \tilde{b}^{(0)} \nonumber \\
  && \quad \vdots \nonumber \\
  && (\hat{\Gamma}\hat{\vec{b}})^{(t)} = \displaystyle - \sum_{s=0}^{t-1} (-\beta)^{t-s} 
     \biggl( \prod_{\tau=s}^{t-1} G^{(\tau+1,\tau)} \biggr) \tilde{b}^{(s)}, 
\end{eqnarray}
and $G$ has the following form: 
\begin{equation}
  G=\left( 
  \begin{array}{ccccc}
    0 &   &   &   & 0 \\
    G^{(0,-1)} & 0 &   &   &   \\
      & G^{(1,0)} & 0 &   &   \\
      &   & \ddots & \ddots &   \\
    0 &   &   & G^{(t+1,t)} & 0 \\
  \end{array}
  \right). 
\end{equation}
Since $G \in \mathbb{R}^{(t+3) \times (t+3)}$ is a nilpotent matrix, i.e., $G^{t+3}=O$, 
then Onsager reaction term $\Gamma \hat{\vec{b}} \in \mathbb{R}^{t+3}$ can be calculated as 
\begin{eqnarray}
  \Gamma \hat{\vec{b}}
  &=& [ (\vec{1}+\beta G)^{-1} \beta G ] \hat{\vec{b}} \nonumber \\
  &=& [ \beta G - (\beta G)^2 + \cdots + (-1)^{t+1} (\beta G)^{t+2} ] \hat{\vec{b}} \nonumber \\
  % &=&
  % \left( 
  % \begin{array}{rccc}
  %   0 &   &   & 0 \\
  %   \beta G^{(0,-1)} & 0 &  & 0 \\
  %   \vdots & \ddots & \ddots &  \\
  %   (-1)^{t+1} \beta^{t+2} G^{(t+1,t)} \times \cdots \times G^{(0,-1)} & \cdots & G^{(t+1,t)} & 0 \\
  % \end{array}
  % \right). 
  &=&
  \left( 
  \begin{array}{c}
    0 \\
    0 \\
    \displaystyle \beta G^{(1,0)} \tilde{b}^{(0)} \\
    \beta G^{(2,1)} \tilde{b}^{(1)} - \beta^2 G^{(2,1)}G^{(1,0)} \tilde{b}^{(0)} \\
    \vdots \\
    \displaystyle - \sum_{s=0}^{t-1} (-\beta)^{t-s} \biggl( \prod_{\tau=s}^{t-1} G^{(\tau+1,\tau)} \biggr) \tilde{b}^{(s)} \\
    \displaystyle - \sum_{s=0}^{t} (-\beta)^{t+1-s} \biggl( \prod_{\tau=s}^{t} G^{(\tau+1,\tau)} \biggr) \tilde{b}^{(s)} \\
  \end{array}
  \right). \nonumber \\
\end{eqnarray}
If $\hat{\Gamma}\hat{\vec{b}}$ is chosen as 
\begin{equation}
  (\hat{\Gamma}\hat{\vec{b}})^{(t+1)} 
  = - \sum_{s=0}^{t} (-\beta)^{t+1-s} \biggl( \prod_{\tau=s}^{t} G^{(\tau+1,\tau)} \biggr) \tilde{b}^{(s)}, 
\end{equation}
the Onsager reaction term is cancelled. 
When we choose $(\hat{\Gamma} \hat{\vec{b}})^{(t+1)}$ as this, 
the average $\llangle \tilde{b}^{(t+2)} \rrangle$ 
$= \int \mathcal{D} \vec{v}$
$\sgn [\hat{k}^{(t+1)}+v^{(t+1)}+\theta^{(t+1)}]$ 
only includes $\theta^{(t+1)}$ and does not depend on $\theta^{(-1)},\cdots,\theta^{(t)}$. 
Therefore, $G^{(t+2,s)}$ $= \partial \llangle \tilde{b}^{(3)} \rrangle$$/\partial \theta^{(-1)}$ $= 0$ holds 
for $s\in\{-1,0,1,\cdots,t\}$. 
\par
If the claim holds for stage $s \in \{-1,0,1,\cdots,t\}$, it holds for stage $t+1$. 
This proves Proposition \ref{proposition:HowToChooseGammaHat}.

\section{Derivation of Proposition \ref{proposition:BP-based_detector}} \label{app:GFA_BP-basedDetector}
%~~~~~~~~~~~~~~~~~~~~~~~~~~~~~~~~~~~~~~~~~~~~~~~~~~~~~~~~~~~~~~~~~~~~~
\par
One can evaluate the BP-based detector in the same manner as soft-PIC. 
The summation over all messages, which is the argument of $\tanh(\cdots)$ of (\ref{eq:BP-basedDetector}) becomes 
\begin{align}
  u_k^{(t)} =
  & R^{(t)} \sum_{k'=1}^K W_{kk'} b_{k'} + A^{(t)} \tilde{b}_k^{(t)} \nonumber \\
  & - \sum_{s=-1}^t J^{(t,s)} \sum_{k'=1}^K W_{kk'} \tilde{b}_k^{(s)} 
    + R^{(s)} \frac{\sigma_0}{\sqrt{N}} \sum_{\mu=1}^N s_k^\mu n^\mu \nonumber \\
  & + \theta_k^{(s)}
  \label{eq:u_k_BP-basedDetector2}
\end{align}
Substituting (\ref{eq:SystemModel}) and (\ref{eq:h_k}) into (\ref{eq:def_of_SumOverAllMessage}), 
the averaged generating functional $\bar{Z}[\vec{\psi}]$ is represented as 
\begin{align}
  \bar{Z}[\vec{\psi}] 
  =& \mathbb{E}_{\vec{s}_1,\cdots,\vec{s}_K,\vec{n}} \biggl[ \int_{\mathbb{R}^{(t+1)K}} \underline{\rmd}\vec{u} \nonumber \\
  & \times 
    p[\tilde{\vec{b}}^{(-1)}] 
    \biggl( \prod_{s=-1}^{t-1} \frac{\gamma}{\sqrt{2\pi}} 
    e^{-\frac{\gamma^2}2 \{ \tilde{b}_k^{(s+1)}-f(u_k^{(s)}) \}^2} \biggr) 
    \nonumber \\
  & \times \exp \biggl[ -\rmi\sum_{s=-1}^t \tilde{\vec{b}}^{(s)} \cdot \vec{\psi}^{(s)} \biggr]
    \nonumber \\
  & \times \prod_{s=0}^{t-1} \prod_{k=1}^K 
    \delta \biggl( u_k^{(s)} - R^{(t)} \sum_{k'=1}^K W_{kk'} b_{k'} - A^{(t)} \tilde{b}_k^{(t)} \nonumber \\
  & + \sum_{s=-1}^t J^{(t,s)} \sum_{k'=1}^K W_{kk'} \tilde{b}_k^{(s)}  - R^{(s)} 
    \frac{\sigma_0}{\sqrt{N}} \sum_{\mu=1}^N s_k^\mu n^\mu 
    \nonumber \\
  & - \theta_k^{(s)} \biggr) \biggr] , 
  \label{eq:def_barZ_BP-basedDetector}
\end{align}
where $\underline{\rmd}\vec{u}$$:=$$\prod_{s=0}^{t-1} \prod_{k=1}^K \frac{\rmd u_k^{(s)}}{\sqrt{2\pi}}$. 
Without loss of generality we can put $b_k=1 \; (\forall k)$. 
We apply the Fourier integral form of Dirac's delta function, which gives 
\begin{align}
  \bar{Z}[\vec{\psi}] =
  & \sum_{\hat{\vec{b}}^{(0)},\cdots,\hat{\vec{b}}^{(t)}} p[\hat{\vec{b}}(0)] 
    \int_{\mathbb{R}^{2(t+1)K}} \underline{\rmd}\vec{u}\underline{\rmd}\hat{\vec{u}} \nonumber \\
  & \times \exp \biggl[ \rmi \sum_{s=0}^{t-1} \sum_{k=1}^K 
    \hat{u}_k^{(s)} \{ u_k^{(s)} - A^{(s)} \tilde{b}_k^{(s)} - \theta_k^{(s)} \} \nonumber \\
  & + \sum_{s=0}^t \sum_{k=1}^K \{ \ln \frac{\gamma}{\sqrt{2\pi}} 
    - \frac{\gamma^2}2 [ \tilde{b}_k^{(t+1)}-f(u_k^{(s)}) ]^2 \} \biggr] \nonumber \\
  & \times \mathbb{E}_{ \vec{s}_1, \cdots, \vec{s}_K, \vec{n} } \biggl\{ \exp \biggl[ \nonumber \\
  & - \rmi \sqrt{\beta} \sigma_0 \sum_{\mu=1}^N \sum_{s=0}^{t-1} R^{(s)} 
    \biggl( \frac1{\sqrt{K}} \sum_{k=1}^K s_k^\mu  \hat{u}_k^{(s)} \biggr) n^\mu \nonumber \\
  & - \rmi \beta \sum_{\mu=1}^N \sum_{s=0}^{t-1} \biggl( \frac1{\sqrt{K}} 
    \sum_{k=1}^K s_k^\mu  \hat{u}_k^{(s)} \biggr) \nonumber \\
  & \times \biggl( \frac1{\sqrt{K}} \sum_{k'=1}^K s_{k'}^\mu \biggl\{ R^{(s)} 
    - \sum_{s'=-1}^s J^{(s,s')} \tilde{b}_k^{(s')} \biggr\} \biggr) \biggr] \biggr\} , 
  \label{eq:def_barZ_BP-basedDetector2}
\end{align}
where $\underline{\rmd}\hat{\vec{u}}$$:=$$\prod_{s=-1}^{t-1} \prod_{k=1}^K \frac{\rmd \hat{u}_k^{(s)}}{\sqrt{2\pi}}$. 
To take the average of $\vec{s}_1, \cdots, \vec{s}_K$, we introduce the following variables: 
\begin{align}
  & v_k^{(s)} := \frac1{\sqrt{K}} \sum_{k=1}^K s_k^\mu \hat{u}_k^{(s)}, \\
  & w_k^{(s)} := \frac1{\sqrt{K}} \sum_{k=1}^K s_k^\mu \biggl\{ R^{(s)} - \sum_{s'=-1}^s J^{(s,s')} \tilde{b}_k^{(s')} \biggr\}. 
\end{align}
It should be noted that the averaged generating functional $\bar{Z}[\vec{\psi}]$ 
of (\ref{eq:def_barZ}) only includes $\vec{s}_1, \cdots, \vec{s}_K$ 
in terms of these variables $v_\mu^{(s)}$ and $w_\mu^{(s)}$. 
Due to this, the random variables $\vec{s}_1, \cdots, \vec{s}_K$ are isolated 
and their average is able to be taken. 
Introducing $v_\mu^{(s)}$ and $w_\mu^{(s)}$ 
into the term $\mathbb{E}_{ \vec{s}_1, \cdots, \vec{s}_K, \vec{n} } \{\cdots\}$ in (\ref{eq:def_barZ}), 
one obtains 
\begin{align}
  & \int_{\mathbb{R}^{2tN}} \underline{\rmd}\vec{v}\underline{\rmd}\vec{w} 
    \mathbb{E}_{ \vec{s}_1, \cdots, \vec{s}_K, \vec{n} } \biggl\{ \nonumber \\
  & \exp \biggl[ 
      - \rmi \sqrt{\beta} \sigma_0 \sum_{\mu=1}^N \sum_{s=-1}^{t-1} R^{(s)} v_\mu^{(s)} n^\mu 
      - \rmi \beta \sum_{\mu=1}^N \sum_{s=-1}^{t-1} v_\mu^{(s)} w_\mu^{(s)} 
    \biggr] \biggr\} \nonumber \\
  & \times \delta \biggl( v_k^{(s)} - \frac1{\sqrt{K}} \sum_{k=1}^K s_k^\mu \hat{u}_k^{(s)} \biggr) \nonumber \\
  & \times \delta \biggl( w_k^{(s)} -\frac1{\sqrt{K}} \sum_{k=1}^K s_k^\mu \biggl\{ R^{(s)} 
    - \sum_{s'=-1}^s J^{(s,s')} \tilde{b}_k^{(s')} \biggr\} \biggr) \nonumber \\
  =
  & \int_{\mathbb{R}^{4tN}} \underline{\rmd}\vec{v}\underline{\rmd}\hat{\vec{v}}\underline{\rmd}\vec{w}\underline{\rmd}\hat{\vec{w}} \nonumber \\
  & \times \exp \biggl[ \rmi \sum_{\mu=1}^N \sum_{s=-1}^{t-1} \{ \hat{v}_\mu^{(s)} v_\mu^{(s)} 
    + \hat{w}_\mu^{(s)} w_\mu^{(s)} - \beta v_\mu^{(s)} w_\mu^{(s)} \} \biggr] \nonumber \\
  & \times \mathbb{E}_{ \vec{n} } \biggl\{ \exp \biggl[ 
    - \rmi \sqrt{\beta} \sigma_0 \sum_{\mu=1}^N \sum_{s=-1}^{t-1} R^{(s)} v_\mu^{(s)} n^\mu \biggr] \biggr\} \nonumber \\
  & \times \mathbb{E}_{ \vec{s}_1, \cdots, \vec{s}_K } \biggl\{ \exp \biggl[ 
      - \rmi \frac1{\sqrt{K}} \sum_{\mu=1}^N \sum_{s=-1}^{t-1} \biggl(
        \hat{v}_\mu^{(s)} \sum_{k=1}^K s_k^\mu \hat{u}_k^{(s)} \nonumber \\
  &   + \hat{w}_\mu^{(s)} \sum_{k=1}^K s_k^\mu ( R^{(s)} - \sum_{s'=-1}^s J^{(s,s')} \tilde{b}_k^{(s)}) 
    \biggr) \biggr] \biggr\}, 
  \label{eq:DisorderedAverage_BP-basedDetector}
\end{align}
where $\underline{\rmd}\vec{v} := \prod_{\mu=1}^{N} \prod_{s=-1}^{t-1} \frac{\rmd v_\mu^{(s)}}{\sqrt{2\pi}}$, 
$\underline{\rmd}\hat{\vec{v}} := \prod_{\mu=1}^{N} \prod_{s=-1}^{t-1} \frac{\rmd \hat{v}_\mu^{(s)}}{\sqrt{2\pi}}$, 
$\underline{\rmd}\vec{w} := \prod_{\mu=1}^{N} \prod_{s=-1}^{t-1} \frac{\rmd w_\mu^{(s)}}{\sqrt{2\pi}}$ 
and $\underline{\rmd}\hat{\vec{w}} := \prod_{\mu=1}^{N} \prod_{s=-1}^{t-1} \frac{\rmd \hat{w}_\mu^{(s)}}{\sqrt{2\pi}}$. 
We here again use the Fourier integral form of Dirac's delta. 
Term $\mathbb{E}_{ \vec{n} } \{\cdots\}$ in (\ref{eq:DisorderedAverage_BP-basedDetector}) is given by 
\begin{align}
  & \mathbb{E}_{ \vec{n} } \biggl\{ \exp \biggl[ - \rmi \sqrt{\beta} \sigma_0 
      \sum_{\mu=1}^N \sum_{s=0}^{t-1} R^{(s)} v_\mu^{(s)} n^\mu \biggr] \biggr\} \nonumber \\
  & = \prod_{\mu=1}^N \exp \biggl[ -\frac12 \beta \sigma_0^2 
      \sum_{s=0}^{t-1} \sum_{s'=0}^{t-1} R^{(s)} R^{(s')} v_\mu^{(s)} v_\mu^{(s')} \biggr]. 
\end{align}
Term $\mathbb{E}_{ \vec{s}_1, \cdots, \vec{s}_K } \{\cdots\}$ 
in (\ref{eq:DisorderedAverage_BP-basedDetector}) becomes 
\begin{align}
  & \mathbb{E}_{ \vec{s}_1, \cdots, \vec{s}_K } \biggl\{ \exp \biggl[ - \rmi \frac1{\sqrt{K}} \sum_{\mu=1}^N \sum_{s=0}^{t-1} 
    \biggl\{ \hat{v}_\mu^{(s)} \sum_{k=1}^K s_k^\mu \hat{u}_k^{(s)} \nonumber \\
  & + \hat{w}_\mu^{(s)} \sum_{k=1}^K s_k^\mu (R^{(s)} - \sum_{s'=-1}^s J^{(s,s')} \tilde{b}_k^{(s')}) 
    \biggr\} \biggr] \biggr\} \nonumber \\
  =
  & \prod_{\mu=1}^N \prod_{k=1}^K \exp \biggl[ - \frac12 
    \biggl( \frac1{\sqrt{K}} \sum_{s=0}^{t-1} \biggl\{ \hat{v}_\mu^{(s)} \hat{u}_k^{(s)} \nonumber \\
  & + \hat{w}_\mu^{(s)} (R^{(s)} - \sum_{s'=-1}^{s} J^{(s,s')} \tilde{b}_k^{(s')}) \biggr\} \biggr)^2 \biggr] \\
  =
  & \prod_{\mu=1}^N \exp \biggl[ - \frac12 \sum_{s=0}^{t-1} \sum_{s'=0}^{t-1} 
    \biggl\{ \hat{v}_\mu^{(s)} \biggl( \frac 1K \sum_{k=1}^K \hat{u}_k^{(s)} \hat{u}_k^{(s')} \biggr) \hat{v}_\mu(s') \nonumber \\
  & + \hat{v}_\mu^{(s)} \biggl( R^{(s')} \biggl[ \frac 1K \sum_{k=1}^K \hat{u}_k^{(s)} \biggr] \nonumber \\
  & - \sum_{\tau'=-1}^{s'} J^{(s',\tau')} \biggl[ \frac 1K \sum_{k=1}^K 
    \tilde{b}_k^{(s')} \hat{u}_k^{(s)} \biggr] \biggr) \hat{w}_\mu(s') \nonumber \\
  & + \hat{w}_\mu^{(s)} \biggl( R^{(s)} \biggl[ \frac 1K \sum_{k=1}^K \hat{u}_k^{(s')} \biggr] \nonumber \\
  & - \sum_{\tau=-1}^s J^{(s,\tau)} \biggl[ \frac 1K \sum_{k=1}^K 
    \tilde{b}_k^{(s)} \hat{u}_k^{(s')} \biggr] \biggr) \hat{v}_\mu(s') \nonumber \\
  & + \hat{w}_\mu^{(s)} \biggl( R^{(s)} R^{(s')} - R^{(s)} \sum_{\tau'=-1}^{s'} 
    J^{(s',\tau')} \biggl[ \frac 1K \sum_{k=1}^K \tilde{b}_k^{(s)} \biggr] \nonumber \\
  & - R^{(s')} \sum_{\tau=-1}^s J^{(s,\tau)} \biggl[ \frac 1K \sum_{k=1}^K 
    \tilde{b}_k^{(s')} \biggr] \nonumber \\
  & + \sum_{\tau=-1}^s \sum_{\tau'=-1}^{s'} J^{(s,\tau)} J^{(s',\tau')} 
    \biggl[ \frac 1K \sum_{k=1}^K \tilde{b}_k^{(s)} \tilde{b}_k^{(s')} \biggr] \biggr) \hat{v}_\mu(s') \biggr\} \biggr] . 
  \label{eq:SpreadSequenceAverage_BP-basedDetector}
\end{align}
for finite $t$. 
The relevant one-stage and two-stage values, which are equal to those of soft-PIC's analysis, 
appeared in (\ref{eq:SpreadSequenceAverage_BP-basedDetector}) are separated 
by introducing (\ref{eq:def_mu}) -- (\ref{eq:def_L}). 
The term of $\delta$ in (\ref{eq:SpreadSequenceAverage}) is again rewritten 
by applying the Fourier integral form of Dirac's delta function 
as well as the case of soft-PIC's analysis. 
\par
Since the initial probability, which is given 
as $p[\hat{\vec{b}}^{(-1)}] := \prod_{k=1}^K \delta(\tilde{b}_k^{(-1)})$, is factorized, 
the averaged generating functional $\bar{Z}[\vec{\psi}]$ factorizes into single-user contributions. 
The averaged generating functional is therefore simplified to 
$\bar{Z}[\vec{\psi}] = \int 
\rmd \vec{\eta} \rmd \hat{\vec{\eta}}
\rmd \vec{k} \rmd \hat{\vec{k}}
\rmd \vec{q} \rmd \hat{\vec{q}}
\rmd \vec{Q} \rmd \hat{\vec{Q}}
\rmd \vec{L} \rmd \hat{\vec{L}}
\exp [ K(\Phi+\Psi+\Omega)+O(\ln K) ]$, 
in which functions $\Phi$, $\Psi$, $\Omega$ are given by 
\begin{align}
  \Phi :=
  & \rmi \sum_{s=-1}^{t-1} \{ \hat{\eta}^{(s)} \eta^{(s)} + \hat{k}^{(s)}k^{(s)} \} \nonumber \\
  & + \rmi \sum_{s=-1}^{t-1} \sum_{s'=-1}^{t-1} 
    \{   \hat{q}^{(s,s')} q^{(s,s')} \nonumber \\
  &    + \hat{Q}^{(s,s')} Q^{(s,s')} 
       + \hat{L}^{(s,s')} L^{(s,s')} 
    \}, 
\end{align}
\begin{align}
  \Psi :=
  & \frac 1K \sum_{k=1}^K \ln \biggl\{ \int_{\mathbb{R}^{t+2}} 
    \biggl( \prod_{s=-1}^{t-1} \rmd \tilde{b}^{(s)} \biggr) p[\tilde{b}^{(-1)}] \int \underline{\rmd}u \underline{\rmd}\hat{u} \nonumber \\
  & \times \exp \biggl[ \sum_{s=-1}^{t-1} 
    \{ \ln \frac{\gamma}{\sqrt{2\pi}} - \frac{\gamma^2}2 [\tilde{b}^{(s+1)}-f(u^{(s)})]^2 \} \nonumber \\
  & - \rmi \sum_{s=-1}^{t-1} \sum_{s'=-1}^{t-1} \{ \hat{q}^{(s,s')} \tilde{b}^{(s)} \tilde{b}^{(s')} \nonumber \\
  & + \hat{Q}^{(s,s')} \hat{u}^{(s)} \hat{u}^{(s')} + \hat{L}^{(s,s')} \tilde{b}^{(s)} \hat{u}^{(s')} \} \nonumber \\
  & + \rmi \sum_{s=-1}^{t-1} \hat{u}^{(s)} \{ u^{(s)} - A^{(s)} \tilde{b}^{(s)} - \theta_k^{(s)} - \hat{k}^{(s)} \} \nonumber \\
  & - \rmi \sum_{s=-1}^{t-1} \tilde{b}^{(s)} \hat{\eta}^{(s)} - \rmi \sum_{s=-1}^t \tilde{b}^{(s)} \psi_k^{(s)} \biggr], 
\end{align}
\begin{align}
  \Omega :=
  & \frac 1K \ln \int \underline{\rmd}\vec{v}\underline{\rmd}\hat{\vec{v}}\underline{\rmd}\vec{w}\underline{\rmd}\hat{\vec{w}} \nonumber \\
  & \times \exp \biggl[ \rmi \sum_{\mu=1}^N \sum_{s=-1}^{t-1} 
     \{ 
         \hat{v}_\mu^{(s)} v_\mu^{(s)} + \hat{w}_\mu^{(s)} w_\mu^{(s)} 
       - \beta v_\mu^{(s)} w_\mu^{(s)} 
     \} \nonumber \\
  & - \frac 12 \sum_{\mu=1}^N \sum_{s=-1}^{t-1} \sum_{s'=-1}^{t-1} 
    \biggl\{ 
        \beta \sigma_0^2 R^{(s)} R^{(s')} v_\mu^{(s)} v_\mu^{(s')} \nonumber \\
  &   + \hat{v}_\mu^{(s)} Q^{(s,s')} \hat{v}_\mu^{(s')} \nonumber \\
  &   + \hat{v}_\mu^{(s)} \biggl( R^{(s')} k^{(s)} - \sum_{\tau'=-1}^{s'} J^{(s', \tau')} L^{(s',s)} \biggr) \hat{w}_\mu^{(s')} \nonumber \\
  &   + \hat{w}_\mu^{(s)} \biggl( R^{(s)} k^{(s')} - \sum_{\tau =-1}^{s } J^{(s, \tau)}   L^{(s,s')} \biggr) \hat{v}_\mu^{(s')} \nonumber \\
  &   + \hat{w}_\mu^{(s)} \biggl( R^{(s)} R^{(s')} \nonumber \\
  &   - R^{(s)}  \sum_{\tau'=-1}^{s'} J^{(s', \tau')} \eta^{(\tau')} 
      - R^{(s')} \sum_{\tau =-1}^{s } J^{(s , \tau )} \eta^{(\tau )} \nonumber \\
  &   + \sum_{\tau=-1}^{s} \sum_{\tau'=-1}^{s'} J^{(s, \tau)} J^{(s', \tau')} q(\tau,\tau') \biggl) 
    \hat{w}_\mu^{(s')} \biggr\} \biggr] .
\end{align}
% where $\underline{\rmd}u := \prod_{s=0}^{t-1} \frac{du^{(s)}}{\sqrt{2\pi}}$ 
% $\underline{\rmd} \hat{u} := \prod_{s=0}^{t-1} \frac{d\hat{u}^{(s)}}{\sqrt{2\pi}}$, 
% $\rmd\vec{\eta} := \prod_{s=0}^{t-1} d\eta^{(s)}$, 
% $\rmd\hat{\vec{\eta}} := \prod_{s=0}^{t-1} d\hat{\eta}^{(s)}$, 
% $\rmd\vec{k} := \prod_{s=0}^{t-1} dk^{(s)}$, 
% $\rmd\hat{\vec{k}} := \prod_{s=0}^{t-1} d\hat{k}^{(s)}$, 
% $\rmd\vec{q} := \prod_{s=0}^{t-1} \prod_{s'=0}^{t-1} dq^{(s,s')}$, 
% $\rmd\hat{\vec{q}} := \prod_{s=0}^{t-1} \prod_{s'=0}^{t-1} d\hat{q}^{(s,s')}$, 
% $\rmd\vec{Q} := \prod_{s=0}^{t-1} \prod_{s'=0}^{t-1} dQ^{(s,s')}$, 
% $\rmd\hat{\vec{Q}} := \prod_{s=0}^{t-1} \prod_{s'=0}^{t-1} d\hat{Q}^{(s,s')}$, 
% $\rmd\vec{L} := \prod_{s=0}^{t-1} \prod_{s'=0}^{t-1} dL^{(s,s')}$ 
% and $\rmd\hat{\vec{L}} := \prod_{s=0}^{t-1} \prod_{s'=0}^{t-1} d\hat{L}^{(s,s')}$. 
\par
One can deduce the meaning of macroscopic parameters 
by differentiation of the averaged generating functional $\bar{Z}[\vec{\psi}]$ 
with respect to $\theta_k^{(s)}$ and $\psi_k^{(s)}$. 
The averaged generating functional $\bar{Z}[\vec{\psi}]$ is dominated by a saddle-point for $K\to\infty$. 
\par
Applying the same way as the derivation of soft-PIC, 
we again obtain (\ref{eq:other_field_dervatives1}) -- (\ref{eq:other_field_dervatives3}), 
%\begin{eqnarray}
%  & & \overline{\langle \tilde{b}_k^{(s)}\rangle}
%      = \langle \tilde{b}^{(s)} \rangle_k, 
%      \label{eq:other_field_dervatives1_BP-basedDetector} \\
%  & & \overline{\langle\tilde{b}_k^{(s)}\tilde{b}_{k'}^{(s')}\rangle} 
%      = \delta_{k,k'} \langle \tilde{b}^{(s)}\tilde{b}^{(s')}\rangle_k \nonumber \\
%  & & \qquad \qquad \qquad     + (1-\delta_{k,k'}) \langle \tilde{b}^{(s)}\rangle_k \langle \tilde{b}^{(s')}\rangle_k, 
%      \label{eq:other_field_dervatives2_BP-basedDetector} \\
%  & & \frac{\partial \overline{\langle\tilde{b}_k^{(s)}\rangle}}{\partial \theta_{k'}^{(s')}} 
%      = - \rmi \delta_{k,k'} \langle \tilde{b}^{(s)}\hat{u}^{(s')}\rangle_k, 
%      \label{eq:other_field_dervatives3_BP-basedDetector}
%\end{eqnarray}
%where $\langle \; \rangle_k$ denotes average as 
%\begin{equation}
%  \langle f(\{\tilde{b},u,\hat{u}\}) \rangle_k := 
%  \frac
%  {\displaystyle \sum_{\tilde{b}^{(0)},\cdots,\tilde{b}^{(t)}} \int \underline{\rmd}u \underline{\rmd} \hat{u} w_k(\{\tilde{b},u,\hat{u}\}) f(\{\tilde{b},u,\hat{u}\})}
%  {\displaystyle \sum_{\tilde{b}^{(0)},\cdots,\tilde{b}^{(t)}} \int \underline{\rmd}u \underline{\rmd} \hat{u} w_k(\{\tilde{b},u,\hat{u}\})}
%\end{equation}
with the average (\ref{eq:average_over_singleUserMeasure}) $\langle \; \rangle_k$ which has the single-user measure 
\begin{eqnarray}
  & & w_k(\{\tilde{b},u,\hat{u}\}) \nonumber \\
  & & := \delta [\tilde{b}^{(-1)}] \exp \biggl[ \sum_{s=-1}^{t-1} 
      \{ \ln \frac{\gamma}{\sqrt{2\pi}} - \frac{\gamma^2}2 [\tilde{b}^{(s+1)}-f(u^{(s)})]^2 \} \nonumber \\
  & & - \rmi \sum_{s=0}^{t-1} \sum_{s'=0}^{t-1} \{ \hat{q}^{(s,s')} \tilde{b}^{(s)} \tilde{b}^{(s')} 
      + \hat{Q}^{(s,s')} \hat{u}^{(s)} \hat{u}^{(s')} \nonumber \\
  & & + \hat{L}^{(s,s')} \tilde{b}^{(s)} \hat{u}^{(s')} \} \nonumber \\
  & & + \rmi \sum_{s=0}^{t-1} \hat{u}^{(s)} \{ u^{(s)} - A^{(s)} \tilde{b}^{(s)} - \theta_k^{(s)} - \hat{k}^{(s)} \} \nonumber \\
  & & - \rmi \sum_{s=0}^{t-1} \tilde{b}^{(s)} \hat{\eta}^{(s)} \biggr] \biggl. \biggr|_{\mathrm{saddle}}, 
  \label{eq:def_Single-UserMeasure_BP-basedDetector}
\end{eqnarray}
instead of the single-use measure (\ref{eq:average_over_singleUserMeasure}) for soft-PIC. 
%Note that these are identical to those in soft-PIC's analysis 
%except for the single-user measure (\ref{eq:def_Single-UserMeasure_BP-basedDetector}). 
\par
The integral in the averaged generating functional $\bar{Z}[\vec{\psi}]$ will be evaluated 
by the dominating saddle-point of the exponent $\Phi+\Psi+\Omega$ in the large system limit $K\to\infty$. 
% We next derive the saddle-point equations by differentiation with respect to integral variables 
% $\eta^{(s)}$, $\hat{\eta}^{(s)}$, 
% $k^{(s)}$, $\hat{k}^{(s)}$, 
% $q^{(s,s')}$, $\hat{q}^{(s,s')}$, 
% $Q^{(s,s')}$, $\hat{Q}^{(s,s')}$, 
% $L^{(s,s')}$ and $\hat{L}^{(s,s')}$. 
% These saddle-point equations will involve the average overlap $m^{(s)}$ (which measures the bit error rate), 
% the average single-user correlation $C^{(s,s')}$ and the average single-user response function $G^{(s,s')}$. 
Using the identities (\ref{eq:field_dervatives1}) -- (\ref{eq:field_dervatives1}) and 
(\ref{eq:other_field_dervatives1}) -- (\ref{eq:other_field_dervatives3}) with (\ref{eq:def_Single-UserMeasure_BP-basedDetector}), 
the differentiation of $\Phi+\Psi+\Omega$ with respect to 
$\eta^{(s)}$, $\hat{\eta}^{(s)}$, 
$k^{(s)}$, $\hat{k}^{(s)}$, 
$q^{(s,s')}$, $\hat{q}^{(s,s')}$, 
$Q^{(s,s')}$, $\hat{Q}^{(s,s')}$, 
$L^{(s,s')}$, and $\hat{L}^{(s,s')}$ 
leads us to the following saddle-point equations (\ref{eq:spe_mhat}) -- (\ref{eq:spe_L}) with (\ref{eq:def_Single-UserMeasure_BP-basedDetector}), respectively. 
Comparing these saddle-point equations 
with (\ref{eq:def_average_overlap}) -- (\ref{eq:def_average_SU_response_function}), 
we again find the following relationships: 
$\eta^{(s)} = m^{(s)}$, $q^{(s,s')} = C^{(s,s')}$, and $L^{(s,s')} = \rmi G^{(s,s')}$. 
\par
The integral in $\Omega$ with respect to $\hat{\vec{v}}$ and $\hat{\vec{w}}$ is given by 
$\Omega =
-\frac 1{2\beta} \{ \ln |\hat{\vec{D}}| 
+ \ln |\vec{Q}+ (\beta^{-1} \vec{1} - \hat{\vec{B}})^\top 
\hat{\vec{D}}^{-1} (\beta^{-1} \vec{1} - \hat{\vec{B}})| \}$, 
where matrices $\hat{\vec{B}}$ and $\hat{\vec{D}}$ whose $(s,s')$ elements are given by 
\begin{align}
  \hat{B}^{(s,s')} := 
  & -\rmi R^{(s)} k^{(s)} - \sum_{\tau=-1}^s J^{(s,\tau)} G^{(\tau,s)} , \\
  \hat{D}^{(s,s')} := 
  & \frac{\sigma_0^2}\beta R^{(s)} R^{(s')} + R^{(s)} R^{(s')} \nonumber \\
  & -R^{(s )} \sum_{\tau'=-1}^{s'} J^{(s', \tau')} m^{(\tau')} \nonumber \\
  & -R^{(s')} \sum_{\tau =-1}^{s } J^{(s , \tau )} m^{(\tau )} \nonumber \\
  & + \sum_{\tau =-1}^{s } \sum_{\tau'=-1}^{s'} J^{(s , \tau )} J^{(s', \tau')} C^{(\tau,\tau')}, 
\end{align}
respectively. 
\par
The saddle-point equations including $\Omega$ are evaluated as follows. 
One finds $\hat{\eta}^{(s)} =0$ and $\hat{q}^{(s)} =0$. 
We put 
\begin{align}
  \hat{\vec{U}} := \beta^{-1} \vec{1} + \hat{\vec{B}}, 
\end{align}
for light notations. 
Noting that $\hat{\vec{U}}$ includes $\vec{G}$ via $\hat{\vec{B}}$ 
and using the cofactor expansion of $|\hat{\vec{U}}|$, we find 
\begin{align}
  \hat{L}^{(s,s')} 
  =& \frac{\partial \Omega}{\partial G^{(s,s')}} \nonumber \\
  =& -\frac 1\beta \frac\partial{\partial G^{(s,s')}} \ln |U| \nonumber \\
  =& -\frac 1{\beta|U|} \frac{\partial}{\partial G^{(s,s')}} 
     \sum_{s=-1}^{t-1} U^{(s,s')} \tilde{U}^{(s,s')} \nonumber \\
  =& -\frac 1{\beta |U|} \sum_{s=-1}^{t-1} J^{(s,s')} \tilde{U}^{(s,s')}, 
\end{align}
where $\tilde{\hat{U}}^{(s,s')}$ denotes a cofactor of the $(s,s')$ element of the matrix $\hat{\vec{U}}$, 
which does not include $G^{(s,s')}$. % cofactor = YO-INSHI
The $(t+1) \times (t+1)$ matrix $\hat{\vec{L}}=(\hat{L}^{(s,s')})$ then becomes 
\begin{align}
  \hat{\vec{L}}
  =& -\frac 1{\beta |\hat{\vec{U}}|} \biggl[ \nonumber \\
   & \left(
     \begin{array}{ccc}
       J^{(-1,-1)} \tilde{\hat{U}}^{(-1,-1)} & \cdots & J^{(-1,-1)} \tilde{\hat{U}}^{(-1,t-1)} \\
       J^{(0,0)} \tilde{\hat{U}}^{(0,-1)} & \cdots & J^{(0,0)} \tilde{\hat{U}}^{(0,t-1)} \\
       \vdots & & \vdots \\
       J^{(t-1,t-1)} \tilde{\hat{U}}^{(t-1,-1)} & \cdots & J^{(t-1,t-1)} \tilde{\hat{U}}^{(t-1,t-1)} \\
     \end{array}
     \right) \nonumber \\
   & + \cdots + \nonumber \\
   & \left(
     \begin{array}{ccc}
       J^{(t-1,t-1)} \tilde{\hat{U}}^{(t-1,-1)} & \cdots & J^{(t-1,t-1)} \tilde{\hat{U}}^{(t-1,t-1)} \\
       0 & \cdots & 0 \\
       \vdots & & \vdots \\
       0 & \cdots & 0 \\
     \end{array}
     \right) \biggr] \nonumber \\
  =& -\frac 1{\beta |\hat{\vec{U}}|} \biggl[ 
     \left(
     \begin{array}{ccc}
       J^{(-1,-1)} & \cdots & J^{(-1,-1)} \\
       J^{(0,0)} & \cdots & J^{(0,0)} \\
       \vdots & & \vdots \\
       J^{(t-1,t-1)} & \cdots & J^{(t-1,t-1)} \\
     \end{array}
     \right) 
     \otimes \vec{\Delta} \mathrm{adj} \hat{\vec{U}^\top} \nonumber \\
   & + \cdots \nonumber \\
   & + \left(
     \begin{array}{ccc}
       J^{(t-1,t-1)} & \cdots & J^{(t-1,t-1)} \\
       0 & \cdots & 0 \\
       \vdots & & \vdots \\
       0 & \cdots & 0 \\
     \end{array}
     \right) 
     \otimes \vec{\Delta}^t \mathrm{adj} \hat{\vec{U}^\top} 
     \biggr] \nonumber \\
  =& - \vec{J}_0 \otimes (\vec{U}^\top)^{-1} 
     - \vec{J}_1 \otimes \vec{\Delta} (\vec{U}^\top)^{-1} \nonumber \\
   & - \cdots 
     - \vec{J}_t \otimes \vec{\Delta}^t (\vec{U}^\top)^{-1} ,
\end{align}
where $\vec{U} := \beta \hat{\vec{U}}$ and $\mathrm{adj} \vec{A}$ denotes an adjoint matrix of $\vec{A}$. 
The definition of $\vec{J}_s$ is given by (\ref{eq:def_Js}). 
\par
Due to $\vec{Q}=\vec{O}$, $\Omega$ can be expanded 
as $\ln |\vec{A}+\vec{Q}|={\rm Tr} \ln \vec{A} + {\rm tr}\, \vec{A}^{-1} \vec{Q}$. 
The $\hat{Q}^{(s,s')}$ is obtained as 
$\hat{Q}^{(s,s')} = - \frac \rmi{2\beta} M^{(s',s)}$, 
and we then have 
$\hat{\vec{Q}} 
= -\rmi \frac 1{2\beta} M^\top 
= -\rmi \frac 12 (\vec{1}+\beta G)^{-1} D (\vec{1}+\beta G^\top)^{-1}$, 
where $\vec{M} := (\beta^{-1} \vec{1} + \vec{G})^{-1} \hat{\vec{D}} (\beta^{-1} \vec{1} + \vec{G}^\top)^{-1}$ 
and $D^{(s,s')} := \beta \hat{D}$. 
Noting that $|\vec{1} - \beta \hat{\vec{B}}^\top|$ only contains $k^{(s)}$ in a single row, 
$k^{(s)}$ is obtained as $\hat{k}^{(s)} = |\Lambda_{[s]}|$, 
where $\vec{B} := \hat{\vec{B}}|_{\vec{k}=\vec{0}} =- \sum_{\tau=-1}^s J^{(s,\tau)} G^{(\tau,s)}$ 
and the definition of $\Lambda_{[s]}$ is given by (\ref{eq:def_Lambda}). 
\par
Some macroscopic parameters are found to vanish in the saddle-point: $k^{(s)}=Q^{(s,s')}=0$. 
In the case of the BP-base detector, the remaining macroscopic parameters can all be expressed in terms of three observables, 
i.e., 
the average overlap $m^{(s)}$, 
the average single-user correlation $C^{(s,s')}$, 
and the average single-user response function $G^{(s,s')}$, 
which are defined by (\ref{eq:def_average_overlap}) -- (\ref{eq:def_average_SU_response_function}). 
The averaged generating functional $\bar{Z}[\vec{\psi}]$ is dominated by a saddle-point in the large system limit. 
We then arrive at Proposition \ref{proposition:BP-based_detector}.

%=====================================================================
\section*{Acknowledgment}
\par
The authors would also like to thank the associate editor, anonymous reviewers and Arian Maleki for their valuable comments. 
This work was partially supported by Grants-in-Aid 
for Scientific Research on Priority Areas No. 14084212, 
for Scientific Research (B) No. 25289114 and ,
for Scientific Research (C) Nos. 16500093, 22500136 and 25330264,
and for Encouragement of Young Scientists (B) No. 18700230 
from the Ministry of Education, Culture, Sports, Science and Technology of Japan (MEXT).

\ifCLASSOPTIONcaptionsoff
  \newpage
\fi

\end{document}